\documentclass[a4paper,11pt]{article}
\pdfoutput=1 

\usepackage{jcappub} 

\usepackage[T1]{fontenc} 
\usepackage[british]{babel}
\usepackage{latexsym,amsmath,amssymb}
\usepackage{graphicx}
\usepackage{booktabs}
\usepackage{multirow}
\usepackage{lpic}
\usepackage{color}
\usepackage{soul}
\usepackage{pdflscape}
\usepackage[table]{xcolor}
\usepackage{array}

\usepackage{babel}
\definecolor{red}{rgb}{1,0,0} 
\definecolor{magenta}{cmyk}{0,1,0,0} 
\definecolor{darkgreen}{rgb}{0,0.6,0}
\def\comment#1{\textcolor{red}{#1}}

\def\revtext#1{\textcolor{blue}{#1}}

\usepackage{tabularx}
\usepackage{adjustbox}
\usepackage{booktabs}

\usepackage{array}
\newcolumntype{$}{>{\global\let\currentrowstyle\relax}}
\newcolumntype{^}{>{\currentrowstyle}}
\newcommand{\rowstyle}[1]{\gdef\currentrowstyle{#1}%
    #1\ignorespaces
}

\newcommand{\obs}{\mathrm{obs}}
\newcommand{\pred}{\mathrm{pred}}
\newcommand{\eff}{\mathrm{eff}}
\newcommand{\intt}{\mathrm{int}}
\newcommand{\LMC}{\mathrm{LMC}}
\newcommand{\MW}{\mathrm{MW}}
\newcommand{\MAnd}{\mathrm{M31}}
\newcommand{\Cepheid}{\mathrm{Cepheid}}
\newcommand{\Anchors}{\mathrm{Anchors}}
\newcommand{\SNe}{\mathrm{SNe\,Ia}}
\newcommand{\km}{\mathrm{km}}
\newcommand{\second}{\mathrm{s}}
\newcommand{\Mpc}{\mathrm{Mpc}}
\newcommand{\NGC}{\mathrm{NGC4258}}
\newcommand{\magn}{\mathrm{mag}}

\definecolor{LightCyan}{rgb}{0.88,1,1}

\title{\boldmath Determining $H_0$ with Bayesian hyper-parameters}

\author[a]{Wilmar Cardona}
\author[a]{Martin Kunz}
\author[b]{Valeria Pettorino}

\affiliation[a]{D\'epartement de Physique Th\'eorique and Center for Astroparticle Physics, Universit\'e de Gen\`eve, 24 Quai Ernest Ansermet, 1211 Gen\`eve 4, Switzerland}
\affiliation[b]{Institut f\"{u}r Theoretische Physik, Universit\"{a}t Heidelberg, Philosophenweg 16, D-69120 Heidelberg, Germany}

\emailAdd{wilmar.cardona@unige.ch}
\emailAdd{Martin.Kunz@unige.ch}
\emailAdd{valeria.pettorino@thphys.uni-heidelberg.de}

\abstract{We re-analyse recent Cepheid data to estimate the Hubble parameter $H_0$ by using Bayesian hyper-parameters (HPs). We consider the two data sets from Riess et al 2011 and 2016 (labelled R11 and R16, with R11 containing less than half the data of R16) and include the available anchor distances (megamaser system NGC4258, detached eclipsing binary distances to LMC and M31, and MW Cepheids with parallaxes), use a weak metallicity prior and no period cut for Cepheids. 
We find that part of the R11 data is down-weighted by the HPs but that R16 is mostly consistent with expectations for a Gaussian distribution, meaning that there is no need to down-weight the R16 data set.
For R16, we find a value of $H_0 = 73.75 \pm 2.11 \,\km\, \second^{-1} \, \Mpc^{-1}$ if we use HPs for all data points (including Cepheid stars, supernovae type Ia, and the available anchor distances), which is about 2.6 $\sigma$ larger than the Planck 2015 value of $H_0 = 67.81 \pm 0.92 \,\km\, \second^{-1} \, \Mpc^{-1}$ and about 3.1 $\sigma$ larger than the updated Planck 2016 value $66.93 \pm 0.62 \,\km\, \second^{-1} \, \Mpc^{-1}$. If we perfom a standard $\chi^2$ analysis as in R16, we find $H_0 = 73.46 \pm 1.40 \, \mathrm{(stat) \,\km\, \second^{-1} \, \Mpc^{-1}}$. We test the effect of different assumptions, and find that the choice of anchor distances affects the final value significantly. 
If we exclude the Milky Way from the anchors, then the value of $H_0$ decreases. We find however no evident reason to exclude the MW data. The HP method used here avoids subjective rejection criteria for outliers and offers a way to test datasets for unknown systematics.}

\begin{document}
\maketitle
\flushbottom

\section{Introduction}
\label{Section:Introduction}

Pinning down the Hubble constant $H_0$ is crucial for the understanding of the standard model of cosmology. It sets the scale for all cosmological times and distances and it allows to tackle cosmological parameters, breaking degeneracies among them (e.g., the equation of state for dark energy and the mass of neutrinos). The expansion rate of the universe can either be directly measured or inferred for a given cosmological model through cosmological probes such as the cosmic microwave background (CMB). Although accurate direct measurements of the Hubble constant have proven to be difficult (e.g., control of systematic errors, relatively small data sets, consistency of different methods for measuring distances), significant progress has been achieved  
over the past decades \cite{Freedman1996,Freedman:2010xv}. The \textit{Hubble Space Telescope (HST)} Key Project enabled a measurement of $H_0$ with an accuracy of $10\%$ by significantly improving the control of systematic errors \cite{Freedman2001}.  More recently, in the \textit{HST} Cycle $15$, the \textit{Supernovae and $H_0$ for the Equation of State (SHOES)} project has reported measurements of $H_0$ accurate to $4.7\%$ ($74.2 \pm 3.6 \, \km\, \second^{-1}\, \Mpc^{-1}$) \cite{Riess:2009pu}, then to $3.3\%$ ($73.8 \pm 2.4 \, \km\, \second^{-1}\, \Mpc^{-1}$) \cite{Riess:2011yx} (called the `R11' data set hereafter) and very recently to $2.4\%$ ($73.24 \pm 1.74 \, \km\, \second^{-1}\, \Mpc^{-1}$) \cite{Riess:2016jrr} (denoted as `R16' in the following). This remarkable progress has been achieved thanks to an improved and expanded SN Ia Hubble diagram, including an enlarged sample of SN Ia host galaxies with Cepheid calibrated distances, a reduction in the systematic uncertainty of the maser distance to NGC4258, and an increase of infrared observations of Cepheid variables in the Large Magellanic Cloud (LMC). Results were consistent with the WMAP data \cite{Hinshaw:2012aka}. 

The 2015 release in temperature, polarization and lensing measurements of the CMB by the \textit{Planck} satellite leads to a present expansion rate of the universe given by $H_0 = 67.81 \pm 0.92 \, \km \, \second^{-1}\, \Mpc^{-1}$ for the base six-parameter $\Lambda$CDM model \cite{Ade:2015xua}. The \textit{Planck} collaboration has recently updated this value to be $H_0 = 66.93 \pm 0.62 \, \km \, \second^{-1}\, \Mpc^{-1}$ \cite{Aghanim:2016yuo}. The derived estimation of $H_0$ from CMB experiments provide indirect and highly model-dependent values of the current expansion rate of the universe (requiring e.g., assumptions about the nature of dark energy, properties of neutrinos, theory of gravity) and therefore do not substitute a direct measurement in the local universe. Moreover, indirect determinations (in a Bayesian approach) rely on prior probability distributions for the cosmological parameters which might have an impact on the results.

The \textit{Planck} Collaboration used a ``conservative'' prior on the Hubble constant ($H_0=70.6\pm3.3\, \km\, \second^{-1}\, \Mpc^{-1}$) derived from a reanalysis of the Cepheid data used in \cite{Riess:2011yx}, done by G. Efstathiou in \cite{Efstathiou:2013via}: in this reanalysis, a different rejection algorithm was used (with respect to that in \cite{Riess:2011yx}) for outliers in the Cepheid period-luminosity relation (the so-called Leavitt Law); in addition, \cite{Efstathiou:2013via} used the revised geometric maser distance to $\NGC$ of \cite{Humphreys:2013eja}. Although consistent with the \textit{Planck} TT estimate at the $1\, \sigma$ level, this determination of $H_0$ assumes that there is no metallicity dependence in the Leavitt Law. Furthermore, it discards data (i) from both Large Magellanic Cloud (LMC) and Milky Way (MW) Cepheid variables (ii) from the sample of Cepheid variables in \cite{Riess:2011yx} using an upper period cut of $60$ days.\footnote{In \cite{Efstathiou:2013via} G. Efstathiou also shows results utilizing the rejection algorithm for outliers used in \cite{Riess:2011yx}, but with the revised geometric maser distance to $\NGC$ \cite{Humphreys:2013eja} which is about $4\%$ higher than that adopted by \cite{Riess:2011yx} in their analysis. Note that in \cite{Riess:2011yx} (see their page 13) the authors provided a recalibration of $H_0$ for each increase of $1\%$ in the distance to $\NGC$: according to this recalibration and the revised geometric maser distance their measurement would be driven downwards from $H_0 = 74.8\, \km \, \second^{-1}\, \Mpc^{-1}$ to $H_0 \approx 73.8 \, \km \, \second^{-1}\, \Mpc^{-1}$ which is higher than all the reported values in table A1 of \cite{Efstathiou:2013via} for the R11 rejection algorithm.}

As discussed in \cite{Freedman:2010xv}, the sensitivity to metallicity of the Leavitt Law is still an open question. In fact, due to changes in the atmospheric metal abundance, a metallicity dependence in the Cepheid period-luminosity is expected. Discarding data involves somehow arbitrary choices (e.g., Chauvenet's criterion, period cut, threshold T in \cite{Efstathiou:2013via}) and might hinder our understanding of the physical basis behind the incompatibility of data sets (if any) \cite{Press:1996fw}. Therefore, neither no metallicity dependence in Leavitt Law nor disregarding data seem to be a priori very conservative assumptions. 

Once systematics are under control (like the presence of unmodeled systematic errors or biases in the outlier rejection algorithm for Cepheid variables), a reliable estimate of $H_0$ is very important also on theoretical grounds. Confirmation of significant discrepancies between direct and indirect estimates of $H_0$ would suggest evidence of new physics. Discrepancies could arise if the local gravitational potential at the position of the observer is not consistently taken into account when measuring the Hubble constant. Nevertheless, an unlikely fluctuation would be required as estimates of $\Delta H_0$ from inhomogeneities are of the order of 1 -- 2  $\km \, \second^{-1}\, \Mpc^{-1}$ only \cite{Marra2013a, Ben-Dayan:2014swa}, although there are claims that observations support the existence of a sufficiently large local departure from homogeneity \cite{Romano:2016utn}. Second-order corrections to the background distance-redshift relation could bias estimations of the Hubble constant derived from CMB \cite{Clarkson:2014pda}. However, it was shown in \cite{Bonvin:2015uha} that those corrections are already taken into account in current CMB analyses.           

It is clear from \cite{Efstathiou:2013via} that the statistical analysis done when measuring $H_0$ plays a part in the final result (for instance, through the outlier rejection algorithm, data sets included, anchors distances included, the period cut on the sample of Cepheid variables, the prior on the parameters of the Period-Luminosity relation). Given the relevance of the Hubble constant for our understanding of the universe, it is necessary to confirm previous results and prove them robust against different statistical approaches.

The goal of this paper is to determine the Hubble constant $H_0$ by using Bayesian hyper-parameters in the analysis of the Cepheid data sets used in both \cite{Riess:2011yx} and \cite{Riess:2016jrr}. In Section \ref{Section:notation-and-method} we explain both our notation and the statistical method employed. We then apply the method to the R11 data set and determine the expansion rate $H_0$ in Section \ref{Section:Application-R11}. In Section \ref{Section:Application} we test the assumptions of our baseline analysis. 
The reader willing to know our \revtext{main} results using the R16 data set can skip Section \ref{Section:Application-R11} and \ref{Section:Application} and go directly to Section \ref{Section:Application-R16}. We conclude and discuss our results in Section \ref{Section:Summary}.


\section{Methodology}
\label{Section:notation-and-method}

\subsection{Distances and standard candles}

Astrophysical objects with a known luminosity -- the so-called standard candles -- are used to probe the expansion rate of the universe. In particular, measuring redshifts and apparent luminosities for supernovae type Ia (SNe Ia) one can establish an empirical redshift-distance relation for these objects. In order to estimate distances to SNe Ia one uses the luminosity distance
\begin{equation}
d_L \equiv \left(\frac{L}{4\pi l} \right)^{1/2} \, , \label{Eq:luminosity-distance-obs}
\end{equation}
where $L$ and $l$ are the absolute luminosity and the apparent luminosity, respectively. For historical reasons the apparent bolometric luminosity $l$ is defined so that 
\begin{equation}\label{Eq:apparent-luminosity}
l = 10^{-2m/5}\times 2.52 \times 10^{-5}\, \mathrm{erg/ cm^2}\, \second
\end{equation}
where $m$ is the apparent bolometric magnitude \cite{Weinberg:1972kfs}. Similarly, one can define the absolute bolometric magnitude $M$ as the apparent bolometric magnitude a source would have at a distance 10 $\mathrm{pc}$
\begin{equation}\label{Eq:absolute-luminosity}
L = 10^{-2M/5} \times 3.02 \times 10^{35}\, \mathrm{erg/}\second.
\end{equation}
Combining equations \eqref{Eq:luminosity-distance-obs}--\eqref{Eq:absolute-luminosity} it is possible to express the luminosity distance in terms of the distance modulus $m-M$:
\begin{equation}
\mu_0 \equiv m - M = 5 \log_{10} \left(\frac{d_L}{1 \mathrm{Mpc}} \right) + 25 \, . \label{Eq:distance-modulus}
\end{equation}

One can also compute the luminosity distance $d_L$ of a light source with redshift $z$ in the context of General Relativity. Assuming a flat FLRW metric, one finds  
\begin{equation}\label{Eq:luminosity-distance-the}
d_L(z) = (1+z) c \int_0^z \frac{dz'}{H(z')} \, 
\end{equation}
where $c$ is the speed of light and $H(z)$ is the Hubble function. Since nowadays the empirical curve for $d_L(z)$ is reasonably well known for relatively small redshift, the Hubble function may then be usefully expressed as a power series in Eq.\ \eqref{Eq:luminosity-distance-the}, leading to
\begin{equation}\label{Eq:luminosity-distance-the-small-z}
d_L(z) \equiv \frac{cz}{H_0}  ( 1+\delta(z)) \approx \frac{cz}{H_0} \left\{ 1 + \frac{1}{2} [1-q_0] z - \frac{1}{6} [1-q_0 - 3 q_0^2 + j_0] z^2 + O(z^3) \right\}
\end{equation} 
where $H_0$ is the Hubble constant, $q_0$ is the present acceleration parameter, $j_0$ the present the present jerk parameter. Here $\delta(z)$ defines a function that vanishes as $z\rightarrow 0$ and that for small redshifts, $z\ll 1$, can be approximately expressed as a series expansion in redshift starting with a term linear in $z$. Using an expansion instead of Eq.\ (\ref{Eq:luminosity-distance-the}) with the $\Lambda$CDM expression for $H(z)$ has the advantage that it is more model independent.

We now have $d_L$ as a function of  $m$ and $M$  in \eqref{Eq:distance-modulus} and an expression of $d_L$ in terms of the expansion rate today (\ref{Eq:luminosity-distance-the-small-z}).
Equating Eqs. \eqref{Eq:distance-modulus} and \eqref{Eq:luminosity-distance-the-small-z} we obtain that
\begin{equation}\label{Eq:av-definition}
5 \log_{10} ( c z ( 1+\delta(z) )) - m_X = 5 \log_{10} H_0 - M_X - 25 \equiv 5 a_X \, ,
\end{equation}
where $X$ denotes the use of wavelength band $X$ (e.g., $U$ for ultraviolet, $B$ for blue, and $V$ for visual) and $a_X$ is a constant which defines the intercept of the $\log_{10} cz - 0.2\, m_X$ relation. Defining $\delta(z)$ through $q_0 = -0.55$ and $j_0 = 1$ \cite{Riess:2006fw}, Riess et al. \cite{Riess:2011yx} used the $V$ wavelength band Hubble diagram for 153 nearby SN-Ia in the redshift range $0.023 < z < 0.1$ \footnote{A conservative lower limit in redshift imposed to avoid the possibility of a local, coherent flow biasing the results.}
to measure the intercept $a_V = 0.697 \pm 0.00201$. Using the Hubble diagram for $217$ $\SNe$ in the redshift range $0.023 < z < 0.15$, Riess et al. \cite{Riess:2016jrr} found $a_B = 0.71273 \pm 0.00176$. 

From Eqs. \eqref{Eq:distance-modulus} and \eqref{Eq:av-definition} one can easily express the apparent magnitude $m_X^{\SNe}$ of a SN Ia in terms of its distance modulus $\mu_0$, the Hubble constant $H_0$, and the intercept $a_X$ as
\begin{equation}
m_X^{\SNe} =  5 \log_{10} H_0 + \mu_0 - 5 a_X - 25 \, . \label{Eq:apparent-magnitude-H0}
\end{equation}
Having measured $a_X$ we can find $H_0$ if we know $\mu_0$ for a supernova for which have measured $m_X^{\SNe}$. Unfortunately, we generally do not have a direct measurement of $\mu_0$ for objects that contain supernovae. In this situation Cepheid variables -- another type of standard candles -- come to the rescue. Cepheids are stars whose apparent luminosity is observed to vary more or less regularly with time. They are quite common, so that there exist several galaxies ($19$ up to date \cite{Riess:2016jrr}) which simultaneously host both SNe Ia and Cepheid variables, and in addition several galaxies for which we have measured the distance directly so that $\mu_0$ is known and which also host Cepheids. This is the case for the megamaser system $\NGC$, $\LMC$, $\MAnd$ as well as for individual Cepheids in the Milky Way. Although the sample of MW Cepheid variables with parallax measurements is relatively small ($15$ up to date \cite{Riess:2016jrr}) and mostly dominated by Cepheid stars with periods $P<10$ days, their inclusion helps to further constrain parameters in the period-luminosity relation.

The apparent luminosity of Cepheids and its link to $\mu_0$ is described by the Leavitt Law \cite{1912HarCi.173....1L}. According to the Leavitt Law there is a relation between period and luminosity of Cepheids: in the $i\mathrm{th}$ galaxy, the pulsation equation for the $j\mathrm{th}$ Cepheid star with apparent magnitude $m^{\Cepheid}_{Y,i,j}$ (in the passband $Y$, not necessarily the same as passband $X$ for the supernovae) and period $P_{i,j}$ leads to a relation 
\begin{equation}\label{Eq:P-L-equation}
m_{Y,i,j}^{\Cepheid} = \mu_{0,i} + M_Y^{\Cepheid} + b_Y (\log P_{i,j}-1) + Z_Y \Delta \log[O/H]_{i,j},
\end{equation}
where $Z_Y$ is the metallicity parameter, $b_Y$ is the slope of the period-luminosity relation, and $M^{\Cepheid}_Y$ is the Cepheid zero point which is common to all Cepheids. A fit of Eq.\ (\ref{Eq:P-L-equation}) for the galaxies with known $\mu_0$ that host Cepheids allows us to determine the parameters $b_Y$, $Z_Y$ and especially $M_Y^{\Cepheid}$ which is fully degenerate with $\mu_{0,i}$. Using then the Leavitt law with now known $M_Y^{\Cepheid}$ for the galaxies with Cepheids and supernovae allows to determine the $\mu_0$ of those galaxies, effectively transferring the distance measurements from the `anchors' (with direct distance determinations) to the `supernova hosts'. Knowing $\mu_0$ for the supernova hosts we can finally use Eq.\ (\ref{Eq:apparent-magnitude-H0}) to determine $H_0$. In reality of course we will fit everything simultaneously.

\subsection{Hyper-parameters}
\label{Section:Hyper-parameters}

Astrophysical observations are difficult, and it is not easy to estimate and include the associated errors and uncertainties correctly. Often the data sets show outliers with error bars that are much smaller than the deviation from the expected fit, for reasons that are not well understood or difficult to quantify. An analysis needs to deal with such outliers, typically by removing them based on some rejection rule.
As discussed in \cite{Riess:2009pu,Riess:2011yx,Efstathiou:2013via}, the rejection of outliers on the Cepheid period-luminosity relation may have a non-negligible effect on the determination of the expansion rate of the universe. One can argue that an outlier rejection criterion: i) involves arbitrary choices (e.g., Chauvenet's criterion, period cut) which might bias the results; ii) rejects data, thus increasing error bars and hindering a better understanding of the data sets \cite{Press:1996fw}. The hyper-parameter (HP, hereafter) method offers a Bayesian alternative to {\em{ad hoc}} selection of data points, avoiding problems associated with using incompatible data points \cite{Lahav:1999hu,Hobson:2002zf}. Instead of adopting an a priori rejection criterion  (galaxy-by-galaxy as in \cite{Riess:2009pu,Riess:2011yx} or from a global fit as in \cite{Riess:2016jrr, Efstathiou:2013via}), in this work we analyse {\em all} the available measurements with Bayesian HPs. The latter effectively allow for relative weights in the Cepheid variables, determined on the basis of how good their simultaneous fit to the model is. 

HPs allow to check for unrecognised systematic effects by introducing a rescaling of the error bar of data point $i$, $\sigma_i \rightarrow \sigma_i/\sqrt{\alpha_i}$. Here $\alpha_i$ is a HP associated with the data point $i$ \cite{Lahav:1999hu,Hobson:2002zf}.
In order to explain how HPs work, we start by assuming a Gaussian likelihood for the datum $D_i$,
\begin{equation}\label{Eq:Gaussian-likelihood}
P_G(D_i|\vec{w}) = \tilde{N}_i \, \frac{\exp(-\chi^2_i(\vec{w})/2)}{\sqrt{2\pi}},
\end{equation}
where $\chi^2_i$ and the normalisation constant $\tilde{N}_i$ are given by
\begin{equation}\label{Eq:chi2} 
\chi^2_i \equiv \frac{\left( x_{\obs,i} - x_{\pred,i}(\vec{w}) \right)^2}{\sigma_i^2} \, , \quad
\tilde{N}_i = 1/\sqrt{\sigma_i^2} \, .
\end{equation}
Here for each measurement $x_{\obs,i}$ there is a corresponding error $\sigma_i$ and a prediction $x_{\pred,i}(\vec{w})$, ($\vec{w}$ being the parameters of a given model). Suppose that some errors have been wrongly estimated due to unrecognised (or underestimated) systematic effects and use hyper-parameters  \cite{Lahav:1999hu} to control the relative weight of the data points in the likelihood. For each measurement {$i$} we introduce a HP to rescale $\sigma_i$ as mentioned above. In that case the Gaussian likelihood becomes \cite{Lahav:1999hu}
\begin{equation}\label{Eq:HP-Gaussian-likelihood}
P(D_i|\vec{w},\alpha_i) = \tilde{N}_i \, \alpha_i^{1/2}\, \frac{\exp(-\alpha_i \chi^2_i(\vec{w})/2)}{\sqrt{2\pi}}.
\end{equation}

However, in general we do not know what value of $\alpha_i$ is correct. In order to circumvent this problem, 
we follow a Bayesian approach, introducing the $\alpha_i$ as nuisance parameters and marginalising over them. Given a set of data points $ \lbrace D_i \rbrace$, we can write the probability for the parameters $\vec{w}$  as
\begin{equation}\label{Eq:probability-of-w}
P(\vec{w}|\lbrace D_i \rbrace) = \int \dots \int P(\vec{w},\lbrace \alpha_i \rbrace|\lbrace D_i \rbrace)\, d\alpha_1 \dots d\alpha_N,
\end{equation}
where $N$ is the total number of measurements. Bayes' theorem allows us to write 
\begin{equation}\label{Eq:probability-of-w-alphas-1}
P(\vec{w},\lbrace \alpha_i \rbrace|\lbrace D_i \rbrace) = \frac{P(\lbrace D_i \rbrace|\vec{w},\lbrace \alpha_i \rbrace)\, P(\vec{w},\lbrace \alpha_i \rbrace)}{P(D_1,\dots,D_N)}
\end{equation}
and 
\begin{equation}\label{Eq:probability-of-w-alphas-2}
P(\vec{w},\lbrace \alpha_i \rbrace) = P(\vec{w}|\lbrace \alpha_i \rbrace)\, P(\lbrace \alpha_i \rbrace).
\end{equation}
As in \cite{Lahav:1999hu} we assume:
\begin{subequations}\label{Eq:assumptions}
\begin{align}
\label{Eq:assumptions:1}
P(\lbrace D_i \rbrace|\vec{w},\lbrace \alpha_i \rbrace) & = P(D_1|\vec{w},\alpha_1)\dots P(D_N|\vec{w},\alpha_N),
\\
\label{Eq:assumptions:2}
P(\vec{w}|\lbrace \alpha_i\rbrace) & = \mathrm{constant},
\\
\label{Eq:assumptions:3}
P(\lbrace \alpha_i \rbrace) & = P(\alpha_1)\dots P(\alpha_N).
\end{align}
\end{subequations}

Combining Eqs. \eqref{Eq:probability-of-w-alphas-1}--\eqref{Eq:assumptions:1} the integrand in Eq. \eqref{Eq:probability-of-w} then reads:
\begin{eqnarray}
P(\vec{w},\lbrace \alpha_i \rbrace|\lbrace D_i \rbrace) &=& \frac{P(D_1|\vec{w},\alpha_1)\dots P(D_N|\vec{w},\alpha_N) P(\vec{w}|\lbrace \alpha_i \rbrace)\, P(\lbrace \alpha_i \rbrace)}{P(D_1,\dots,D_N)} \\
&=& \frac{P(D_1|\vec{w},\alpha_1)\dots P(D_N|\vec{w},\alpha_N) P(\vec{w}|\lbrace \alpha_i \rbrace)\, P(\alpha_1)\dots P(\alpha_N)}{P(D_1,\dots,D_N)} \, .
\end{eqnarray}

Thus far the formalism is fairly general and it contains two unspecified functions: the probability distributions for HPs and data points. In this work we will assume uniform priors for HPs ($P(\alpha_i)=1$) and that errors are never actually smaller than the value given (i.e.\ error bars tend to be underestimated, not overestimated, leading to $\alpha_i \in [0,1]$).\footnote{We have examined the more general case of an improper Jeffrey's prior (allowing decreasing as well as increasing error bars). This works well when many data points are associated to the same HP so that the $\chi^2$ never vanishes. But when each data point has its own associated HP then the model curve can pass through that data point so that $\chi^2_i=0$. In this case the likelihood grows without bounds as $\alpha_i\rightarrow 0$, in other words the HP-marginalised likelihood is singular when at least one of the points has $\chi^2=0$ as can also be seen from Eq. (16) in \cite{Lahav:1999hu}.} A low posterior value of the HP indicates that the point has less weight within the fit. This may indicate the presence of systematic effects or the requirement for better modelling. With these assumptions 
the integrand in Eq.\eqref{Eq:probability-of-w} becomes
\begin{equation}
P(\vec{w},\lbrace \alpha_i \rbrace|\lbrace D_i \rbrace) = \mathrm{constant} \times  \frac{P(D_1|\vec{w},\alpha_1)\dots P(D_N|\vec{w},\alpha_N) \, }{P(D_1,\dots,D_N)}
\end{equation}
and Eq.~\eqref{Eq:probability-of-w} now reads
\begin{equation}\label{Eq:probability-of-w-2}
P(\vec{w},\lbrace D_i \rbrace) = \mathrm{constant} \times \frac{P(D_1|\vec{w})\dots P(D_N|\vec{w})}{P(D_1,\dots,D_N)},
\end{equation}
where 
\begin{equation}\label{Eq:probability-of-w-2b}
P(D_i|\vec{w}) \equiv \int_0^1 P(D_i|\vec{w},\alpha_i)\, d\alpha_i.
\end{equation}

The integral in Eq. \eqref{Eq:probability-of-w-2b} can be explicitly evaluated for the Gaussian HP likelihood (\ref{Eq:HP-Gaussian-likelihood}), and gives, for each data point,
\begin{eqnarray}\label{Eq:hyper-likelihood}
P(D_i|\vec{w}\,) = \tilde{N}_i \, \left(\frac{ \mathrm{Erf}\left( \frac{\chi_i(\vec{w})}{\sqrt{2}} \right)  -\sqrt{\frac{2}\pi} \chi_i(\vec{w}) \exp(-\chi^2_i(\vec{w})/2)}{ \chi^3_i(\vec{w})} \right) \equiv \tilde{N}_i \tilde{\chi}^2_i(\chi^2_i(\vec{w})).
\end{eqnarray}
which defines the effective $\chi$ square function $\tilde{\chi}^2_i$. We can now rewrite Eq. \eqref{Eq:probability-of-w-2} as
\begin{equation}\label{Eq:probability-of-w-2c}
\ln P(\vec{w},\lbrace D_i \rbrace) = \sum_i \ln \tilde{N}_i + \ln \tilde{\chi}^2_i, 
\end{equation} 
where constant terms have been omitted. 

Since we have analytically marginalized over the hyper-parameters, they no longer appear in the posterior distribution. Each HP however does have a posterior pdf associated with it, and we could determine it by adding a HP explicitly as a parameter and including it in the sampling procedure. In general this might entail adding thousands of extra parameters, which would make the exploration of the posterior numerically much more demanding.
However, one can easily obtain the most likely value (the peak or mode of the pdf) for each HP by maximizing \eqref{Eq:HP-Gaussian-likelihood} with respect to the HPs at a given set of best fit parameters $\vec{w}$. We find that for each data point the most likely value is given by \begin{subequations}\label{Eq:effective-HPs}
\begin{align}
\label{Eq:effective-HP-1}
\alpha^{\eff}_i & = 1,\quad \mathrm{if} \quad \chi^2_i\leq1
\\
\label{Eq:effective-HP-2}
\alpha^{\eff}_i & = \frac{1}{\chi^2_i},\quad \mathrm{if} \quad \chi^2_i> 1.
\end{align}
\end{subequations}
Although this point-estimate of the hyper-parameter posterior does not contain the full information, we can nonetheless use these effective HP values to flag data points that, if down-weighted, improve the overall likelihood. We consider this a sign that they are not fully compatible with the other data points, for the model adopted.

\begin{figure}[tbp]
\centering
\includegraphics[scale=1]{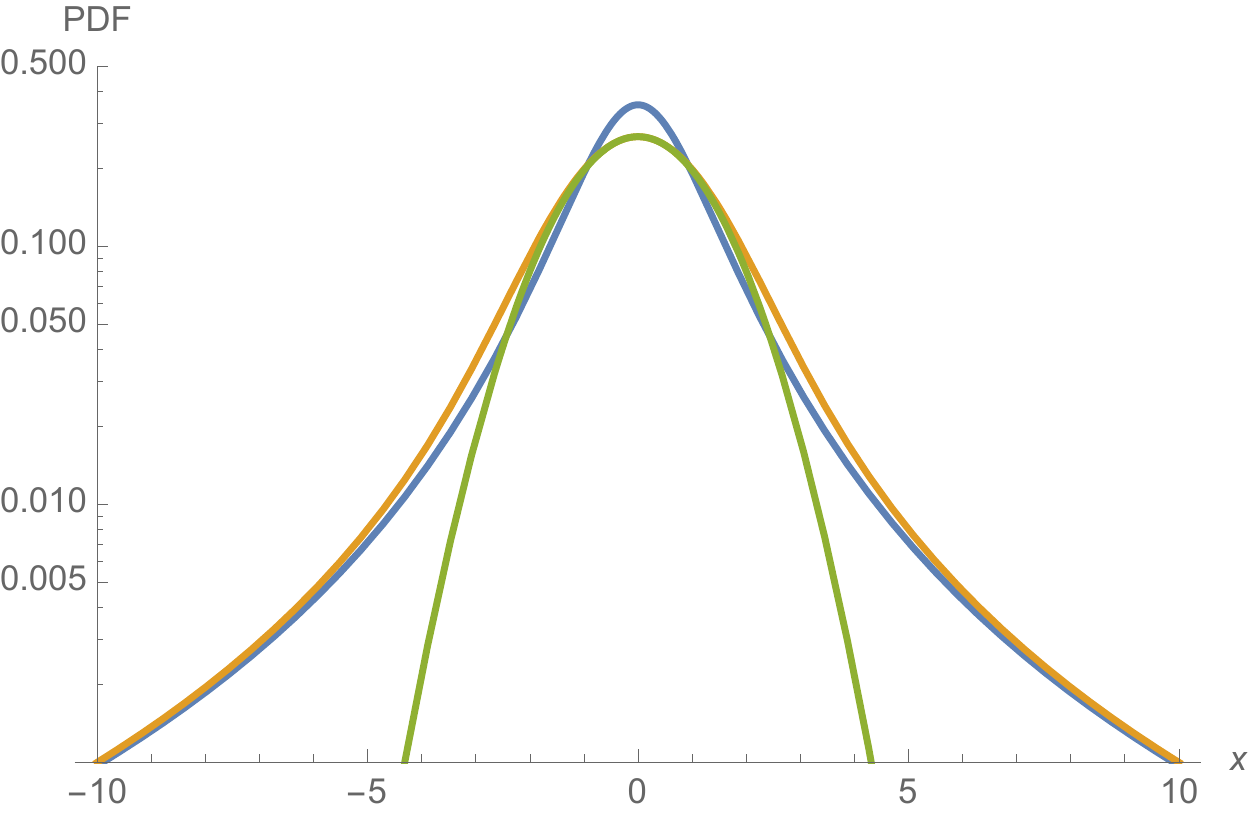}
\caption{The hyper-parameter marginalized probability distribution function (pdf) of Eq.\ (\ref{Eq:hyper-likelihood}) for $\sigma = 1$, in orange. Close to the origin, $x=0$, it is similar to a Gaussian pdf with $\sigma = \sqrt{5/3}$ (green), except that its amplitude at the peak is 16\% lower than a normalised Gaussian with that variance. Asymptotically for $|x|\rightarrow\infty$ it decreases as $1/x^3$ and looks like a student's t distribution with 2 degrees of freedom (blue), but the latter is narrower at small $x$. \comment{} \label{fig:hplike}}
\end{figure}

In Fig.\ \ref{fig:hplike} we show the hyper-parameter marginalized pdf (left hand side of Eq.\ (\ref{Eq:probability-of-w-2c})) for a single datum with standard deviation $\sigma=1$ (orange line) and compare it to a Gaussian curve (green line) with the same curvature at the peak (corresponding to $\sigma=\sqrt{5/3}$) and the same amplitude at zero (which is about 16\% lower than the amplitude of a Gaussian pdf, as the HP-marginalized has wider wings) and to a student's t distribution with two degrees of freedom (blue line). We see that close to the peak the HP marginalised pdf looks like a Gaussian, while it decays as a power law ($\propto x^{-3}$) as we move away from the peak, like the student's t distribution. This is not really a surprise as the student's t distribution arises when marginalising a Gaussian pdf over $\sigma$, which is partially also the case for the HP-marginalised pdf.

As the hyper-parameter marginalized likelihood (orange line) has a lower curvature at the peak and wider tails than the original Gaussian pdf (green line in Fig.\ \ref{fig:hplike}), we expect that it will lead to a less precise determination of parameters for a given, fixed $\sigma$, with respect to the standard analysis which does not include hyperparameters. This is indeed the case: we find numerically that for a large ensemble of $10^6$ data points drawn independently from a univariate Gaussian pdf, the uncertainty on the mean of the HPs posterior distribution is about 40\% larger than for a Gaussian likelihood. If we however estimate the variance $\sigma$ that enters Eq.\ (\ref{Eq:chi2}) simultaneously with the mean then the hyper-parameter marginalised likelihood recovers the mean with the same precision as the Gaussian, while $\sigma$ is $0.606$ (the Gaussian likelihood obviously recovers $\sigma=1$). We conclude from this little test that it is preferable, when using HPs, to leave $\sigma$ as a free parameter rather than fixing it, in order not to overestimate the uncertainty. The main reason why this procedure makes sense is linked to our choice that HPs only increase error bars. The global $\sigma$ on the other hand can be used to reduce the overall error bar size as it does not lead to a singularity when $\sigma\rightarrow 0$ due to its global nature (under the assumption that we have enough non-degenerate data points). The opposite situation, a large $\sigma$ and strong correlations between $\sigma$ and HPs, is prevented by the standard normalization condition.

HPs offer a way to deal with outliers without relying on arbitrary choices; simultaneously, they reduce the sensitivity to biases that can arise when error bars are underestimated. To illustrate this, we artificially turn 10\% of the data points into outliers by randomly (and somewhat arbitrarily) multiplying their value by a factor between two and ten. For $10^6$ such data points, including $10^5$ outliers, we find that a Gaussian likelihood (centered at zero) recovers a mean of $\mu = (-4.7\pm1.0) \times 10^{-3}$ ($\mu = (-4.7\pm2.2) \times 10^{-3}$ and $\sigma=2.2$ when simultaneously estimating $\sigma$), while the HP marginalised likelihood finds $\mu= (-0.8\pm1.5)\times 10^{-3}$ ($\mu = (-0.6\pm1.1)\times10^{-3}$ and $\sigma=0.70$ when simultaneously estimating $\sigma$). At least in this scenario the hyper-parameters successfully remove a bias from the result with only a small increase in the errors. This example also shows that in such a situation the use of HP can actually lead to an increased precision relative to the Gaussian result when estimating both the mean and $\sigma$ simultaneously. We will encounter a case where using HPs leads to smaller error bars later in our analysis, see Fig.\ \ref{fig:sigmaint}. Of course if the data is actually Gaussian distributed then using HPs offers no advantage, it is always best to use the true probability distribution function when this is possible. The advantage of HPs comes from allowing us to deal with situations where the true pdf is uncertain.

The use of HPs when combining different data also offers a way to test for internal consistency by checking if the average effective HP values, Eqs.\ (\ref{Eq:effective-HP-1}) and (\ref{Eq:effective-HP-2}), are lower than expected. To this end we define the normalised weight of $j\mathrm{th}$ kind of data  as 
\begin{equation}
|| \alpha^{j} || \equiv \frac{\sum_{i=1}^{K_j} \alpha_{i,j}}{K_j},
\label{Eq:normalised-weights}
\end{equation} 
where $\alpha_{i,j}$ denotes $i\mathrm{th}$ HP of data of kind $j$ ($j=\Cepheid,\,\SNe,\,\Anchors$), and $K_j$ stands for the number of objects of kind $j$. A low normalised weight $|| \alpha^{j} || $ in (\ref{Eq:normalised-weights}) for a given kind $j$ of data is an indicator for problems e.g.\ in the report error bars or the data values, or in the model used to fit the data. However, there are also points outside of the $1\sigma$ region for data drawn from the expected Gaussian distribution, it is straightforward to compute from Eqs.\ (\ref{Eq:effective-HPs}) that we expect $||\alpha|| \approx 0.85$ in this situation. Data sets with a value of $||\alpha||$ close to $0.85$ can be considered reliable. Smaller values indicate too many outliers relative to the given error bars -- the numerical example with 10\% of outliers mentioned above has $||\alpha||=0.79$. It is also possible to use the effective HP to test whether data sets are mutually consistent. If they are not, then the global fit to the combined data may reduce the normalised weights of each data set. We will use this later to decide on some of the choices that need to be made to analyse the distance data.

We will now apply HPs to combine Cepheid variable measurements and determine the current expansion rate of the universe. We explore the parameter space $\vec{w}$ with the help of a Markov Chain Monte Carlo (MCMC) approach, using flat priors if not specified differently. A summary of all the fits done for this work are illustrated in Tables \ref{Table:details-fits}--\ref{Table:Constraints-main-analysis}, where we tested different assumptions. We will now first describe our baseline choice and the resulting constraint on $H_0$ and then, in Section \ref{Section:Application}, describe in detail each of the assumptions.


\section{Expansion rate: applying hyper-parameters to R11 data}
\label{Section:Application-R11}

In this section we describe our baseline HP marginalised analysis of the local expansion rate data. 
As we will discuss in more detail below, we include three different distance
anchors, we use hyper-parameters everywhere (Cepheids, supernovae and anchors), we do not apply a period cut in the Cepheid data and use a strong prior on Cepheid metallicity. 
We consider the resulting constraint on $H_0$ corresponding to this set of `baseline choices' [fit (29) in Tables \ref{Table:details-fits}--\ref{Table:Constraints-main-analysis}] as our `best estimate' for the R11 data set \cite{Riess:2011yx}. In Section \ref{Section:Application} we will then study what happens if we change these
assumptions.

\paragraph{The R11 data set}  It comprehends a sample of 53 Large Magellanic Cloud (LMC) Cepheid variables with $H$-band magnitudes, $m_H$, listed in Table 3 of  \cite{2004AJ:128.2239P} and $V$, $I$ band magnitudes, $m_V$, $m_I$, listed in Table 4 of \cite{2002ApJS:142:71S}; the sample of Cepheid variables in the $8$ SNe Ia hosts of  \cite{Riess:2011yx}; $13$ Milky Way (MW) Cepheid stars with parallax measurements listed in Table 2 of \cite{vanLeeuwen:2007xw} (eliminating Polaris and correcting for Lutz-Kelker bias).

\paragraph{Baseline anchors}  We will show in Section \ref{sec:cepheids} that the period-luminosity parameters independently derived from LMC, MW and NGC 4258 Cepheid variables are in good agreement and therefore we do not see any reason to discard any of the data sets when determining the Hubble constant with hyper-parameters. Here we use the sample of Cepheid variables for the SNe Ia hosts from \cite{Riess:2011yx}, the set of LMC Cepheid variables discussed in Section \ref{Subsection:LMCR11} and the set of MW Cepheid variables discussed in Subsection \ref{Subsection:MW-1}. The MW Cepheid zero-points have been determined with the help of their parallax distance so that they contribute to the absolute distance determinations. We further use both the revised NGC4258 geometric maser distance from \cite{Humphreys:2013eja} and the distance to NGC4258 determined by considering type IIP SNe from \cite{Polshaw:2015ika}. We also make use of the distance to LMC derived from observations of eclipsing binaries from \cite{Pietrzynski:2013gia}. {In Subsection \ref{subsection:Anchors} we study how our method performs for different combinations of anchor distances.

\paragraph{Baseline HPs} We use hyper-parameters everywhere, for Cepheids, supernovae and anchors. In particular, we use hyper-parameters for all Cepheid fits as there are outliers in most of the data sets. As for supernovae, we find that the SNe Ia data show potential inconsistencies in our fits (see Figure \ref{Fig:HP-SNIa-main-analysis} and Table \ref{Table:SNIa-HP-fit-M1a} below), therefore we include hyper-parameters also in supernovae. We further include hyper-parameters in the available distance moduli to both NGC 4258 and LMC. For anchor distances and SNe Ia magnitudes we then assume a Gaussian HP likelihood as in Eq.\ \eqref{Eq:hyper-likelihood}. Hence in order to find the best-fitting parameters $\vec{w}$ we maximize 
\begin{equation}
\ln P(\vec{w},\lbrace D_i \rbrace) = \ln P^{\Cepheid} + \ln P^{\SNe} + \ln P^{\Anchors}. 
\end{equation}
For Cepheid variables we have, applying Eq.\ (\ref{Eq:probability-of-w-2c}),
\begin{subequations}
\begin{equation}
\ln P^{\Cepheid} = \sum_{ij} \ln \tilde{\chi}^{2}(\chi^{2,\Cepheid}_{ij}) + \ln \tilde{N}^{\Cepheid}_{ij},
\end{equation}
where the index $i$ identifies the galaxy and the index $j$ refers to the Cepheid belonging to the $i\mathrm{th}$ galaxy; here
\begin{equation} \label{cepheid_formulae}
\chi^{2,\Cepheid}_{ij} = \frac{(m_{W,ij} - m^P_{W,i,j})^2}{\sigma_{e,ij}^2 + \sigma_{\intt,i}^2}\, , \quad
\tilde{N}^{\Cepheid}_{ij} = \frac{1}{\sqrt{\sigma_{e,ij}^2 + \sigma_{\intt,i}^2}} \, ,
\end{equation}
and the Cepheid magnitude $m_{W,ij}$ is modelled as in Eq.\ \eqref{Eq:P-L-equation}; for the `Wesenheit reddening-free' magnitudes, denoted by $W$, that combine H, V, I bands:
\begin{equation}
m_{W,ij} = m_{H,ij} - 0.410\, (m_{V,ij} - m_{I,ij}) \,\, ;
\label{eq:3c}
\end{equation}
\end{subequations}
finally $\sigma_{\intt,i}$ is the internal scatter for the $i\mathrm{th}$ galaxy. The internal scatter is a common additional dispersion of the data points that is independent of the measurement error and due to variations in the physical mechanism behind the period-luminosity relation \cite{Efstathiou:2013via}. In the R11 data set the internal scatter is not known, it is estimated simultaneously to the other parameters and can then be marginalized over if we are not interested in its distribution. More precisely, we sample in $\log \sigma_{\intt}$ with a flat prior $\ln \sigma_{\intt} \in [-3,-0.7]$. As discussed in Section \ref{Section:Hyper-parameters}, allowing for a free $\sigma$ also permits to avoid artificially inflating the uncertainty on the inferred parameter values when using HPs. Indeed, as we will see in the next section, HPs can even allow to decrease the posterior uncertainty, as they allow to use a smaller $\sigma_{\intt}$ by down-weighting outliers that lead to a non-Gaussian distribution with heavier tails.

For SNe Ia we use the likelihood:
\begin{subequations}
\begin{equation}
 \ln P^{\SNe} = \sum_{i} \ln \tilde{\chi}^{2}(\chi^{2,\SNe}_{i}) + \ln \tilde{N}^{\SNe}_{i} - \frac{(a^{\rm R11}_V - a_V)^2}{2 \sigma_{a_V}^2} - \frac{\ln (2\pi\sigma_{a_V}^2)}{2} - \frac{(a_{\rm cal})^2}{2 \sigma_{a_{\rm cal}}^2} - \frac{\ln (2\pi\sigma_{a_{\rm cal}}^2)}{2}
 \label{eq:ialike}
\end{equation}
where 
\begin{equation}
\chi^{2,\SNe}_{i} = \frac{(m^0_{V,i} - m^{th}_{V,i})^2}{\sigma_{i}^2} \, , \quad
\tilde{N}^{\SNe}_{i} = \frac{1}{\sqrt{\sigma_{i}^2}} \, ,
\end{equation}
\begin{equation}
m^{th}_{V,i} = \mu_{0,i} + 5 \log H_0 - 25 - 5 a_V \, ,
\end{equation}
\end{subequations}
and $m^0_{V,i}$, $\sigma_i$ are taken from the table 3 in \cite{Riess:2011yx}. Here $a_V$ is the intercept of the SNe Ia magnitude-redshift relation, and \cite{Riess:2011yx} gives its value as $a_V = 0.697\pm0.00201$. We call the mean value 
in the above expression for the likelihood $a^{\rm R11}_V = 0.697$, the uncertainty $\sigma_{a_V} = 0.00201$ and assume that $a_V$ itself has a Gaussian pdf given by these quantities. If we were dealing with Gaussian likelihood for $m^0_{V,i}$ then we could marginalize analytically over $a_V$, which would then contribute a fully correlated error to the covariance matrix for the $m^0_{V,i}$. But as we are using HPs, we instead add $a_V$ as an explicit (nuisance) parameter in Eq.\ (\ref{eq:ialike}), together with its associated Gaussian likelihood, and sample from it numerically. Similarly, we take into account the calibration error, $\sigma_{a_{\rm cal}}$, between the ground based and the WFC3 photometry by introducing a nuisance parameter $a_{\rm cal}$. We assume it has a Gaussian pdf with zero mean and $\sigma_{a_{\rm cal}}=0.04$. 

Finally, motivated by the inconsistencies of distance anchors found by G. Efstathiou in  \cite{Efstathiou:2013via}, we include hyper-parameters also when dealing with the available distance moduli:
\begin{eqnarray} \label{obs_def}
\ln P^{\Anchors} & = & \ln \tilde{\chi}^{2}\left( \frac{(\mu_{\rm 0,4258} - \mu^{\obs_1}_{\rm 0,4258})^2}{\sigma^2_{\obs_1,\rm 4258}}\right) - \frac{\ln (2\pi\sigma_{\obs_1,\rm 4258}^2)}{2} \nonumber \\ 
& + & \ln \tilde{\chi}^{2}\left( \frac{(\mu_{\rm 0,4258} - \mu^{\obs_2}_{\rm 0,4258})^2}{\sigma^2_{\obs_2,\rm 4258}}\right) - \frac{\ln (2\pi\sigma_{\obs_2,\rm 4258}^2)}{2} \nonumber \\
& + & \ln \tilde{\chi}^{2}\left(\frac{(\mu_{0,\LMC} - \mu^\obs_{0,\LMC})^2}{\sigma^2_{0,\LMC}}\right) - \frac{\ln (2\pi\sigma_{0,\LMC}^2)}{2}
\end{eqnarray}
where the numerical values for $\mu_{0,\LMC}^\obs$ are given by Eq. \eqref{Eq:LMC-measured-distance-modulus} and  $\mu^{\obs_1}_{\rm 0,4258}$,  $\mu^{\obs_2}_{\rm 0,4258}$ in equations \eqref{Eq:NGC4258-measured-distance-modulus-2013} and \eqref{Eq:NGC4258-measured-distance-modulus-2015}, respectively.

\paragraph{Baseline Cepheid period} At this point we have assembled all ingredients necessary to determine the Hubble parameter, using HPs rather than a rejection algorithm. A few additional choices are however necessary. The first one concerns the period cut for the Cepheids: \cite{Riess:2011yx} uses periods up to 
205 days (effectively equivalent to no period cut), while \cite{Efstathiou:2013via} limits himself to periods shorter than 60 days. As we will discuss in the next section, we see no significant trend for the LMC and MW Cepheid stars that would justify a tighter cut when using the R11 data. For our choice of the baseline, discussed in this section, we perform no cut in the period (equivalent to choosing a $205$ days cut as \cite{Riess:2011yx}); in the next section, we also perform the analysis for both cuts and report the results in Table \ref{Table:Constraints-main-analysis-thesis-PC}.
The difference between using data with $P<60$ or $P<205$ is smaller than $0.5\, \km \, \second^{-1} \, \Mpc^{-1}$ in $H_0$ (with a somewhat larger impact on $b_W$). For this reason we use the larger data set, $P<205$, for our
final numbers.

\paragraph{Baseline metallicity in Leavitt Law} The second choice concerns the treatment of a metallicity dependence in the Cepheid fits. As we will discuss in Subsection \ref{Subsection:Zw-dependence}, this is still an open question. The combined Cepheid data used here is unable to significantly constrain $Z_W$, instead we find a strong degeneracy with $M_W$ (see Fig. \ref{Fig:Main-analysis-fitM1a}). From a Bayesian model comparison point of view, there is no significant preference for specific priors or $Z_W=0$. However, looking at fits $(29)$--$(39)$ in Tables \ref{Table:details-fits}--\ref{Table:Constraints-main-analysis} we see that also this choice has no significant impact on $H_0$. From theoretical
arguments there is probably some dependence on metallicity, but as we cannot strongly constrain it, we have decided to use the `strong' prior, $Z_W = 0 \pm 0.02$, as our baseline model.

\paragraph{Baseline result} These choices correspond to the fit (29) of Tables \ref{Table:details-fits}--\ref{Table:Constraints-main-analysis}. The resulting constraint on the Hubble parameter is:
\begin{equation}\label{Eq:H0-value-standard-analysis}
	H_0 = 75.0 \pm 3.9 \, \km \, \second^{-1} \, \Mpc^{-1} \, \,\,\,\,\,\,\,\,\, (\rm{using\,\,\, R11 \,\,\,data}).
\end{equation}

\paragraph{Discussion} The last three columns in Tables \ref{Table:details-fits}--\ref{Table:Constraints-main-analysis} show the normalised weights (\ref{Eq:normalised-weights}).  We  also use these normalised weights to asses compatibility of the whole data set.
 For our particular case of three kinds of data, and as long as the normalised weight is not larger than expected for Gaussian data, $\sum_j|| \alpha^{j} ||\lesssim 3\times 0.85 = 2.55$ (see the discussion after Eq.\ \eqref{Eq:normalised-weights}), the most internally consistent fit would be that with $\max(\sum_j|| \alpha^{j} ||)$. This happens to be the case for our primary fit $(29)$, for which  $\sum_j|| \alpha^{j}||=2.32$. This value is our baseline result for the R11 data set and it is compared to results from previous analyses in Fig.\ \ref{Fig:H0bestfits} (bottom panel). 
 Our `baseline analysis' and its variants (see fits $(29)$--$(39)$ in Tables \ref{Table:details-fits}--\ref{Table:Constraints-main-analysis}  and Fig. \ref{Fig:H0-values-3-anchors}) agree with previous measurements (\cite{Riess:2011yx} and \cite{Efstathiou:2013via}) when using the same data set and choice of anchors. While analyses using HPs agree with the most recent direct local measurement of $H_0$ \cite{Riess:2016jrr}, the H0LiCOW time delay based measurement \cite{Bonvin:2016crt} and the WMAP 2009 indirect determination \cite{Hinshaw:2012aka}, our result is only marginally compatible with the indirect determination by the Planck collaboration \cite{Ade:2015xua}. Although we have not found reasons to discard any of the data sets, in Subsection \ref{subsection:Anchors} we have carried out analyses simultaneously including only one or two anchors. Those cases suggest that i) inclusion of MW Cepheid variables drives $H_0$ to higher values independently of both prior on the metallicity parameter and period cut for period-luminosity relation ii) a strong prior on the metallicity parameter when including distance modulus to LMC also drives $H_0$ to higher values. We discuss this with more details in the next section. 

\begin{figure}[t]
\centering
{\includegraphics[width=0.63 \textwidth]{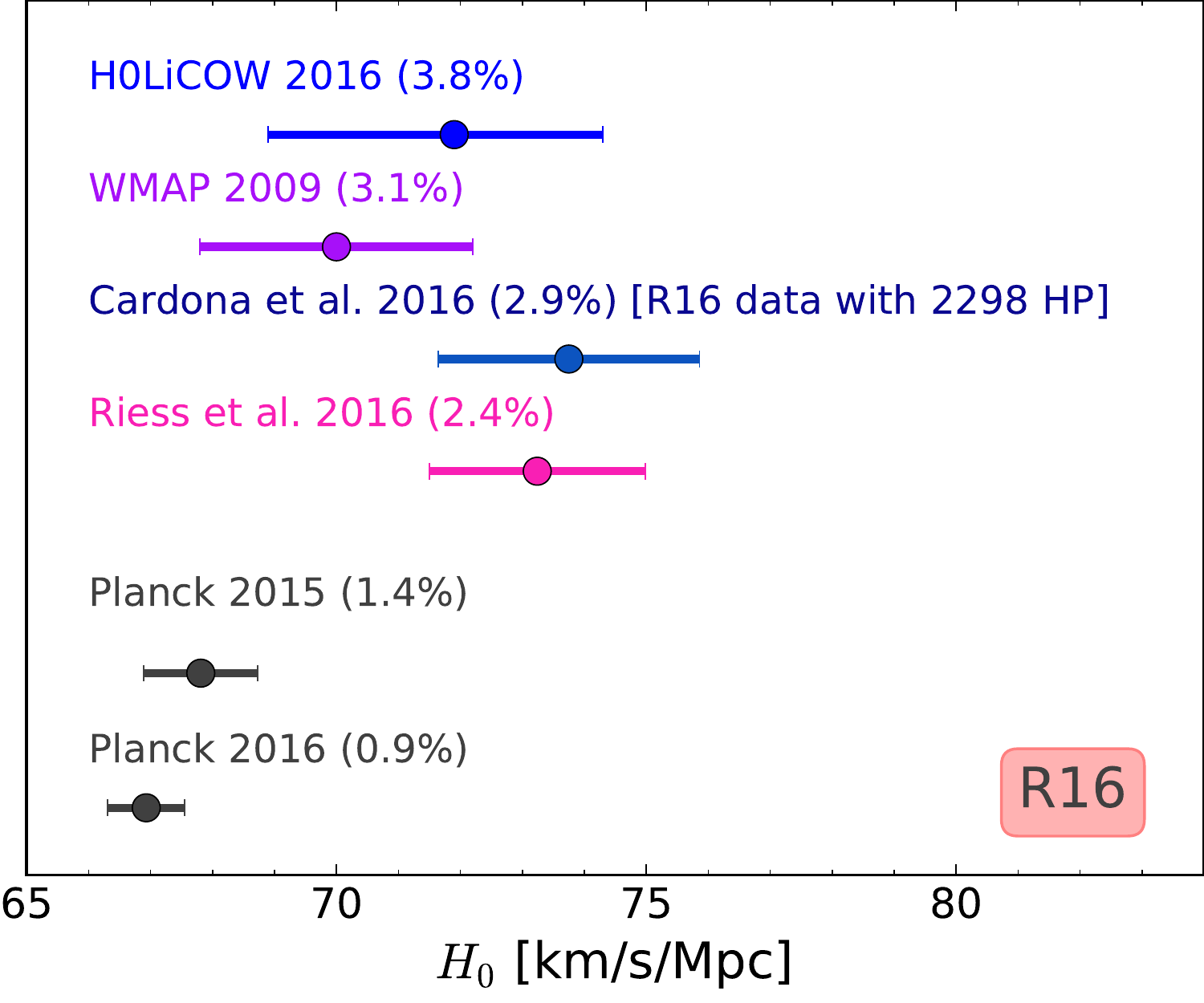} 
 \vspace{0.5cm}
 
\includegraphics[width=0.63 \textwidth]{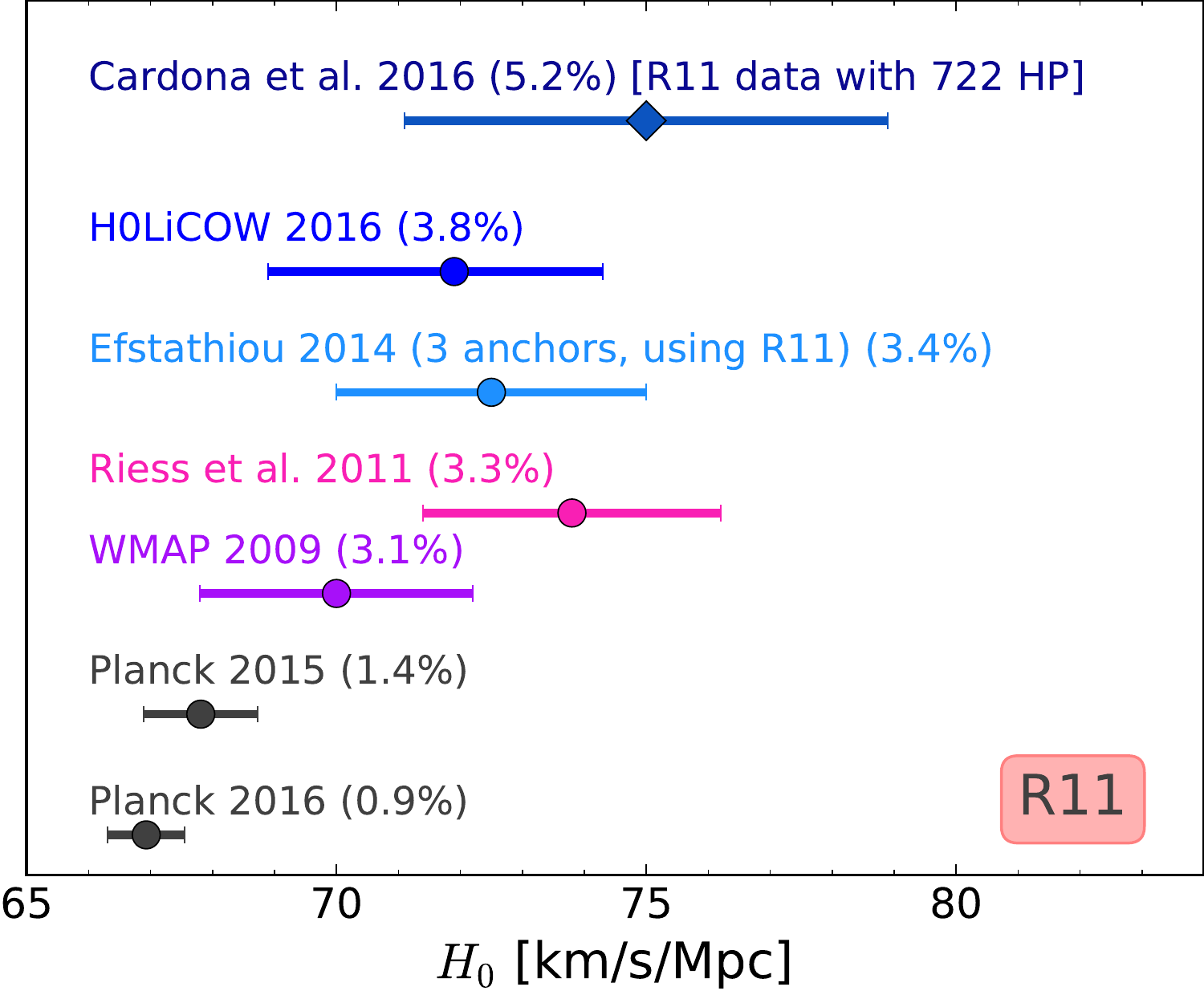}}
\caption{Different determinations of the Hubble constant. The top panel includes direct measurements that used R16 data; indirect measurements are also shown. From top to bottom (less precise to more precise):  the H0LiCOW distance measurement \cite{Bonvin:2016crt}, the indirect determination by the WMAP team \cite{Hinshaw:2012aka}, our `baseline analysis' for the R16 data set in Eq. \eqref{Eq:primary-best-fit-R16}, the Riess et al. measurement \cite{Riess:2016jrr} (R16), the indirect measurement by the Planck collaboration \cite{Ade:2015xua} 
and the revised indirect measurement by the Planck collaboration \cite{Aghanim:2016yuo}.
The bottom panel includes direct measurements that used R11 and the same indirect measurements as in the top panel. From less precise to more precise: our `baseline analysis' for the R11 data set in Eq. \eqref{Eq:H0-value-standard-analysis}, 
H0LiCOW,  Efstathiou's measurement using three anchor distances and a $60$ days period cut-off \cite{Efstathiou:2013via}, measurement by Riess et al.\cite{Riess:2011yx} (R11), WMAP 2009, Planck 2015, Planck 2016.
 }\label{Fig:H0bestfits}
\end{figure}

\begin{figure}[hbtp]
\centering
\includegraphics[width=1 \textwidth]{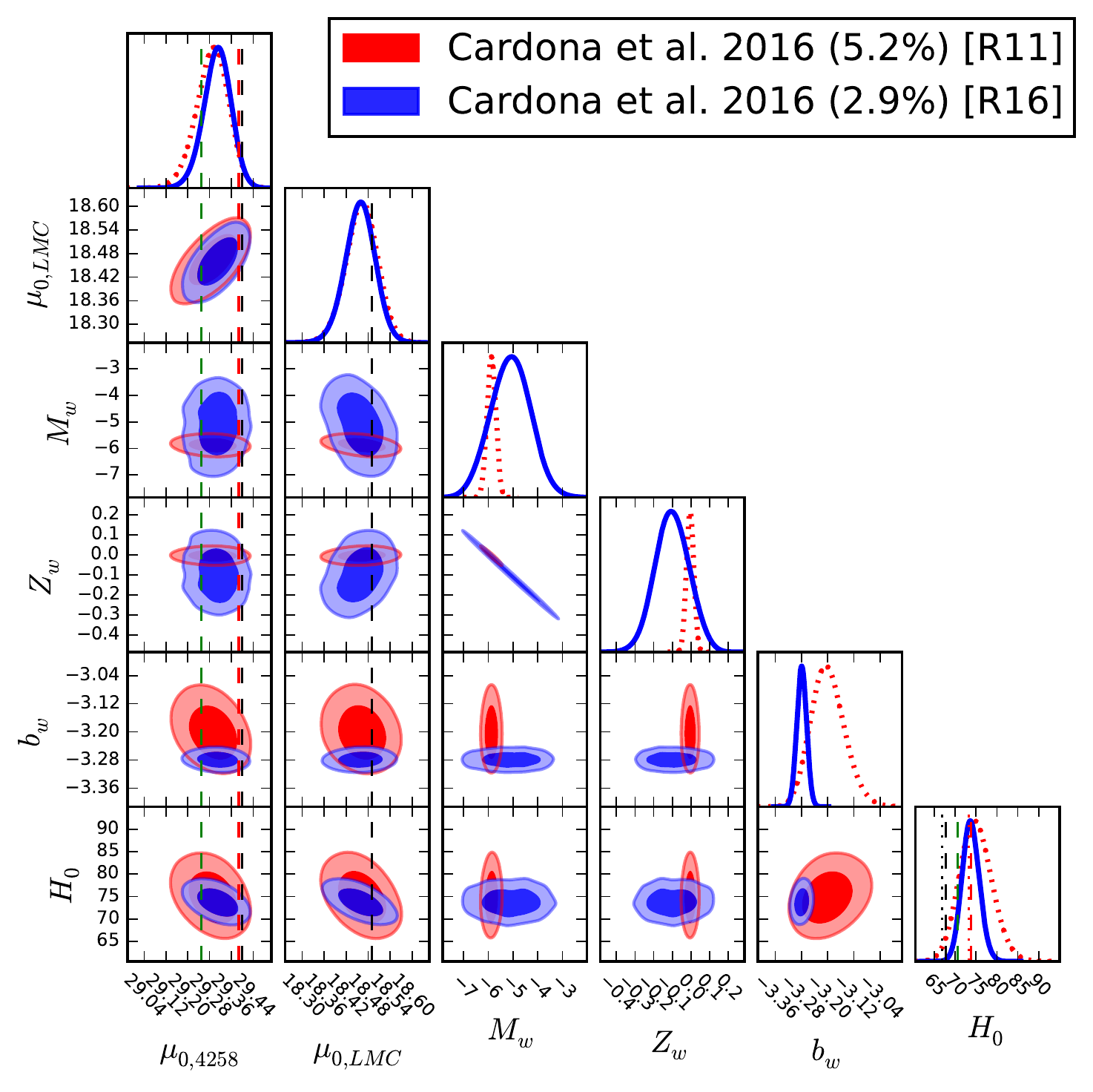}
\caption{Posterior constraints for fit $(29)$ (red contours) and fit $(44)$ (blue contours), marginalizing over HPs. The analysis with R11 data uses a strong prior on the metallicity $Z_w$ but is otherwise weaker because it uses less data than R16. Green, red and black vertical dashed lines in $\mu_{0,4258}$ column indicate NGC 4258 distance modulus from \cite{Polshaw:2015ika}, \cite{Riess:2016jrr} and \cite{Humphreys:2013eja}, respectively. The black dashed vertical line in $\mu_{0,\LMC}$ column shows the LMC distance modulus from \cite{Pietrzynski:2013gia}. The black dotted vertical line in the $H_0$ column indicates the updated Planck 2016 value for the base six-parameter $\Lambda$CDM model \cite{Aghanim:2016yuo}. Black, green, and red dashed vertical lines in $H_0$ column respectively indicate the values derived by the Planck collaboration for the base six-parameter $\Lambda$CDM model \cite{Ade:2015xua}, Efstathiou's value \cite{Efstathiou:2013via} used by the Planck collaboration as a prior, and the $3\%$ measurement reported by \cite{Riess:2011yx}; The red dotted vertical line indicates the best estimate from the analysis in \cite{Riess:2016jrr}. 
The numbers of HPs is 722 for fit $(29)$ (red contours here) and 2298 for fit $(44)$ (blue contours here).
}
\label{Fig:Main-analysis-fitM1a}
\end{figure}

\begin{figure}[hbtp]
\centering
\includegraphics[scale=.75]{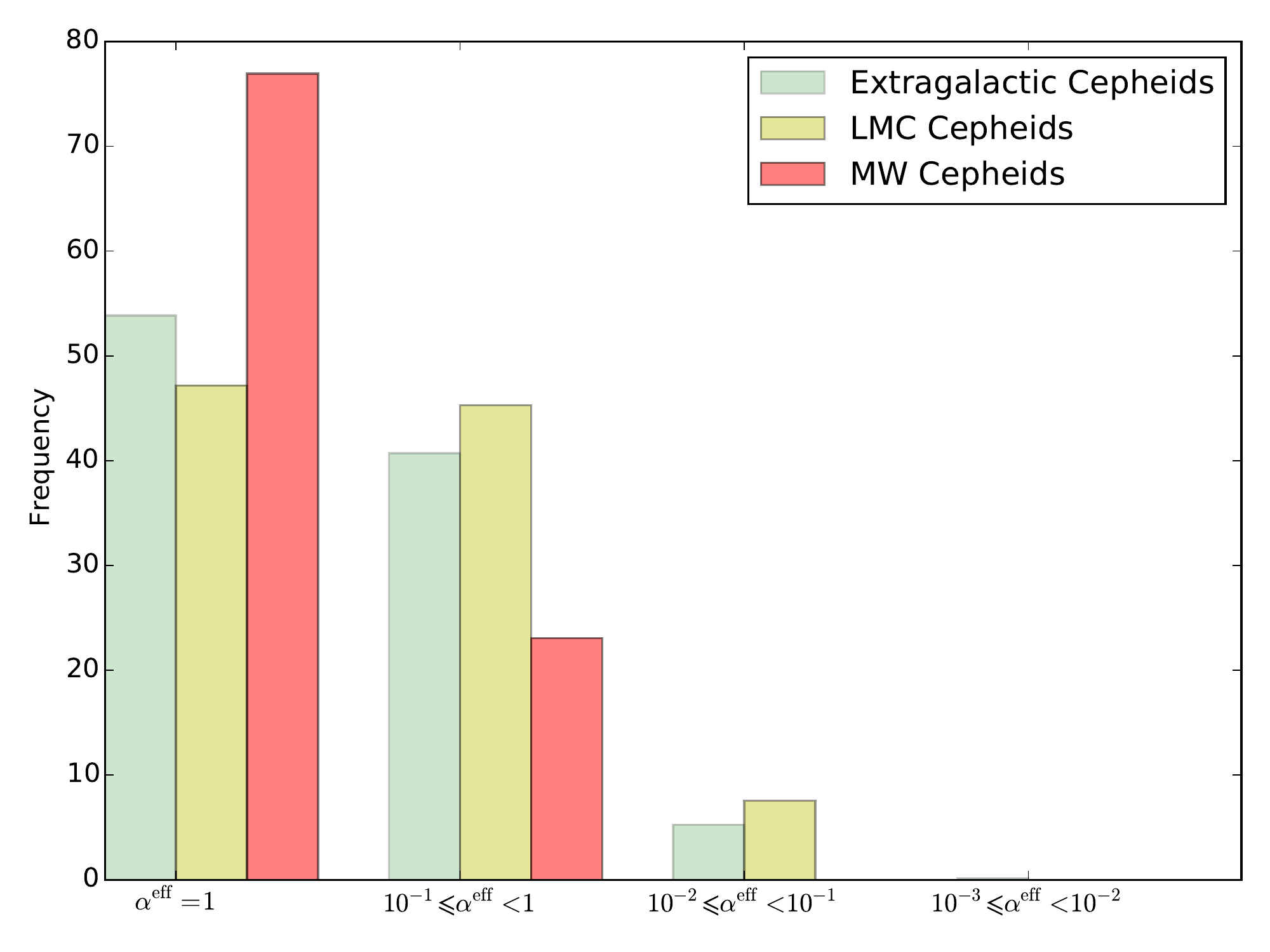}
\caption{Effective hyper-parameters for the R11 Cepheid sample used in fit $(29)$. Out of the 646 Cepheid variables in the 8 $\SNe$ host galaxies and in the $\NGC$ megamaser system, 
 348 have $\alpha^{\eff}=1$, 263 $10^{-1}\leq \alpha^{\eff} < 1$, 34  $10^{-2}\leq \alpha^{\eff} < 10^{-1}$, and 1 $10^{-3} \leq \alpha^{\eff} < 10^{-2}$. Out of the 53 LMC Cepheid variables, 25 have $\alpha^{\eff}=1$, 24 $10^{-1}\leq \alpha^{\eff} < 1$, 4  $10^{-2}\leq \alpha^{\eff} < 10^{-1}$. Out of the 13 MW Cepheid stars, 
  10 stars have $\alpha^{\eff}=1$ and 3 stars with $10^{-1}\leq \alpha^{\eff} < 1$ (compare with Fig. \ref{Fig:MW-Cepheid-variables}). Overall, $23\%$ of the MW Cepheids are down-weighted; the fraction reaches $46\%$ for extragalactic Cepheids in \cite{Riess:2011yx}; as for the LMC Cepheid variables, the analysis down-weights $53\%$ of the stars.}
\label{Fig:effective-HP-fitM1a}
\end{figure}

\begin{figure}[hbtp]
\centering
\includegraphics[scale=0.75]{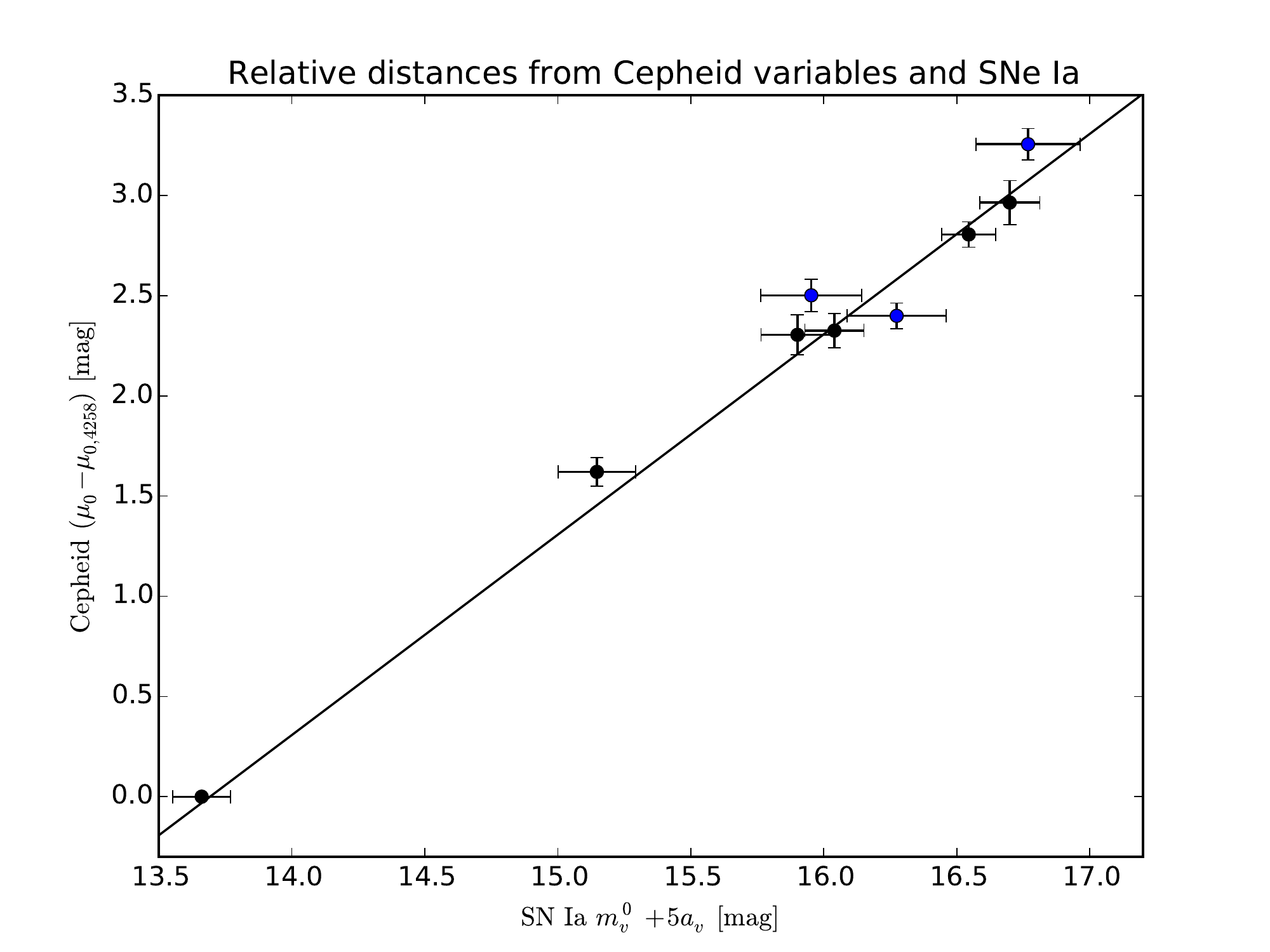}
\caption{Relative distances from Cepheids and SNe Ia. We plot the peak apparent visual magnitudes of each $\SNe$ (from Table 3 in \cite{Riess:2011yx}) with error bars rescaled by HPs (colour code is the same as in Fig. \ref{Fig:LMC-Cepheid-variables-fit-c}) against the relative distances between hosts determined from the `baseline analysis' fit $(29)$ in Tables \ref{Table:details-fits}--\ref{Table:Constraints-main-analysis}. The solid line shows the corresponding best fit. The first point on the left corresponds to the expected reddening-free, peak magnitude of an $\SNe$ appearing in the megamaser system NGC 4258 which is derived from the fit $(29)$. }
\label{Fig:HP-SNIa-main-analysis}
\end{figure}

\begin{table}[tbp]
\centering
\begin{tabular}{@{}lccccr}
\hline
\multicolumn{6}{c}{Distance parameters} \\
\hline
Host & SN Ia & $\mu_{0,i}-\mu_{0,4258}$ & $\mu_{0,i} $ best & $\alpha_{\eff}$ & $\sigma_{\intt,i}$\\
\hline

 n4536 & SN 1981B & $1.620\,(0.071)$&$30.91\,(0.08)$ &$1$ & $0.1$ \\

 n4639 & SN 1990N & $2.325\,(0.085)$& $31.61\,(0.09)$& $1 $ & $0.03$\\

 n3370 & SN 1994ae & $2.805\,(0.063)$& $32.09\,(0.07)$& $1 $ & $0.02$ \\
 
 n3982 & SN 1998aq & $2.502\,(0.081)$& $31.79\,(0.08)$& $ 0.23$ & $0.03$\\
  
 n3021 & SN 1995al & $2.964\,(0.110)$& $32.25\,(0.11)$& $ 1$ & $0.03$\\
    
 n1309 & SN 2002fk & $3.255\,(0.079)$& $32.54\,(0.08)$& $ 0.28$ & $0.03$\\

 n5584 & SN 2007af & $2.399\,(0.064)$& $31.69\,(0.07)$& $ 0.43$ & $0.03$\\
       
 n4038 & SN 2007sr & $2.304\,(0.099)$& $31.59\,(0.11)$& $ 1$ & $0.03$\\
       
\hline
\end{tabular}
\caption{\label{Table:SNIa-HP-fit-M1a} Distance parameters for the SNe Ia hosts corresponding to our `baseline analysis', fit $(29)$. Numbers in brackets indicate the standard deviation. The last two columns correspond to the effective HP for each SN Ia host and the corresponding internal scatter, respectively.}
\end{table}
We show the parameter constraints for the baseline case in Fig.\ \ref{Fig:Main-analysis-fitM1a}, where we compare our analysis done on R11 and R16 data. The dashed vertical lines in the 1D marginal posterior for $H_0$ of Fig.\ \ref{Fig:Main-analysis-fitM1a} indicate (from left to right) the mean values of the Planck $\Lambda$CDM determinations of $H_0$ ($66.93 \pm 0.62 \, \km \, \second^{-1}\, \Mpc^{-1}$ in 2016 from temperature and polarisation data, and $67.8\pm0.9\, \km \, \second^{-1} \, \Mpc^{-1}$ in 2015 from temperature and lensing data \cite{Aghanim:2016yuo,Ade:2015xua}) 
and of the analyses of \cite{Efstathiou:2013via} ($70.6\pm3.3 \, \km \, \second^{-1} \, \Mpc^{-1}$), \cite{Riess:2016jrr} ($73.24\pm1.74\, \km \, \second^{-1} \, \Mpc^{-1}$), and \cite{Riess:2011yx} ($73.8\pm2.4\, \km \, \second^{-1} \, \Mpc^{-1}$ ). Our result here agrees best with the latter two (although our width is somewhat larger due to the use of hyper-parameters), but even the Planck value lies within our $95\%$ credible interval. Note that as the HP likelihoods have wide wings and are very non-Gaussian, it could be the case that the likelihood for $H_0$ also is very
non-Gaussian. We found however that it is relatively close to a normal pdf.

As we mentioned above, the available distance moduli are included with HPs in our `baseline analysis'. The resulting effective HPs are: $\alpha^{\eff}_{\LMC}=1$, $\alpha^{\eff,\rm obs1}_{\rm 4258}=0.6$, $\alpha^{\eff,\rm obs2}_{\rm 4258}=1$ as defined in (\ref{obs_def}). The last two effective HPs correspond to the observed $\NGC$ distance moduli that we introduce in the context of the R16 analysis in \eqref{Eq:NGC4258-measured-distance-modulus-2013} and \eqref{Eq:NGC4258-measured-distance-modulus-2015}, respectively. This shows that the geometric maser distance estimate to the active galaxy NGC 4258 from \cite{Humphreys:2013eja} is slightly down-weighted in our analysis.  As can be seen from the vertical, red, dashed line in Figure \ref{Fig:Main-analysis-fitM1a}, the revised maser distance to NGC 4258 from \cite{Riess:2016jrr} is now closer to our $68\%$ confidence region than the previously used maser distance \cite{Humphreys:2013eja} (black vertical dashed line).

Finally, we also note that the difference between the supernova distance \cite{Polshaw:2015ika} and
the maser distance \cite{Humphreys:2013eja} to NGC 4258 is comparable to the posterior uncertainty on $\mu_{0,4258}$ shown in Fig.\ \ref{Fig:Main-analysis-fitM1a}. Looking at the marginal 2D posterior for $\mu_{0,4258}$ and $H_0$ gives the impression that the difference may be limiting the precision on $H_0$ currently achievable, and that the very recent supernova distance would prefer a higher $H_0$. Note, however, that our `baseline analysis' is almost unaffected by the inclusion of the supernova distance \cite{Polshaw:2015ika} to the megamaser system NGC 4258. Fit $(30)$ includes only the geometric maser distance \cite{Humphreys:2013eja} and we find negligible changes in $H_0$ and $M_W$ w.r.t our `baseline analysis'. A significantly lower $H_0$ only results when leaving out the MW Cepheid distance anchor and/or using fewer hyper-parameters, as mentioned above.

Figure \ref{Fig:effective-HP-fitM1a} shows a histogram for HPs in the sample of Cepheid variables used in our `baseline  analysis' (extragalactic, LMC and MW Chepheids). Whereas in R11 \cite{Riess:2011yx} about 20\% of the Cepheids are rejected by the Chauvenet's criterion (with rejection of datum $i$ being equivalent to setting $\alpha_i=0$ in our scheme), our analysis finds that about 46\% of the Cepheid variables in \cite{Riess:2011yx} are down-weighted when using HPs ($\alpha^\eff_i <1$). A 20\% rejection corresponds to an average $\alpha$  of less than $0.8$, as defined in (\ref{Eq:normalised-weights}),  while our baseline fit has $||\alpha^{\rm{Cepheid}} || = 0.72$. The analysis in R16 \cite{Riess:2016jrr}, also using a Chauvenet's criterion, finds an outlier fraction of $2-5\%$ in a larger sample of Cepheid variables.

Table \ref{Table:SNIa-HP-fit-M1a} shows the distance parameters for the SNe Ia hosts of our baseline analysis, together with the corresponding effective HP and internal scatter for each host. The entries with $\alpha^{\eff}<1$ point to the presence of possible outliers among the sample of SNe Ia hosts, justifying our use of HPs in the apparent visual magnitudes of each SN Ia. This is also visible in Figure \ref{Fig:HP-SNIa-main-analysis} where the three `blue' supernovae have their error bars enlarged in order to remain consistent with the global best fit (the diagonal solid line).
This could be a hint of unaccounted systematics in the light-curve fits for those SNe Ia. Note that R16 uses a different light-curve fitting algorithm (SALT-II) to that utilised in R11
(MLCS2k2) finding no evidence for any of their 19 SNe Ia hosts to be an outlier.


\section{Testing assumptions}
\label{Section:Application}

In this section we look at the choices that we made for the baseline analysis described in the previous section. Specifically, in Subsection \ref{sec:cepheids} we look at the consistency of the Cepheid fits and the distance anchors. In Subsection \ref{Subsection:Zw-dependence} we consider the metallicity dependence of the Cepheid fits. In Subsections \ref{Subsection:period-cut} and \ref{Subsection:HP-variants} we study to what extent both the period cut-off and inclusion of different kinds of data with HPs make a difference in our analysis. For this section we use as reference the R11 data. In Section \ref{Section:Application-R16} we will then highlight how our `baseline analysis' changes when analysing the R16 data set with respect to our analysis of R11 data set.

\subsection{Consistency tests for Cepheid distances, using hyper-parameters\label{sec:cepheids}}

In this subsection we take a closer look at the Cepheids in the LMC, the MW, and the megamaser system NGC 4258. This will allow us to show in a simple example how the HPs affect the analysis, to check whether the Cepheids and anchor distances of these systems are in agreement, and also to compare our outcome with results in \cite{Efstathiou:2013via}.

\subsubsection{The LMC Cepheid variables}
\label{Subsection:LMCR11}

We start out our analysis by applying HPs to the set of 
53 Large Magellanic Cloud (LMC) Cepheid variables with $H$-band magnitudes, $m_H$, listed in Table 3 of  \cite{2004AJ:128.2239P} and $V$, $I$ band magnitudes, $m_V$, $m_I$, listed in Table 4 of \cite{2002ApJS:142:71S}. Following \cite{Riess:2011yx,Efstathiou:2013via,Riess:2016jrr}, we rely primarily on (near-infrared) NIR `Wesenheit reddening-free' magnitudes, defined as
\begin{equation}\label{Eq:Wesenheit-reddening-free}
m_{W,i} = m_{H,i} - R (m_{V,i} - m_{I,i}),
\end{equation}
where $R$ is a constant defining the reddening law; when analysing the R11 data set, we use $R=0.410$ as G. Efstathiou did \cite{Efstathiou:2013via}; when utilising the R16 data set, we study the sensitivity of our results to variations in $R$. For the purpose of comparing with \cite{Efstathiou:2013via} we neglect for now metallicity dependence and fit the data with a period-luminosity relation
\begin{equation}\label{Eq:P-L-LMC}
m^{\Cepheid}_W(P) \equiv m^P_W = A + b_W (\log P - 1)
\end{equation}
where $A = \mu_{0,\LMC} + M_W$ in notation of Eq. \eqref{Eq:P-L-equation} and $P$ is the period. In order to apply HPs we use (\ref{Eq:chi2}) as done already in (\ref{cepheid_formulae}) and define 
\begin{equation}\label{Eq:chi2-LMC}
\chi^{2,\LMC}_{i} = \frac{(m_{W,i} - m^P_{W})^2}{\sigma_i^2 + \sigma^{2,\LMC}_{\intt}},
\end{equation}
where $\sigma_i$ is the observational error on $m_{W,i}$ and $\sigma_{\intt}^{\LMC}$ is the `internal scatter', discussed in Section \ref{Section:Application-R11} after Eq.\ (\ref{eq:3c}). 
Maximizing  
\begin{equation}
\label{Eq:likelihood-HP}
\ln P^{\LMC}(A,b_W,\lbrace D_i \rbrace) = \sum_{i} \ln \tilde{\chi}_i^{2}(\chi^{2,\LMC}_{i}) + \ln \tilde{N}^{\LMC}_{i},
\end{equation}
where 
\begin{equation}
\label{Eq:normalization}
\tilde{N}^{\LMC}_i = \frac{1}{\sqrt{\sigma_i^2 +\sigma^{2,\LMC}_{\intt}}},
\end{equation}
we find the best-fitting parameters A, $b_{\rm{W}}$, $\sigma_{\rm{int}}$ of the period-luminosity relation \eqref{Eq:P-L-LMC}. Table \ref{Table:LMC-fits} shows results for the LMC Cepheid variables and different period cuts [fits (a),(b),(c)] as opposed to the standard $\chi^2$ minimization, without HPs [fit (d)]. 
Using the direct distance determination to the LMC \cite{Pietrzynski:2013gia},
\begin{equation}
\mu_{0,\LMC}^{\obs} = 18.49 \pm 0.05,
\label{Eq:LMC-measured-distance-modulus}
\end{equation}
and the result for $A$ from Table \ref{Table:LMC-fits} based on fit (c) gives a Cepheid zero point
\begin{equation}\label{Eq:zero-point-LMC-R11}
M_W = -5.93 \pm 0.07 .
\end{equation}
We will comment on the influence of the period cut later in this section (Subsection \ref{Subsection:period-cut}). For now we show in Figure \ref{Fig:LMC-Cepheid-variables-fit-c} the period-luminosity relation for the LMC Cepheid variables and the best fits of cases (c) (solid black line) and (d) (red dotted line) in Table \ref{Table:LMC-fits}. 
In Figure \ref{Fig:LMC-Cepheid-variables-fit-c} we show the colour-coded effective HP values for all the data points in the LMC Cepheid sample, based on (\ref{Eq:effective-HP-1}) and (\ref{Eq:effective-HP-2}). Green points are outliers: the lower panel, where we plot the residuals with respect to the best fit (c), shows how outliers have a lower effective hyper-parameter and thus less weight. If instead we used a standard $\chi^2$ fit to all of these points [i.e.\ without period cut, fit (d)] then the fit would be pulled towards a steeper slope (higher $b_W$) by the green points near the extrema of $P$ as shown by the red dotted line in the upper panel of Fig. \ref{Fig:LMC-Cepheid-variables-fit-c}. In other words, while a standard minimization would be influenced by outliers, HPs downgrade their relevance.

Fig.\ \ref{fig:sigmaint} shows the posterior probability distribution function (pdf) of the parameters A, $b_{\rm{W}}$, $\sigma_{\rm{int}}$ for the fits (c) and (d) of Table \ref{Table:LMC-fits}, i.e. with and without use of HPs.

\begin{table}[tbp]
\centering
\begin{tabular}{@{}ccccc}
\hline
\multicolumn{5}{c}{LMC Cepheid variables} \\
\hline
Fit & $A$ & $b_W$ & $\sigma_{\intt}^{\LMC}$ & Period cut \\
\hline
 a & $12.570\,(0.035)$ & $-3.32\,(0.10)$ & $0.06$ & $10<P<60$ \\
  
 b & $12.562\,(0.016)$&$-3.30\,(0.05)$ & $0.06$ & $P<60$ \\

 c & $12.562\,(0.016)$& $-3.31\,(0.05)$& $0.06$ & $P<205$ \\

 d & $12.555\,(0.019)$& $-3.24\,(0.06)$& $0.12$ & $P<205$ \\
\hline
\end{tabular}
\caption{\label{Table:LMC-fits} Mean value and standard deviation (in brackets) for the parameters in the period-luminosity relation. Fits (a), (b), and (c) use HPs whereas fit (d) is a standard $\chi^2$ minimization as done in \cite{Efstathiou:2013via}.}
\end{table}

\begin{figure}[tbp]
\centering 
\includegraphics[scale=0.75]{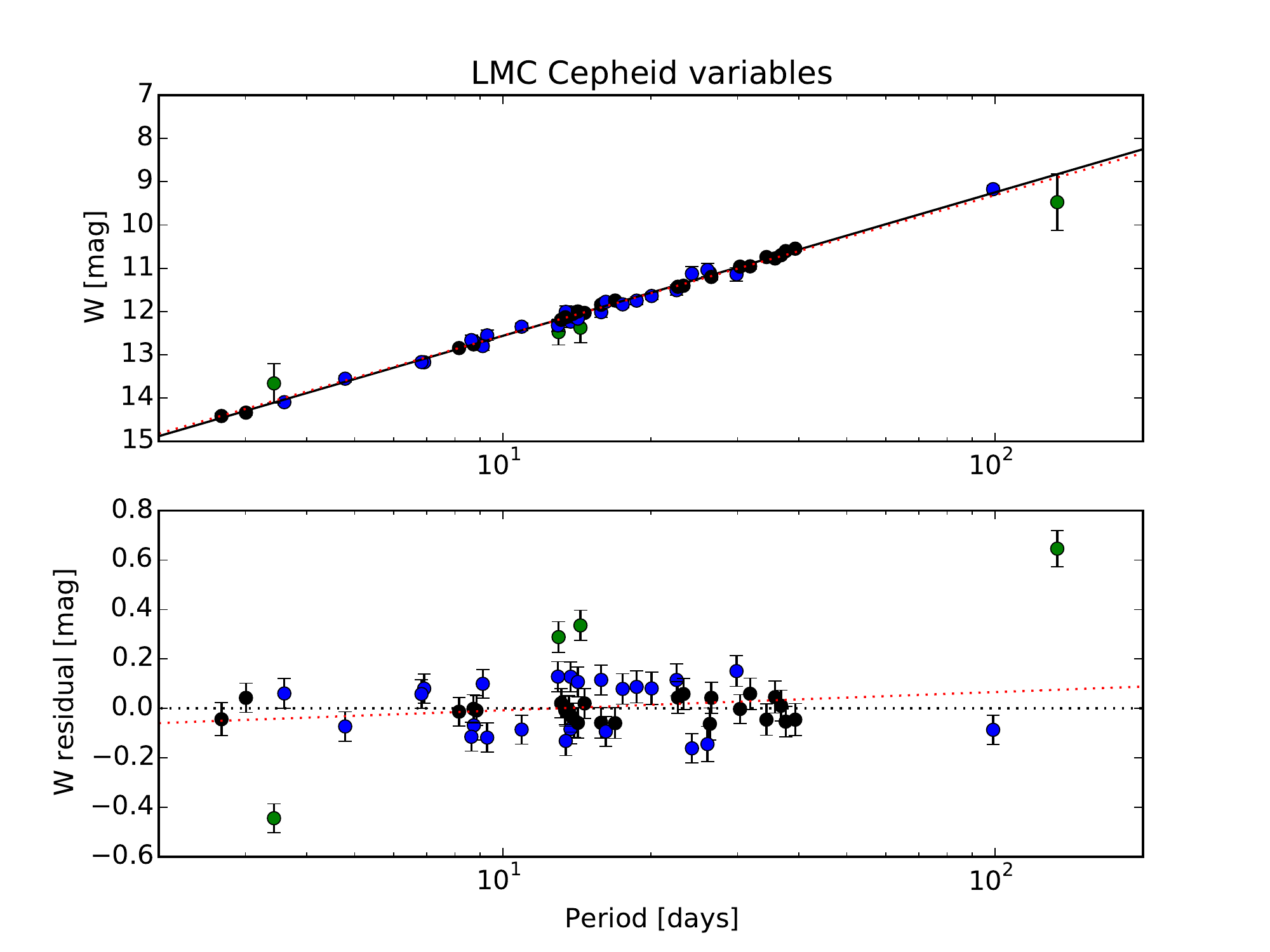} 
\caption{Period-luminosity relation for the LMC Cepheid variables. The upper panel shows the best fit of both the case (c) (solid black line) and case (d) (red dotted line) in Table \ref{Table:LMC-fits}. Error bars have been rescaled with corresponding effective HPs which are colour-coded as follows: black if $\alpha^{\eff} = 1$, blue if $10^{-1}\leq \alpha^{\eff} < 1$, green if $10^{-2}\leq \alpha^{\eff} < 10^{-1}$. The lower panel shows magnitude residuals with respect to the best fit (c); error bars are not rescaled in the lower panel. The red dotted line in the lower panel is the difference between the red and black lines in the top panel. Note that the y axis in the top panel is decreasing from bottom to top.}
\label{Fig:LMC-Cepheid-variables-fit-c}
\end{figure}

\begin{figure}[hbtp]
\centering
\includegraphics[scale=1]{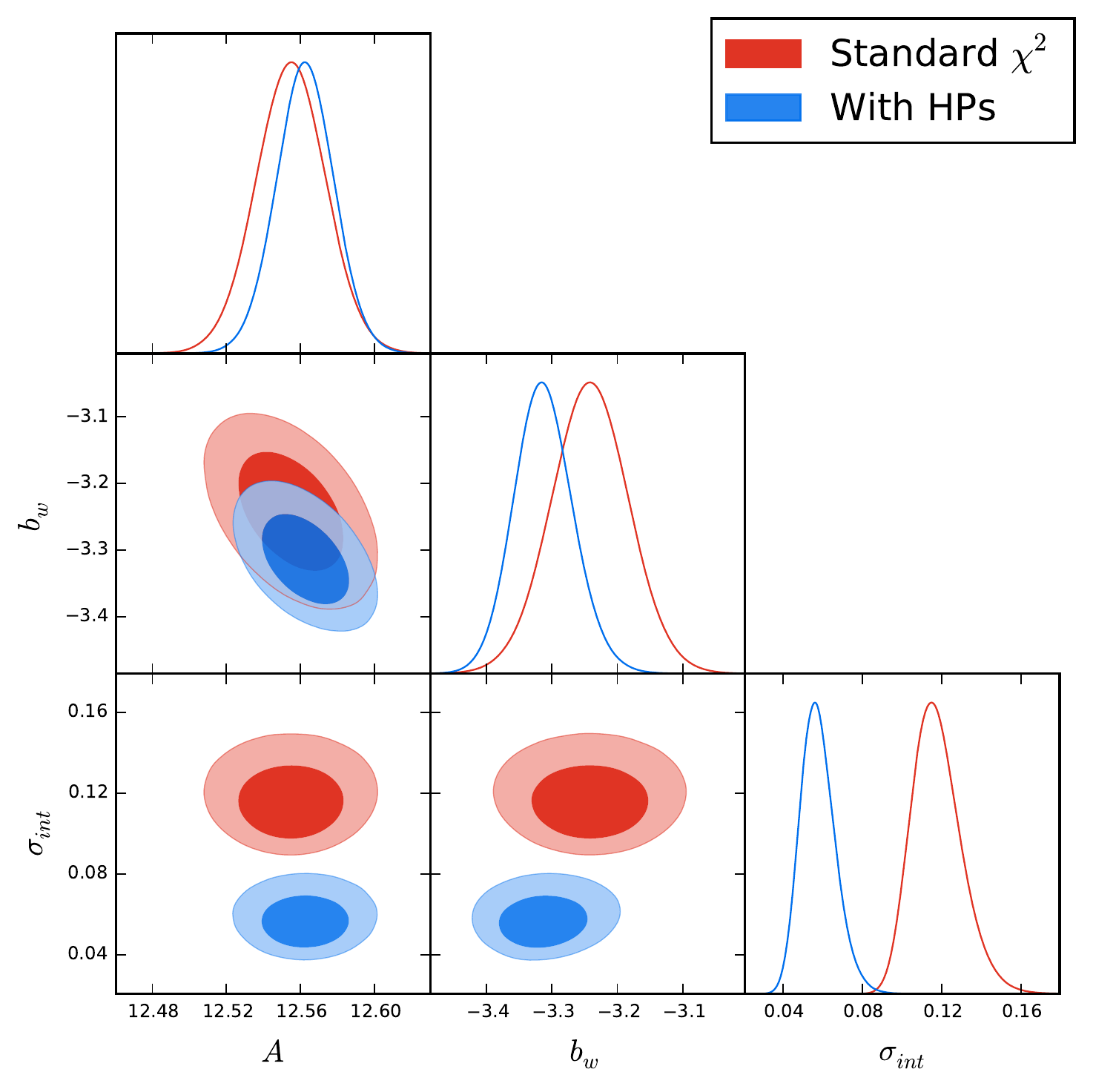}
\caption{Two- and 1D posteriors for the parameters in the period-luminosity relation of LMC Cepheid stars. Red shows a standard $\chi^2$  minimisation [fit (d)] and blue uses HPs [fit (c)].}
\label{fig:sigmaint}
\end{figure}

The LMC Cepheids are also treated as illustrative example in section 2 of \cite{Efstathiou:2013via}. This allows us to compare the approach used there, which does not use hyper-parameters, with the results found here. In our approach, the posterior mean values of $A$ and $b_W$ always lie in between (and agree with) the values given in fits (4a) and (4b) of \cite{Efstathiou:2013via} where period cuts of $P<60$ and $10<P<60$ are used. Our results do not depend significantly on the period cut (except that the error becomes larger for the most restrictive cut, $10<P<60$). We conclude that our treatment performs reasonably well when compared to the standard $\chi^2$ approach. We will investigate this conclusion further as we add more data.

The LMC Cepheids provide a nice illustration of how hyper-parameters can lead to smaller error bars. If we just naively fit the data shown in Fig.\ \ref{Fig:LMC-Cepheid-variables-fit-c} then we need to use a large $\sigma_\intt \approx 0.12$ in order to obtain a reasonable goodness of fit. When using hyper-parameters however, the outliers are down-weighted and $\sigma_\intt \approx 0.06$ suffices. The fact that we can use a smaller $\sigma_\intt$ with HPs then leads to smaller uncertainties also on the other parameters, as can be seen in Fig.\ \ref{fig:sigmaint} and Table \ref{Table:LMC-fits}.

\subsubsection{The Milky Way Cepheid variables}
\label{Subsection:MW-1}

We discuss next the set of $13$ Milky Way (MW) Cepheid stars with parallax measurements listed in Table 2 of \cite{vanLeeuwen:2007xw} (eliminating Polaris as in \cite{Efstathiou:2013via} and correcting for Lutz-Kelker bias). We consider the MW Cepheids separately here because, as we will see, the MW data pushes the inferred value of $H_0$ to higher values, and it is thus important to check whether there is a reason to discard this data set (as done in \cite{Efstathiou:2013via}) or not.

The period-luminosity relation for MW Cepheid stars in terms of the absolute Wesenheit reddening-free luminosity $M_W$ for a Cepheid with period $P$ is given by
\begin{equation}\label{Eq:P-L-MW}
M^P_W = M_W + b_W (\log P - 1)
\end{equation}
where we again neglected the metallicity dependence ($Z_W=0$) in Eq.\ \eqref{Eq:P-L-equation}. For the MW Cepheids $M^P_W = m^P_W - \mu_\pi$ and $\mu_\pi$ is the distance modulus derived from parallaxes. Again here we define 
\begin{equation}\label{Eq:chi2-MW}
\chi^{2,\MW}_{i} = \frac{(M_{W,i} - M^P_{W})^2}{\sigma_i^2 + \sigma^{2,\MW}_{\intt}},
\end{equation}
where $\sigma_i$ is the observational error on $M_{W,i}$ and $\sigma_{\intt}^{\MW}$ is the internal dispersion (that we again include as a free, marginalised parameter as in the previous section on the LMC Cepheids). Maximizing  
\begin{equation}
\label{Eq:likelihood-HP}
\ln P^{\MW}(M_W,b_W,\lbrace D_i \rbrace) = \sum_{i} \ln \tilde{\chi}_i^{2}(\chi^{2,\MW}_{i}) + \ln \tilde{N}^{\MW}_{i},
\end{equation}
where 
\begin{equation}
\label{Eq:normalization}
\tilde{N}^{\MW}_i = \frac{1}{\sqrt{\sigma_i^2 +\sigma^{2,\MW}_{\intt}}},
\end{equation}
we find the best-fitting parameters of the period-luminosity relation \eqref{Eq:P-L-MW}, given by
\begin{equation}\label{Eq:MW-bestfit}
M_W = -5.88 \pm 0.07  , \qquad b_W = -3.30 \pm 0.26  , \qquad \sigma_{\intt}^{\MW} = 0.02 ,
\end{equation}
in good agreement with fits for the LMC Cepheids shown in Table \ref{Table:LMC-fits} and with \cite{Efstathiou:2013via}. 
Figure \ref{Fig:MW-Cepheid-variables} shows the period-luminosity relation for the MW Cepheid variables and the best fit \eqref{Eq:MW-bestfit}.

\begin{figure}[tbp]
\centering 
\includegraphics[scale=0.75]{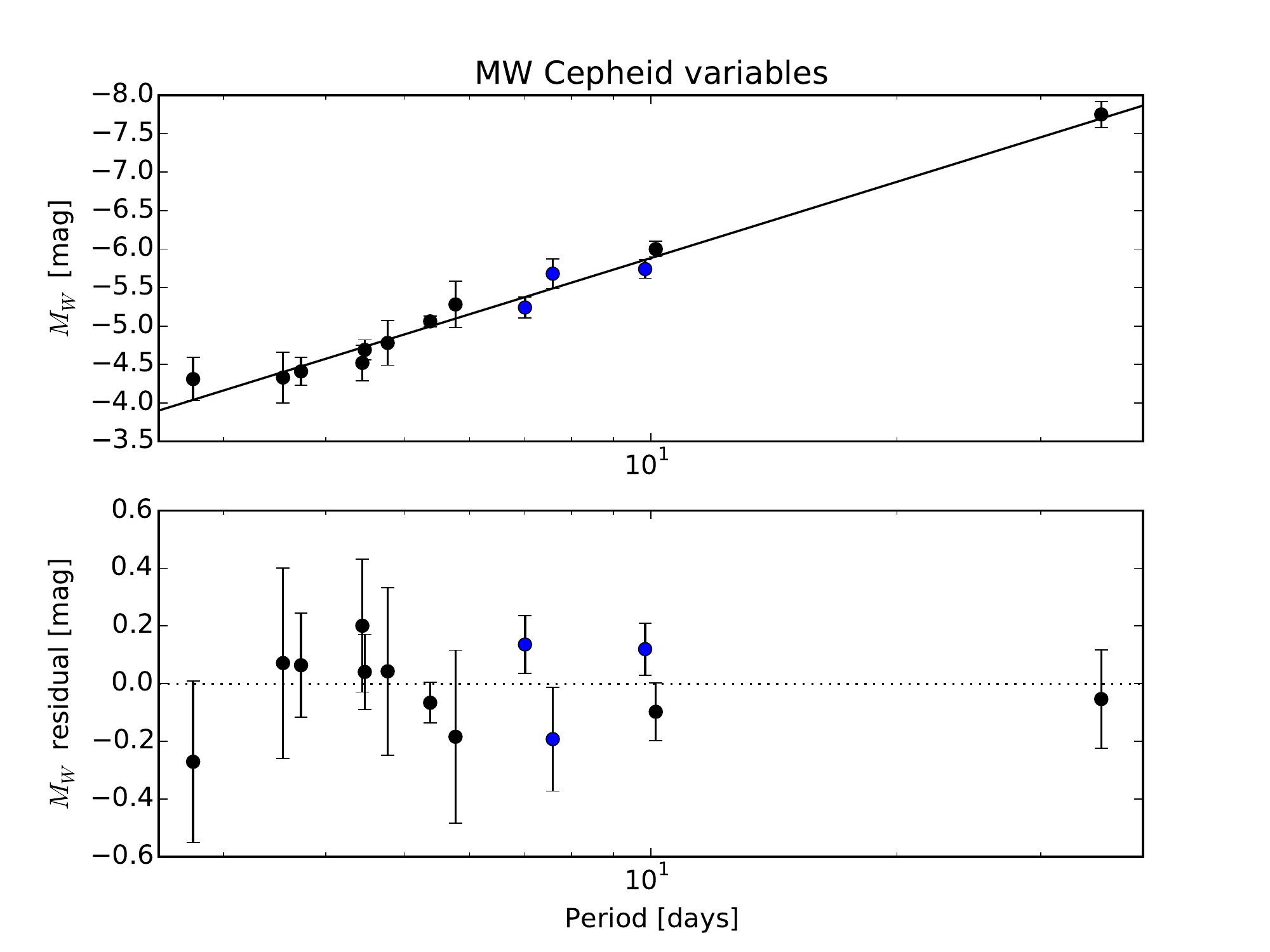} 
\caption{Period-luminosity relation for the MW Cepheid variables. In the upper panel the best fit is shown (black solid line). Error bars have been rescaled with corresponding effective HPs which are colour-coded as in Figure \ref{Fig:LMC-Cepheid-variables-fit-c}: black if $\alpha^{\eff} = 1$, blue if $10^{-1}\leq \alpha^{\eff} < 1$. In the lower panel  we show magnitude residuals with respect to the best fit \eqref{Eq:MW-bestfit}; error bars are not rescaled and colours correspond to those in the upper panel.}
\label{Fig:MW-Cepheid-variables}
\end{figure}

The consistency in both the Cepheid zero point $M_W$ \eqref{Eq:zero-point-LMC-R11} and the slope $b_W$ between the MW and the LMC Cepheid data, as well as the lack of marked outliers visible in Fig.\ \ref{Fig:MW-Cepheid-variables} provides no argument for excluding the MW data, at least based on the Cepheid stars. For this reason we include the MW data in our `baseline analysis' discussed in Section \ref{Section:Application-R11}. We also note that the MW Cepheid data have no preference for a non-zero $\sigma^{\MW}_{\intt}$, although including it does not change the fit significantly.

\subsubsection{Cepheid variables in the megamaser system NGC 4258}
\label{Subsection:NGC4258}

In this subsection we use the set of NGC 4258 Cepheid variables included in the sample of \cite{Riess:2011yx} to fit the period-luminosity relation \eqref{Eq:P-L-LMC} setting now $A = \mu_{0,\rm 4258} + M_W$ and neglecting metallicity dependence, as done for the LMC and MW cases. We find
\begin{figure}[tbp]
\centering 
\includegraphics[scale=0.75]{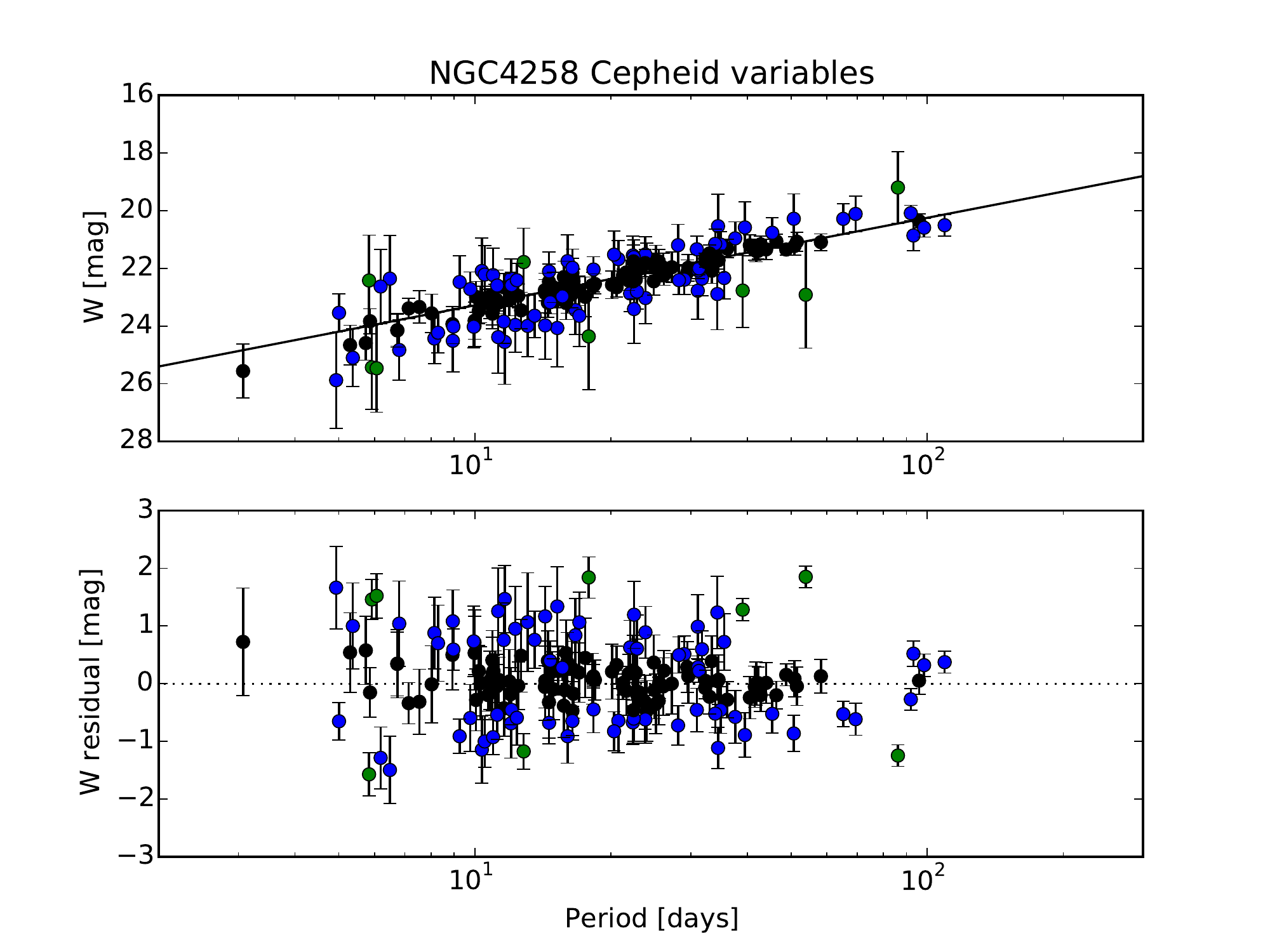} 
\caption{Period-luminosity relation for the NGC 4258 Cepheid variables. The upper panel shows the best fit (solid black line); error bars have been rescaled with corresponding effective HPs which are colour-coded as in Fig. \ref{Fig:LMC-Cepheid-variables-fit-c} (black if $\alpha^{\eff} = 1$, blue if $10^{-1}\leq \alpha^{\eff} < 1$, green if $10^{-2}\leq \alpha^{\eff} < 10^{-1}$). The lower panel shows magnitude residuals; error bars are not rescaled and colours correspond to those in upper panel.}
\label{Fig:NGC4258-Cepheid-variables}
\end{figure}

\begin{equation}\label{Eq:NGC4258-bestfit}
A = 23.281 \pm 0.078  , \qquad b_W = -3.02 \pm 0.17  , \qquad \sigma_{\intt}^{\rm 4258} = 0.12 .
\end{equation}
Fig.\ \ref{Fig:NGC4258-Cepheid-variables} shows the period-luminosity relation and residuals for the NGC 4258 Cepheid variables. Note that the slope $b_W$ for the fit \eqref{Eq:NGC4258-bestfit} is about $1.7\sigma$ away from the fit (c) in Table \ref{Table:LMC-fits} and the internal scatter $\sigma_{\intt}^{\rm 4258}=2\sigma_{\intt}^{\LMC}$. The NGC 4258 distance modulus derived from the geometric maser distance estimate to the active galaxy NGC 4258 \cite{Humphreys:2013eja} is 
\begin{equation}
\mu_{0,\rm 4258}^{\obs1} = 29.40 \pm 0.07,
\label{Eq:NGC4258-measured-distance-modulus-2013}
\end{equation}
which leads to a Cepheid zero point $M_W=-6.12 \pm 0.15$, a value about $1.6\sigma$ away from that in fit \eqref{Eq:MW-bestfit}. The standardised candle method for type IIP SNe \cite{Polshaw:2015ika} provides an alternative determination of the NGC 4258 distance modulus
\begin{equation}
\mu_{0,\rm 4258}^{\obs2} = 29.25 \pm 0.26,
\label{Eq:NGC4258-measured-distance-modulus-2015}
\end{equation}
which in turn gives $M_W=-5.97 \pm 0.34$, in good agreement with \eqref{Eq:MW-bestfit}. 
\subsubsection{Summarizing results}

In Subsections \ref{Subsection:LMCR11}--\ref{Subsection:NGC4258} we used Cepheid stars in the galaxies having direct distance determinations (i.e., $\LMC$, $\MW$ and $\NGC$) and constrained the parameters in the period-luminosity relation for each galaxy separately. We summarise our findings in Table \ref{Table:fits-section-4-1}. Although we did not consider a metallicity dependence in the Leavitt law, we find good agreement for both the Cepheid zero point and the slope of the period-luminosity relation. This compatibility provides no argument to exclude any of these data when determining the current expansion rate of the universe. In the next subsection we will study whether or not a metallicity dependence in the period-luminosity relation improves this agreement further.

\begin{table}[tbp]
\centering
\begin{tabular}{@{}ccccc}
\hline
\multicolumn{5}{c}{Consistency of period-luminosity relation} \\
\hline
Fit & Galaxy & $M_W$ & $b_W$ & $\sigma_{\intt}$ \\
\hline
 c & $\LMC$ & $-5.93\,(0.07)$& $-3.31\,(0.05)$& $0.06$  \\
 
 e & $\MW$ & $-5.88\,(0.07)$ & $-3.30\,(0.26)$ & $0.02$  \\
  
 f & $\NGC$ & $-6.12\,(0.15)$&$-3.02\,(0.17)$ & $0.12$  \\

\hline
\end{tabular}
\caption{\label{Table:fits-section-4-1} Mean values for the parameters in the period-luminosity relation. Numbers in brackets indicate the standard deviation. Fit (c) was derived using the distance modulus in Eq. \eqref{Eq:LMC-measured-distance-modulus}, fit (e) corresponds to Eq. \eqref{Eq:MW-bestfit}, and fit (f) was derived using the distance modulus in Eq. \eqref{Eq:NGC4258-measured-distance-modulus-2013}.}
\end{table}

\subsection{Metallicity dependence in the period-luminosity relation}\label{Subsection:Zw-dependence}
\begin{table}[tbp]
\centering
\begin{tabular}{@{}ccccc}
\hline
\multicolumn{5}{c}{Sample of Cepheid variables} \\
\hline
galaxy & $A$ & $b_W$ & $Z_W$ & $\sigma_{\intt}$ \\
\hline
 LMC & $12.699\,(2.139)$ & $-3.31\,(0.05)$ & $-0.016\,(0.252)\,[W]$ & $0.06$ \\

LMC & $12.562\,(0.170)$ & $-3.31\,(0.05)$ & $0.000\,(0.020)\,[S]$ & $0.06$ \\
  
 MW & $-5.88\,(2.23)$ & $-3.30\,(0.26)$ & $0.000\,(0.250)\,[W]$ & $0.02$ \\

 MW & $-5.88\,(0.19)$ & $-3.30\,(0.26)$ & $0.000\,(0.020)\,[S]$ & $0.02$ \\
 
 NGC4258 & $25.175\,(1.957)$ & $-3.00\,(0.17)$ & $-0.214\,(0.221)\,[W]$& $0.12$ \\

 NGC4258 & $23.293\,(0.193)$ & $-3.02\,(0.17)$ & $-0.001\,(0.020)\,[S]$& $0.12$ \\
 
 n4536 & $24.620\,(1.866)$ & $-2.85\,(0.17)$ & $ 0.021\,(0.214)\,[W]$ & $0.07$ \\

 n4536 & $24.803\,(0.207)$ & $-2.84\,(0.17)$ & $ 0.000\,(0.020)\,[S]$ & $0.07$ \\
 
 n4639 & $26.572\,(1.846)$ & $-2.46\,(0.42)$ & $-0.147\,(0.210)\,[W]$ & $0.03$ \\
 
 n4639 & $25.303\,(0.302)$ & $-2.49\,(0.42)$ & $-0.001\,(0.020)\,[S]$ & $0.04$ \\
  
 n3982 & $26.591\,(1.724)$ & $-3.28\,(0.42)$ & $-0.083\,(0.200)\,[W]$ & $0.03$ \\
  
 n3982 & $25.888\,(0.283)$ & $-3.27\,(0.42)$ & $ 0.000\,(0.020)\,[S]$ & $0.03$ \\
   
 n3370 & $28.317\,(1.696)$ & $-2.99\,(0.20)$ & $-0.260\,(0.196)\,[W]$ & $0.02$ \\
 
 n3370 & $26.098\,(0.224)$ & $-3.03\,(0.20)$ & $-0.003\,(0.020)\,[S]$ & $0.02$ \\
  
 n3021 & $28.226\,(1.983)$ & $-2.86\,(0.51)$ & $-0.239\,(0.231)\,[W]$ & $0.03$ \\

 n3021 & $26.211\,(0.325)$ & $-2.99\,(0.50)$ & $-0.002\,(0.020)\,[S]$ & $0.03$ \\
  
 n1309 & $26.788\,(2.032)$ & $-2.09\,(0.42)$ & $-0.105\,(0.225)\,[W]$ & $0.03$ \\

 n1309 & $25.857\,(0.354)$ & $-2.08\,(0.42)$ & $-0.000\,(0.020)\,[S]$ & $0.03$ \\
   
 n4038 & $24.200\,(2.171)$ & $-2.47\,(0.27)$ & $0.092\,(0.243)\,[W]$ & $0.03$ \\

 n4038 & $25.011\,(0.291)$ & $-2.45\,(0.27)$ & $0.000\,(0.020)\,[S]$ & $0.03$ \\
    
 n5584 & $25.428\,(1.782)$ & $-2.83\,(0.24)$ & $0.013\,(0.204)\,[W]$ & $0.03$ \\   

 n5584 & $25.541\,(0.240)$ & $-2.83\,(0.24)$ & $0.000\,(0.020)\,[S]$ & $0.03$ \\   
 
\hline
\end{tabular}
\caption{\label{Table:Zw-dependence-of-PL-relation} Mean values and standard deviation (in brackets) in the period-luminosity relation parameters for the sample of Cepheid variables used in this section, with different metallicity priors. The period range used to fit the data is $P < 205$, the same used in fit (c) of Table \ref{Table:LMC-fits}. $[W]$ stands for a Gaussian prior with $\bar{Z_W}=0$ and $\sigma_{Z_W}=0.25$. $[S]$ stands for a Gaussian prior with $\bar{Z_W}=0$ and $\sigma_{Z_W}=0.02$. The mode of the marginalized internal dispersion for each galaxy is shown in the last column.}
\end{table}
Although the Leavitt law is expected to depend at some level on the Cepheid metal abundance \cite{Freedman:2010xv}, thus far we have neglected this effect. Here we will study how an additional degree of freedom in the period-luminosity relation (i.e., $Z_W \neq 0$) impacts the fits we have presented. As in \cite{Efstathiou:2013via} we consider a mean metallicity $\Delta \log[O/H]=8.5$ for the LMC and $\Delta \log[O/H]=8.9$ for the MW. For all other galaxies we use the metallicity reported in Table 2 of \cite{Riess:2011yx}; Cepheid variables in those galaxies have a mean metallicity close to that of the Cepheid variables in the MW, $\Delta \log[O/H] \approx 8.9$.

Table \ref{Table:Zw-dependence-of-PL-relation} shows the fit for the period-luminosity relation \eqref{Eq:P-L-equation} for all the galaxies containing Cepheid variables (using the offset $A=\mu_{0,i} + M_W$ as for most galaxies the distance modulus is not known). We notice that the effect of metallicity on both the slope $b_W$ and the Cepheid zero point $M_W$ (through its dependence on $A$) is never greater than $4.3\%$ (n3021) and $8.1\%$ (NGC4028), respectively. The metallicity parameter $Z_W$ is compatible with zero in all galaxies, its main effect being a small shift and a potentially large increase in the standard deviation of the Cepheid zero point due to a degeneracy between these two parameters (see Fig. \ref{Fig:Main-analysis-fitM1a}). 

Another important point to note from results in Table \ref{Table:Zw-dependence-of-PL-relation} is the fact that about half of the host galaxies in the sample of \cite{Riess:2011yx} have a less steep slope $b_W$ than that of the LMC Cepheid variables, the shift being $\gtrsim 2\sigma$ for n4536, n4639, n1309, n4038, and n5584. This difference in the slope is however not improved by leaving more freedom in the metallicity dependence.

The Cepheid zero point $M_W$ derived from MW Cepheids is insensitive to including metallicity dependence in the period-luminosity relation. For both LMC and NGC4258 Cepheid variables, a small dependence on metallicity (strong prior) brings the Cepheid zero point in slightly better agreement with that derived from MW Cepheids. In particular, for the NGC 4258 distance modulus in Eq. \eqref{Eq:NGC4258-measured-distance-modulus-2013} and using a strong prior on $Z_W$, we obtain $M_W=-6.11 \pm 0.26$.

Because of the $M_W - Z_W$ degeneracy, and since the additional freedom in the metallicity dependence does not bring the different Cepheid data sets into better agreement,
we use the `strong' prior on the metallicity, $Z_W = 0 \pm 0.02$, as our baseline choice for R11, illustrated in Section  \ref{Section:Application-R11}. For our final result, applied to R16, we will instead use a weak prior.

\subsection{Period cut}
\label{Subsection:period-cut}
Another assumption within the analysis is the cut on the Cepheid period. In Table \ref{Table:Constraints-main-analysis-thesis-PC} we compare our `baseline analysis' for the R11 data set, fit $(29)$ with fit $(34)$. Differently to the `baseline analysis', the latter uses a tighter period cut-off $P<60$ days. The constraints on $H_0$ and the period-luminosity parameters are almost unaffected, when changing the period cut. The compatibility of the data sets is slightly improved when using no period cut-off for Cepheid stars.
\begin{table}
\centering
\resizebox{\textwidth}{0.8cm}{
\begin{tabular}{cccccccc}

\hline
Fit & $H_0$ & $M_W$ & $b_W$ & $Z_W$ &$|| \alpha^{\Cepheid}||$ & $|| \alpha^{\SNe}||$ & $|| \alpha^{\Anchors}||$ \\
\hline

$29$ & $75.0\,(3.9)$ & $-5.88\,(0.18)$ & $-3.20\,(0.05)$ & $-0.005\,(0.020) $ & $ 0.72 $ & $ 0.74 $ & $ 0.79$\\
$34$ & $75.4\,(3.7)$ & $-5.85\,(0.18)$ & $-3.27\,(0.04)$ & $-0.007\,(0.020) $ & $ 0.72 $ & $ 0.73$ & $0.64 $\\

\hline
\end{tabular}}
\caption{\label{Table:Constraints-main-analysis-thesis-PC} Constraints for the baseline analysis of the R11 data set using different period cuts. Fit $(29)$ uses $P<205$ days, whereas fit $(34)$ uses $P<60$ days. Fits correspond to those in Tables \ref{Table:details-fits}--\ref{Table:Constraints-main-analysis}. Numbers in brackets indicate the standard deviation.}
\end{table}

\subsection{Hyper-parameters ensemble}
\label{Subsection:HP-variants}
In principle we can choose to apply HPs to different sets of data, rather than to all of them, if for some reason we believe that some sets are more reliable than others. In Table \ref{Table:Constraints-main-analysis-thesis-HPe} and Fig. \ref{Fig:H0-values-3-anchors} we show what happens when we change our `baseline HPs' choice (see Section \ref{Section:Application-R11}) (which includes all kinds of data with a Gaussian HP likelihood) and only apply HPs to part of the data. As expected, the measurement using HPs everywhere -- our `baseline analysis' -- is the one with larger uncertainties. The measurement including only Cepheid stars with a Gaussian HP likelihood is of comparable precision as the measurement by Riess et al. \cite{Riess:2011yx} which uses a standard $\chi^2$ minimisation and an outlier rejection algorithm. The results show that including more HPs increases the inferred value of $H_0$, so this is an important choice that affects the result. The normalised weights in Table \ref{Table:Constraints-main-analysis-thesis-HPe} do not show full compatibility of the data sets, so a HP analysis of the R11 data set seems to be justified.

\begin{table}
\centering
\resizebox{\textwidth}{1.2cm}{
\begin{tabular}{cccccccc}

\hline
Fit & $H_0$ & $M_W$ & $b_W$ & $Z_W$ &$|| \alpha^{\Cepheid}||$ & $|| \alpha^{\SNe}||$ & $|| \alpha^{\Anchors}||$ \\
\hline

$29$ & $75.0\,(3.9)$ & $-5.88\,(0.18)$ & $-3.20\,(0.05)$ & $-0.005\,(0.020) $ & $ 0.72 $ & $ 0.74 $ & $ 0.79$\\
$31$ & $73.2\,(2.5)$ & $-5.89\,(0.18)$ & $-3.19\,(0.05)$ & $-0.004\,(0.020) $ & $ 0.72$ & - & $ 0.92$\\
$32$ & $74.1\,(3.7)$ & $-5.89\,(0.18)$ & $-3.21\,(0.05)$ & $-0.005\,(0.020) $ & $ 0.72 $ & $ 0.81$ & -\\
$33$ & $72.4\,(2.2)$ & $-5.90\,(0.17)$ & $-3.20\,(0.05)$ & $-0.004\,(0.020) $ & $ 0.71$ & - & -\\

\hline
\end{tabular}}
\caption{\label{Table:Constraints-main-analysis-thesis-HPe} Constraints for different variants of our `baseline HPs' choice. While fit $(29)$ applies HPs to all kinds of data, the other fits select only one or two different sets. Details of fits are indicated in Tables \ref{Table:details-fits}--\ref{Table:Constraints-main-analysis}. Numbers in brackets indicate the standard deviation.}
\end{table}

\begin{figure}[t]
\centering
\includegraphics[width=\textwidth]{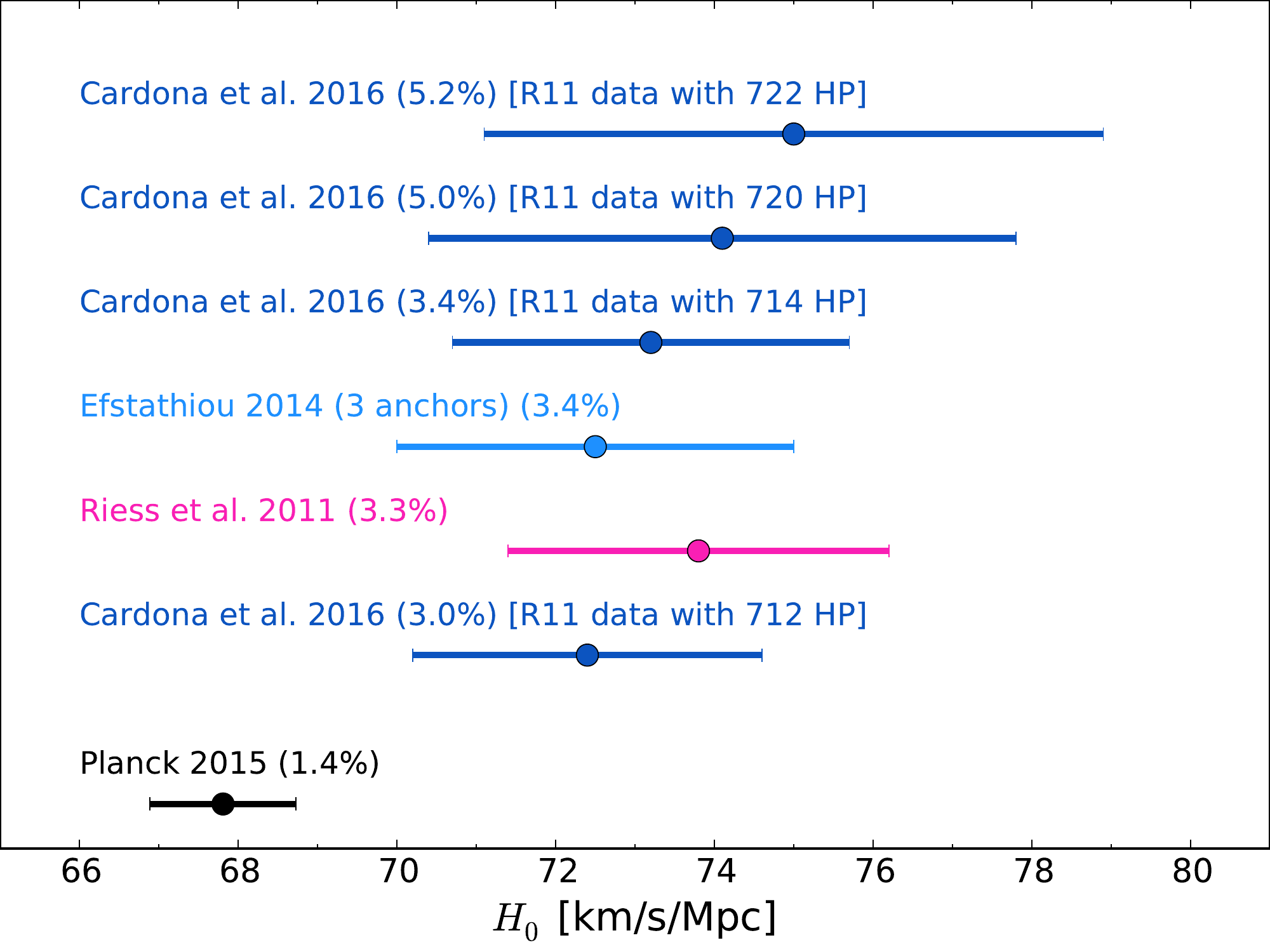}
\caption{Same as Fig.\ \ref{Fig:H0bestfits}, but showing the effect of using hyper-parameters on different combinations of Cepheids, $\SNe$, anchors.  Black shows the indirect determination by the Planck collaboration \cite{Ade:2015xua}. Direct measurements, all using three distance anchors (NGC 4258 distance modulus, LMC distance modulus and MW Cepheid variables), are shown in light blue, magenta and dark blue. Light blue shows Efstathiou's measurement \cite{Efstathiou:2013via} which uses a $60$ days period cut. Magenta corresponds to measurement by Riess et al. \cite{Riess:2011yx}. These two measurements use outlier rejection algorithms and a usual $\chi^2$ minimisation. Dark blue points are the results of this paper and correspond from top to bottom to fits $(29)$, $(31)$, $(32)$, $(33)$ in Tables \ref{Table:details-fits}--\ref{Table:Constraints-main-analysis}. Fit $(29)$ includes Cepheids, $\SNe$, and anchors with the HP Gaussian PDF in Eq. \eqref{Eq:hyper-likelihood}; fit $(31)$ includes both Cepheids and anchors with the HP Gaussian PDF, and $\SNe$ data is included with a usual Gaussian PDF \eqref{Eq:Gaussian-likelihood}; fit $(32)$ includes both $\SNe$ and Cepheid variables with the HP Gaussian PDF, and includes anchor distances with a Gaussian PDF; fit $(33)$ includes Cepheid stars with the HP Gaussian PDF, and includes both $\SNe$ and anchor distances with a Gaussian PDF. The number of HPs is given in brackets for these fits. Within error bars all analyses agree, and more HPs lead to a larger uncertainty on $H_0$.} \label{Fig:H0-values-3-anchors}
\end{figure}

\subsection{Anchors}
\label{subsection:Anchors}
We have seen in Section \ref{Section:Application-R11} that results including all the available distance anchors are robust against the statistical method utilised in the analysis. In order to better understand both the data sets and the results in our `baseline analysis' we investigate below the application of our method to different combination of distance anchors excluding one or two of them. The $H_0$ measurements corresponding to Subsections \ref{Subsection:A-NGC4258}--\ref{Subsection-A-Combining-distance-anchors} are shown in Figure \ref{Fig:single-combined-anchor} and Table \ref{Table:Constraints-different-anchor-choice}. 

\begin{figure}[hbtp]
\centering
\includegraphics[scale=.8]{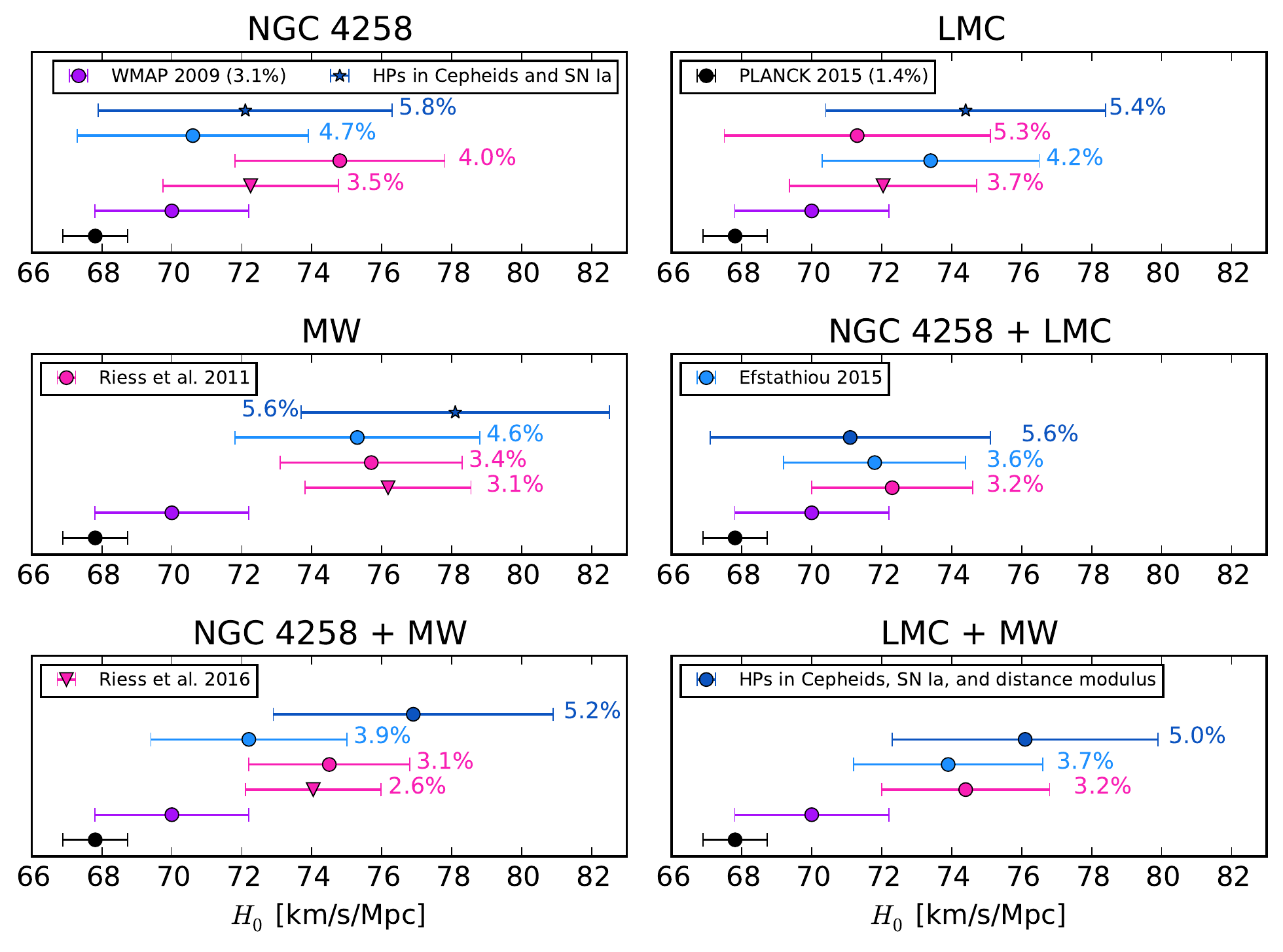}
\caption{Different measurements of the Hubble constant using different combinations of anchor distances. Each panel shows measurements arranged from less precise (top) to more precise (bottom). Colours are as in Figure \ref{Fig:H0bestfits}. In this Figure, all points refer to the R11 dataset except Riess et al 2016 (pink triangle), also shown for reference, and the indirect determinations of WMAP and Planck.}
\label{Fig:single-combined-anchor}
\end{figure}

\begin{table}[tbp]
\centering
\resizebox{\textwidth}{2.cm}{
\begin{tabular}{cccccccc}

\hline
Fit & $H_0$ & $M_W$ & $b_W$ & $Z_W$ &$|| \alpha^{\Cepheid}||$ & $|| \alpha^{\SNe}||$ & $|| \alpha^{\Anchors}||$ \\
\hline

$4$ & $70.8\,(4.2)$& $-6.01\,(0.19)$ & $-3.17\,(0.06)$ & $-0.006\,(0.020) $ & $ 0.72 $ & $ 0.72 $ & - \\

$10$ & $74.4\,(4.0)$& $-5.90\,(0.18)$& $-3.17\,(0.06)$& $-0.006\,(0.020)$& $ 0.72$ & $ 0.78 $ & - \\

$16$ & $77.1\,(4.1)$& $-5.81\,(0.18)$& $-3.26\,(0.05)$& $-0.008\,(0.020)$ & $ 0.74 $ & $ 0.77 $ & - \\

$20$ &$73.9\,(4.0)$ & $-5.89\,(0.18)$& $-3.26\,(0.05)$& $-0.008\,(0.020)$ & $ 0.72 $ & $ 0.78$ & $1 $ \\

$22$ & $76.9\,(4.0)$&$-2.13\,(1.27)$ &$-3.27\,(0.04) $ &$-0.425\,(0.143) $ & $ 0.73 $ & $ 0.70 $ & $ 0.48 $\\

$28$ &$76.1\,(3.8)$ &$-5.84\,(0.18)$ &$-3.27\,(0.04) $ &$-0.007\,(0.020) $ & $0.72 $ & $ 0.71$ & $1 $\\

$29$ & $75.0\,(3.9)$ & $-5.88\,(0.18)$ & $-3.20\,(0.05)$ & $-0.005\,(0.020) $ & $ 0.72 $ & $ 0.74 $ & $ 0.86$\\

$ 44 $ & $73.75\,(2.10)$ & $-5.10\,(0.79)$ & $-3.28\,(0.01)$ & $-0.10\,(0.09) $ & $0.85 $ & $ 0.83$ & $ 1 $\\

\hline
\end{tabular}}
\caption{\label{Table:Constraints-different-anchor-choice} Constraints for different choices of anchor distances. Details of fits in Tables \ref{Table:details-fits}--\ref{Table:Constraints-main-analysis}. The last line, fit (44), corresponds to our baseline analysis of the R16 data.}

\end{table}

\subsubsection{Megamaser system NGC 4258 distance modulus}
\label{Subsection:A-NGC4258}

In this subsection we apply the hyper-parameter method to the sample of Cepheid variables and SNe Ia hosts in \cite{Riess:2011yx}. For the purpose of comparing with \cite{Efstathiou:2013via} and \cite{Riess:2016jrr}, we restrict ourselves to a single anchor distance: the distance modulus to the megamaser system NGC 4258 from \cite{Humphreys:2013eja}. We include LMC Cepheid variables, but exclude MW Cepheid stars.  Results are shown in fits $(1)$--$(6)$ of Tables \ref{Table:details-fits}--\ref{Table:Constraints-main-analysis}, with fit (4) also included in the top panel of Figure \ref{Fig:single-combined-anchor}.

The $H_0$ values in fits $(1)$--$(6)$ of Tabless \ref{Table:details-fits}--\ref{Table:Constraints-main-analysis} are shifted downwards with respect to our `baseline analysis' value (Eq. \ref{Eq:H0-value-standard-analysis}) by $3-6 \%$. Fits with a tighter period cut prefer larger values of $H_0$: a $2\%$ change with respect to those fits without period cut (independently of the metallicity dependence). Those changes are due to both $M_W-Z_W$ and $M_W-H_0$ degeneracies: a tighter period cut enhances metallicity dependence driving the Cepheid zero point to higher values, which in turn prefers higher $H_0$. Allowing a metallicity dependence of the period-luminosity relation without prior on $Z_W$ we found a departure from $Z_W=0$ which is larger than $2\sigma$; this assumption, however, makes the Cepheid zero point $M_W$ less compatible with the value measured from MW Cepheid variables (see Eq. \eqref{Eq:MW-bestfit}). When HPs are used in Cepheid variables, SN Ia hosts, and distance modulus we obtain a measurement of $H_0$ with a relative error of $8\%$}: the exclusion of two anchor distances in our `baseline analysis' makes the measurement more uncertain. Since for fits $(1)$--$(3)$, and $(5)$ the HP for the NGC 4258 distance modulus is equal to one, we have also studied two cases, fits $(4)$ and $(6)$, where only Cepheid variables and SNe Ia are included with HPs. As a result we obtain a measurement of $H_0$ with a relative error of  $6\%$ which is still more uncertain than the corresponding case using the three anchor distances.

In the upper left panel of Figure \ref{Fig:single-combined-anchor} we show $H_0$ value from fit $(4)$ in Tables \ref{Table:details-fits}--\ref{Table:Constraints-main-analysis}, together with the values found by Riess et al. \cite{Riess:2011yx} ($H_0=74.8\pm 3.0 \, \km\, \second^{-1}\, \Mpc^{-1}$) which did not use the revised distance modulus from \cite{Humphreys:2013eja}, Efstathiou \cite{Efstathiou:2013via} ($H_0 = 70.6 \pm 3.3 \, \km\, \second^{-1}\, \Mpc^{-1}$), Riess et al. \cite{Riess:2016jrr} ($H_0=72.25\pm 2.51 \, \km\, \second^{-1}\, \Mpc^{-1}$) which used a redetermination of the distance modulus to NGC 4258. Our value is compatible with all previous direct determinations of $H_0$ and also with the indirect determinations from Planck and WMAP. The measurement using HPs is the most uncertain among the direct measurements due to inclusion of SNe Ia with HPs. The Planck collaboration used Efstathiou's value as a prior for $H_0$ when combining CMB measurements and local measurements of the Hubble constant. The choice of $H_0$ prior is important as a discrepancy between high and low redshifts in the $\Lambda$CDM model could point to new physics and change conclusions for the extended models discussed e.g.\ in the {\em Planck} publications \cite{Ade:2015xua,2016A&A...594A..14P}.

\subsubsection{LMC distance modulus}
\label{Subsection:LMC-anchor}

In this subsection we apply our method to the sample of Cepheid variables and SNe Ia hosts in \cite{Riess:2011yx}, but restrict ourselves to a single anchor distance: the distance modulus to the LMC from \cite{Pietrzynski:2013gia}. We also include NGC4258 Cepheid variables, but omit MW Cepheid variables. Results are shown in fits $(7)$--$(12)$ of Tables \ref{Table:details-fits}--\ref{Table:Constraints-main-analysis} and fit (10) is shown in the top-right panel of Fig.\ \ref{Fig:single-combined-anchor}. We have examined different period cuts and different assumptions for the metallicity dependence in the Leavitt Law.

The fits $(7)$--$(12)$ in Tables \ref{Table:details-fits}--\ref{Table:Constraints-main-analysis} show a $0-7\%$ shift in $H_0$ w.r.t the value in Eq. \eqref{Eq:H0-value-standard-analysis}. A period cut shifts $H_0$ values by $0.7-2\%$ (w.r.t. fits without period cut), whereas the use of HPs in LMC distance modulus changes $H_0$ values by $\lesssim 0.4\%$.
Fits $(7)$--$(12)$ in Tables \ref{Table:details-fits}--\ref{Table:Constraints-main-analysis} represent  $5-7\%$ measurements of the Hubble constant. In upper right panel of Figure \ref{Fig:single-combined-anchor} we show $H_0$ value of fit $(10)$ along with measurements by Riess et al. \cite{Riess:2011yx} ($H_0=71.3\pm 3.8 \, \km\, \second^{-1}\, \Mpc^{-1}$), Efstathiou \cite{Efstathiou:2013via} ($H_0 = 73.4 \pm 3.1 \, \km\, \second^{-1}\, \Mpc^{-1}$), Riess et al. \cite{Riess:2016jrr} ($H_0=72.04\pm 2.67 \, \km\, \second^{-1}\, \Mpc^{-1}$). Our measurement is almost as precise as that of \cite{Riess:2011yx} and is in good agreement with all the other direct determinations of the Hubble constant. While the WMAP value agrees at $1\sigma$ level with ours, the Planck value agrees at $2\sigma$ level.

\subsubsection{Parallax measurements of Cepheid variables in the Milky Way}
\label{Subsection:MW}

Parallax measurements of Cepheid variables (see Subsection \ref{Subsection:MW-1}) in our galaxy are used in this section as the sole anchor distance scale. We include Cepheid variables in both the megamaser system NGC 4258 and those in the LMC. We show the resulting constraints in fits $(13)$--$(16)$ of Tables \ref{Table:details-fits}--\ref{Table:Constraints-main-analysis}. In addition to different period cuts and different assumptions for the metallicity dependence in the Leavitt Law, we have included some cases (see Table \ref{Table:MW-fits}) where those hosts galaxies whose slope $b_W$ departs from the LMC value (fit (c) in Table \ref{Table:LMC-fits}) by $\gtrsim 2\sigma$ are excluded from the fit.

Fits $(13)$--$(16)$ in Tables \ref{Table:details-fits}--\ref{Table:Constraints-main-analysis} are shifted upwards w.r.t our 'baseline analysis' by $3-4\%$. In this case we obtain a $5-6\%$ measurement of the Hubble constant. A tighter period cut in the Leavitt Law shifts $H_0$ values by $0.3-0.4\%$ (w.r.t. fits without period cut). A strong prior on the metallicity parameter $Z_W$ drives downwards $H_0$ by $1-2\%$ (w.r.t. fits with no prior). When including the metallicity dependence without a prior we observe a slight $b_W-H_0$ degeneracy: less negative $b_W$ prefers higher $H_0$ values. Fits $(13^l)$--$(16^l)$ in Table \ref{Table:MW-fits} do not include Cepheid variables in galaxy hosts with a slope $b_W$ differing $\gtrsim 2\sigma$ from the LMC value (see Subsection \ref{Subsection:Zw-dependence}). The main impact of this change is seen in $H_0$ having both mean value and standard deviation increased. Furthermore, those cases without prior on $Z_W$ present an enhancement on their metallicity dependence. As expected the slope $b_W$ is now closer to the LMC value (even without a tighter period cut).

In the middle left panel of Figure \ref{Fig:single-combined-anchor} we show $H_0$ value from fit $(16)$ along with values determined by Riess et al. \cite{Riess:2011yx} ($H_0=75.7\pm 2.6 \, \km\, \second^{-1}\, \Mpc^{-1}$); Efstathiou  \cite{Efstathiou:2013via} ($H_0 = 75.3 \pm 3.5 \, \km\, \second^{-1}\, \Mpc^{-1}$); Riess et al. \cite{Riess:2016jrr} ($H_0=76.18\pm 2.37 \, \km\, \second^{-1}\, \Mpc^{-1}$) that used a bigger sample of MW Cepheid variables. Our measurement is the most uncertain, but it is compatible with all other direct measurements showing that the value is robust against the statistical approach employed. All direct measurements present a $\approx 2\sigma$ disagreement with the Planck value.

\begin{table}[tbp]
\centering
\begin{tabular}{@{}lcccccr}
\hline 
\multicolumn{7}{c}{Milky Way anchor} \\
\hline 
Fit & $H_0$ & $M_W$ & $b_W$ & $Z_W$ & $\sigma_{\intt}^{\MW}$ & $\sigma_{\intt}^{\LMC}$ \\
\hline

$13^{l}$ & $83.9\,(9.7)$& $-2.00\,(1.60)$ & $-3.27\,(0.05)\,[N]$& $-0.436\,(0.179)\,[N]$ & $ 0.02$ & $0.06$ \\

$14^{l}$ & $84.6\,(9.5)$& $-1.56\,(1.66)$ & $-3.27\,(0.05)\,[N]$& $-0.485\,(0.186)\,[N]$ & $0.02$ & $0.06$\\

$15^{l}$ & $80.9\,(8.3)$& $-5.83\,(0.19)$& $-3.28\,(0.05)\,[N]$& $-0.007\,(0.020)\,[S]$ & $0.02$ & $0.06$ \\

$16^{l}$ & $80.7\,(8.2)$& $-5.83\,(0.18)$& $-3.28\,(0.05)\,[N]$& $-0.006\,(0.020)\,[S]$ & $0.02 $ & $0.06$ \\
   
\hline   
\end{tabular}
\caption{\label{Table:MW-fits} Fit details correspond to fits $(13)$--$(16)$ in Table \ref{Table:details-fits}, but here we use a reduced sample of $\SNe$ hosts, see Subsection \ref{Subsection:MW}. }
\end{table}

\subsubsection{Combining two distance anchors}
\label{Subsection-A-Combining-distance-anchors}

It is questionable to argue in favour or against specific choices of distance anchors, which is why we decided to combine all
anchors in the main text. Here we provide the constraints for different combinations of two anchor distances. Results are shown in fits $(17)$--$(28)$ of Tables \ref{Table:details-fits}--\ref{Table:Constraints-main-analysis}. 

Using distance moduli to both NGC 4258 and LMC as anchor distances, fits $(17)$--$(20)$, we see that the effect of the strong prior on the metallicity parameter is less important than in the case where only LMC is utilised as anchor distance. There is also a $\approx 2\sigma$ departure from $Z_W=0$ when no prior on $Z_W$ is used, but the Cepheid zero point is is disagreement with the value measured by MW Cepheid variables alone. As noted before for the case using only LMC as anchor distance, again here the $M_W-H_0$, $M_W-Z_W$, and $Z_W-H_0$ degeneracies drive $H_0$ towards higher values. The change, however, is smaller than in the case where only LMC is used as anchor. In the middle right panel of Figure \ref{Fig:single-combined-anchor} we show $H_0$ value from fit $(20)$ as well as the measurements of Riess et al. ($H_0=72.3\pm 2.3 \, \km\, \second^{-1}\, \Mpc^{-1}$) \cite{Riess:2011yx} and Efstathiou ($H_0=71.8\pm 2.6 \, \km\, \second^{-1}\, \Mpc^{-1}$) \cite{Efstathiou:2013via}.

Fits including both distance modulus to NGC 4258 and MW Cepheid variables as distance anchors, fits $(21)$--$(24)$, follow same trend as fits only including MW Cepheid variables as distance anchor: $H_0$ is pushed downwards when a strong prior on the metallicity parameter is used, and the period cut does not have a big impact on the $H_0$ value. In the lower left panel of Figure \ref{Fig:single-combined-anchor} we show $H_0$ value from fit $(22)$ as well as the measurements of Riess et al. ($H_0=74.5\pm 2.3 \, \km\, \second^{-1}\, \Mpc^{-1}$) \cite{Riess:2011yx}, Efstathiou ($H_0=72.2\pm 2.8 \, \km\, \second^{-1}\, \Mpc^{-1}$) \cite{Efstathiou:2013via}, and Riess et al ($H_0=74.04\pm 1.93 \, \km\, \second^{-1}\, \Mpc^{-1}$) \cite{Riess:2016jrr}.

Fits including both MW Cepheid variables and LMC distance modulus as anchor distances, fits $(25)$--$(28)$,  also follow same trend as for the case where MW Cepheid variables are used as the only anchor distance. In the lower right panel of Figure \ref{Fig:single-combined-anchor} we show measurement of fit $(28)$ along with measurements by Riess et al. ($H_0=74.4\pm 2.4 \, \km\, \second^{-1}\, \Mpc^{-1}$) \cite{Riess:2011yx}, and Efstathiou ($H_0=73.9\pm 2.7 \, \km\, \second^{-1}\, \Mpc^{-1}$) \cite{Efstathiou:2013via}.
 
\subsubsection{Combining all distance anchors}
\label{Subsection-A-Combining-All-distance-anchors}

Fits $(29)$--$(39)$ include as distance anchors the distance moduli to both $\NGC$ and $\LMC$ as well as $\MW$ Cepheid stars. See discussion in Section \ref{Section:Application-R11} and Fig. \ref{Fig:H0-values-3-anchors}.

\subsubsection{Distance anchor summary}

The choice of distance anchors influences the resulting constraints on $H_0$ significantly, as is easily seen from Fig.\ \ref{Fig:single-combined-anchor}. Especially the inclusion of the MW Cepheids pushes $H_0$ to higher values. This is an aspect that should be investigated further -- the huge increase in quantity and quality of the MW Cepheid parallaxes from Gaia promises a significant improvement for the future \cite{2016arXiv160905175C}. For the time being, while we note that the choice of anchor is important, we see no reason to exclude any of the anchors and for this reason use the combination of all of them as our baseline choice. We also note that the analyses of Riess et al \cite{Riess:2011yx,Riess:2016jrr} and Efstathiou \cite{Efstathiou:2013via} that are based on outlier rejection algorithms and the HP based method used here agree well when the same anchors are used.


\section{Final Results: applying hyper-parameters to R16 data}
\label{Section:Application-R16}

\paragraph{The R16 data set}  We now apply HPs to measure the current expansion rate of the Universe $H_0$ by using the R16 data set. It comprehends: 
a larger sample of Cepheids in the LMC ($775$ compared to $53$ in R11 data set); $2$ new $\mathit{HST}$-based trigonometric parallaxes for the MW Cepheids (a total of $15$ MW parallaxes, taking into account the $13$ included in the R11 data set); $11$ new $\SNe$ host galaxies (for a total of $19$, taking into account the $8$ in the R11 data set); $\mathit{HST}$ observations of $372$ Cepheid variables in M31 (which were not in the R11 Cepheid sample); the possibility to use $\MAnd$ as an anchor distance taking advantage of the two detached eclipsing binaries based distances to $\MAnd$; $\NGC$ Cepheid stars observed with the same instrument as those in the $19$ $\SNe$ host galaxies, thus reducing the cross-instrument zeropoint errors.

\paragraph{Baseline anchors}  As we have seen in the previous sections, we do not find any clear reason to discard any of the data sets. In this section we will utilise all available Cepheid data in the R16 data set (including MW Cepheid stars) along with the $\LMC$ distance modulus in Eq. \eqref{Eq:LMC-measured-distance-modulus}, $\MAnd$ distance modulus \cite{Riess:2016jrr} given by 
\begin{equation}
\mu_{0,\MAnd}^{\obs} = 24.36 \pm 0.08\, \magn,
\label{Eq:M31-measured-distance-modulus-2016}
\end{equation}
and the improved $\NGC$ distance modulus\footnote{Although the standardised candle method for type IIP SNe \cite{Polshaw:2015ika} provides an alternative determination of the $\NGC$ distance modulus, see Eq. \eqref{Eq:NGC4258-measured-distance-modulus-2015}, in this section we only include the distance modulus from \cite{Riess:2016jrr} in order to perform a more direct comparison with their results.} \cite{Riess:2016jrr}
\begin{equation}
\mu_{0,\NGC}^{\obs} = 29.387 \pm 0.0568\, \magn.
\label{Eq:NGC4258-measured-distance-modulus-2016}
\end{equation}

\paragraph{Baseline HPs} We have performed the same analysis as for the primary best fit with the R11 data set [fit $(29)$ in Tables \ref{Table:details-fits}--\ref{Table:Constraints-main-analysis}] which includes all types of data with Gaussian HP likelihoods, but also some variations to estimate the impact of different choices for the reddening law, period cut-off and metallicity dependence. Different variants are specified in fits $(40)$--$(55)$ of Table \ref{Table:details-fits} and the corresponding constraints are shown in Table \ref{Table:Constraints-main-analysis}. Changes in the Hubble constant $H_0$ due to different reddening laws (different $R$ in Eq. \eqref{Eq:Wesenheit-reddening-free}) range from $0.33$ up to $0.45\, \km\, \second^{-1}\, \Mpc^{-1}$. Differences in the period cut-off produce changes in $H_0$ ranging from $0.06$ to $0.38\, \km\, \second^{-1}\, \Mpc^{-1}$. Allowing for a strong or weak metallicity dependence in the period-luminosity relation we find differences in $H_0$ which range from $0$ to $0.25\, \km\, \second^{-1}\, \Mpc^{-1}$. The standard deviation for measurements of $H_0$ in fits $(40)$--$(55)$ is $\sigma_{\mathrm{syst}}=0.20\, \km\, \second^{-1}\, \Mpc^{-1} $ which we consider as a systematic error due to changes in the reddening law, period cut-off, and metallicity dependence.

According to the normalised weight criterion discussed in Section \ref{Section:Application-R11}, the most compatible fits for the R16 data set are fits $(52),\, (44)$, and $(42)$ having $\sum_j || \alpha^{j}||=2.71$, $2.68$, and $2.66$ which are higher than that for the primary fit $(29)$ using the R11 data set, and close to the value expected for Gaussian data and a correct model\footnote{Even slightly larger, due to the normalised weight of the distance anchors. This maybe due to random fluctuations because of the small number of anchors, or possibly point to overestimated error bars.}. For sake of comparison with the best estimate of $H_0$ in \cite{Riess:2016jrr}, we choose fit $(44)$, which uses the same period cut-off as the best estimate in \cite{Riess:2016jrr}, as our primary fit for the R16 data set.
We note that $\sum_j || \alpha^{j}||=2.68$ still amounts to effectively rejecting 10\% of the data.

\paragraph{Baseline Cepheid period} As in the analysis of Section \ref{Section:Application-R11}, we do not use any period cut-off for Cepheid stars, which is the same as using a period cut of $P<205$.

\paragraph{Baseline metallicity in Leavitt law} Differently to our analysis of the R11 data set, here we choose a weak metallicity prior $Z_W = 0 \pm 0.25$ for our baseline analysis as, differently from our analysis of the R11 data, the fits with the highest normalised weights are all using a weak metallicity prior. The choice of strong or weak metallicity prior has however nearly no impact on the resulting constraint on $H_0$.

\paragraph{Baseline result} Adding in quadrature the statistical error (quoted in Tables \ref{Table:details-fits}--\ref{Table:Constraints-main-analysis}) and the systematic error estimated above, we find 
\begin{equation}
H_0 = 73.75 \pm 2.11\,\km\, \second^{-1}\, \Mpc^{-1}  \, \,\,\,\,\,\,\,\,\, (\rm{using\,\,\, R16 \,\,\,data}),
\label{Eq:primary-best-fit-R16}
\end{equation}
which is a $2.9\%$ measurement of the Hubble constant. The small change in the uncertainty due to the inclusion of a systematic error shows that HPs are already taking into account most of this contribution to the error budget.

\begin{figure}[hbtp]
\includegraphics[width=\textwidth]{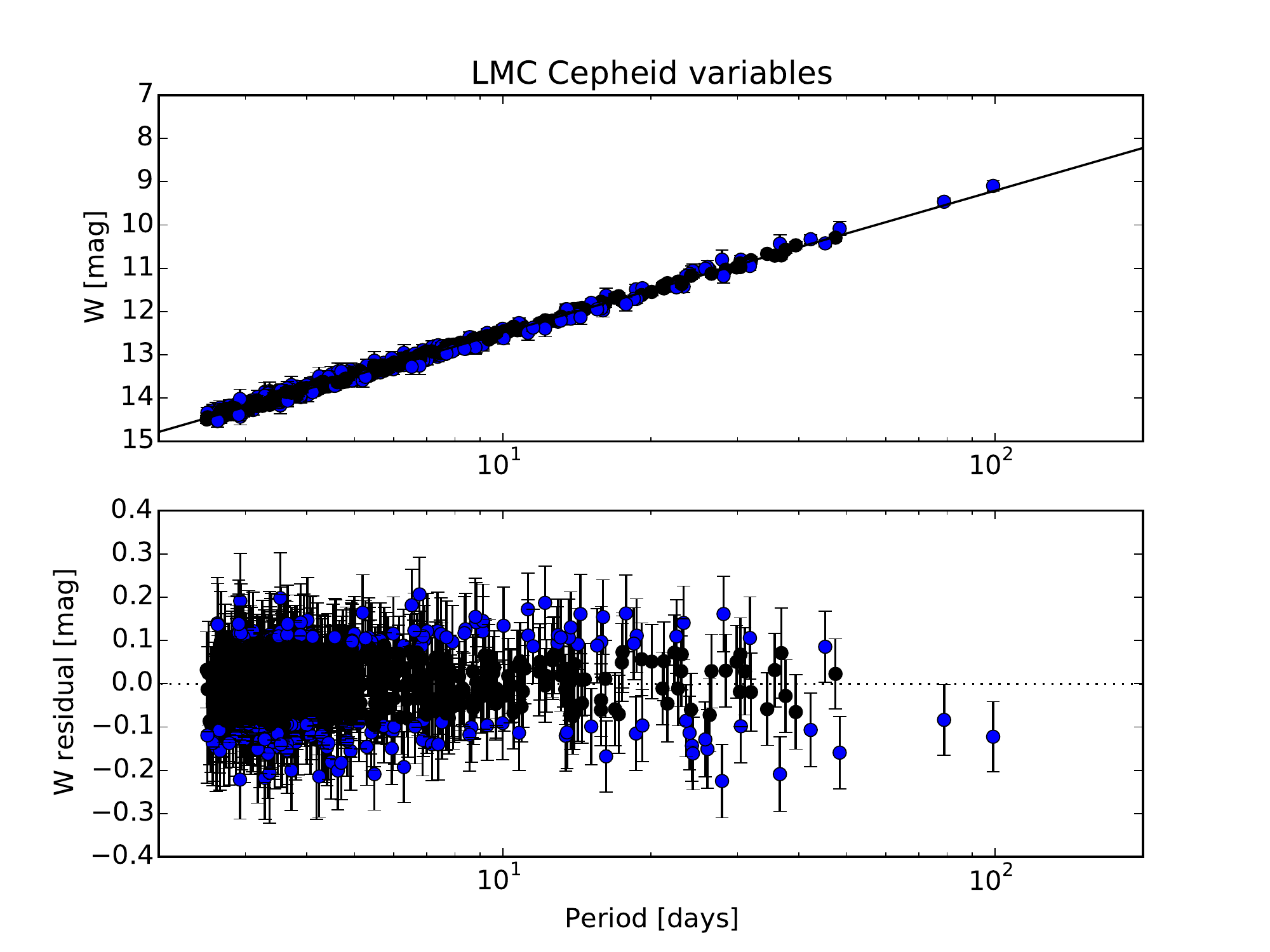}
\caption{Period-luminosity relation (upper panel) and magnitude residuals for the $\LMC$ Cepheid variables in the R16 data set. The solid line shows the best fit of fit $(44)$. In the upper panel we rescale error bars with HPs, in the lower panel we do not. Data are colour-coded as explained in Figure \ref{Fig:LMC-Cepheid-variables-fit-c}.}
\label{Fig:R16-LMC}
\end{figure}

\begin{figure}[hbtp]
\includegraphics[width=\textwidth]{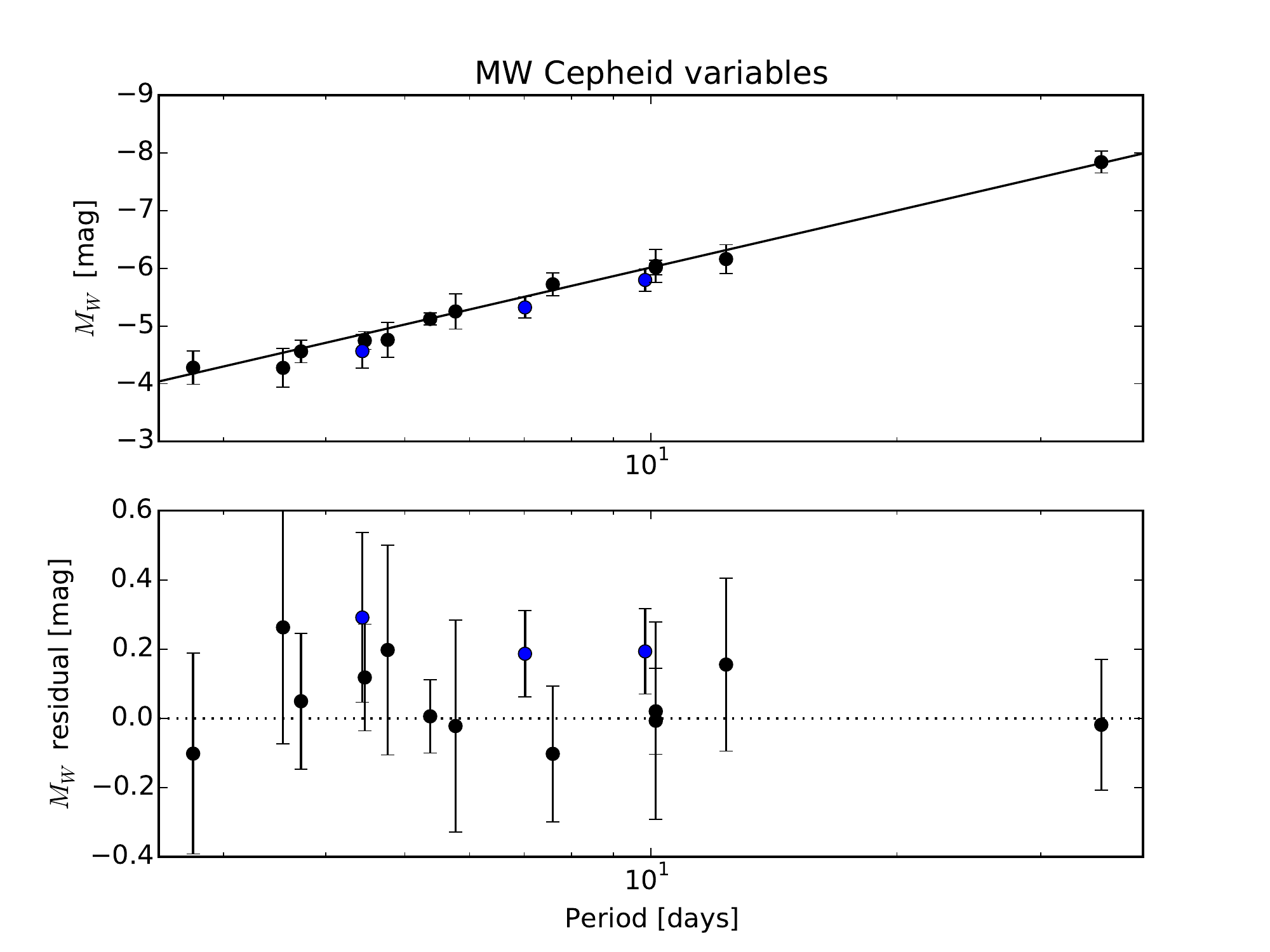}
\caption{Period-luminosity relation (upper panel) and magnitude residuals for the $\MW$ Cepheid variables in the R16 data set. The solid  line shows the best fit of fit $(44)$. In the upper panel we rescale error bars with HPs, in the lower panel we do not. Data are colour-coded as explained in Figure \ref{Fig:LMC-Cepheid-variables-fit-c}.}
\label{Fig:R16-MW}
\end{figure}

\begin{figure}[hbtp]
\includegraphics[width=\textwidth]{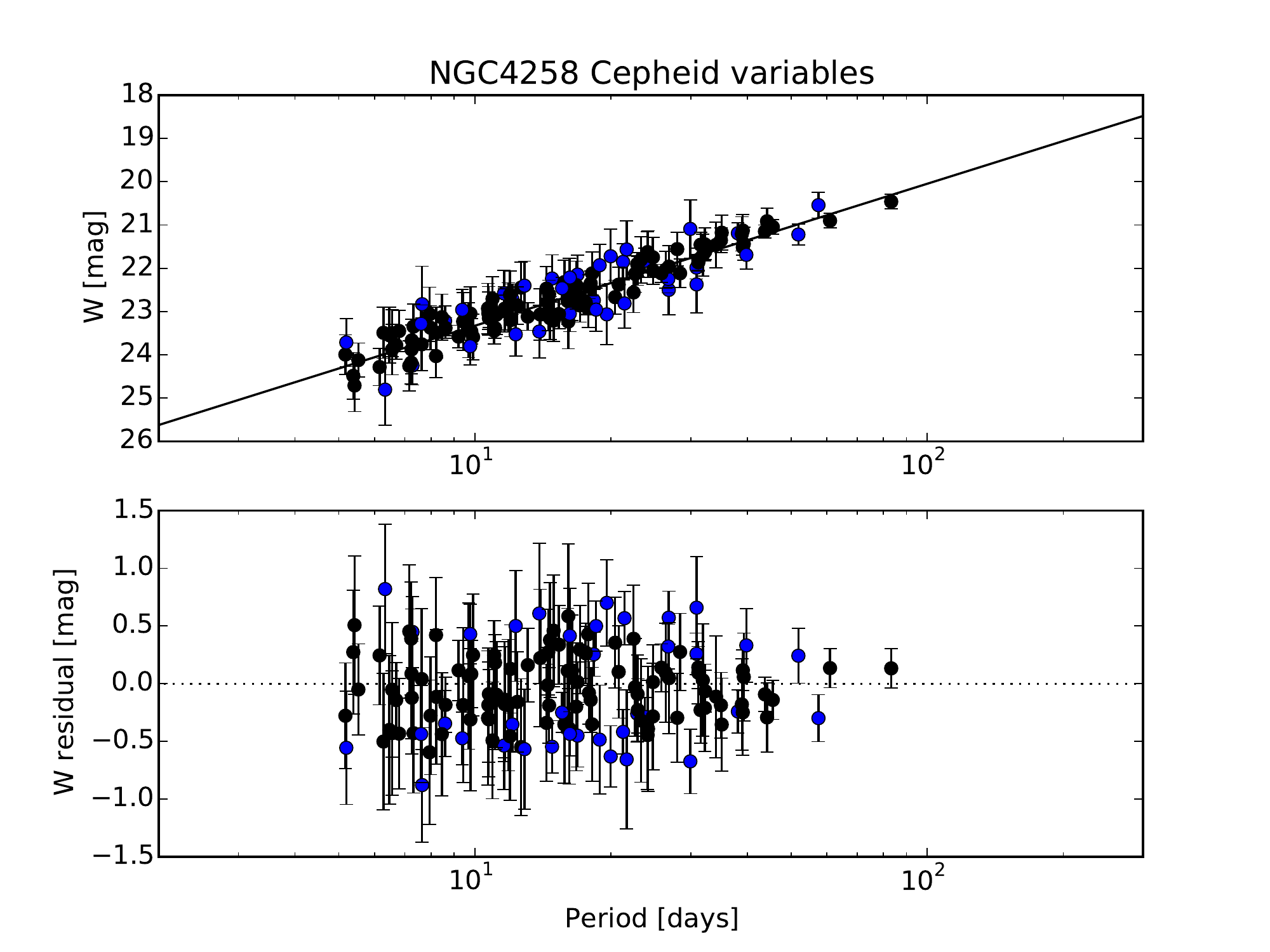}
\caption{Period-luminosity relation (upper panel) and magnitude residuals for the $\NGC$ Cepheid variables in the R16 data set. The solid line shows the best fit of fit $(44)$. In the upper panel we rescale error bars with HPs, in the lower panel we do not. Data are colour-coded as explained in Figure \ref{Fig:LMC-Cepheid-variables-fit-c}.}
\label{Fig:R16-4258}
\end{figure}

\begin{figure}[hbtp]
\centering
\includegraphics[width=\textwidth]{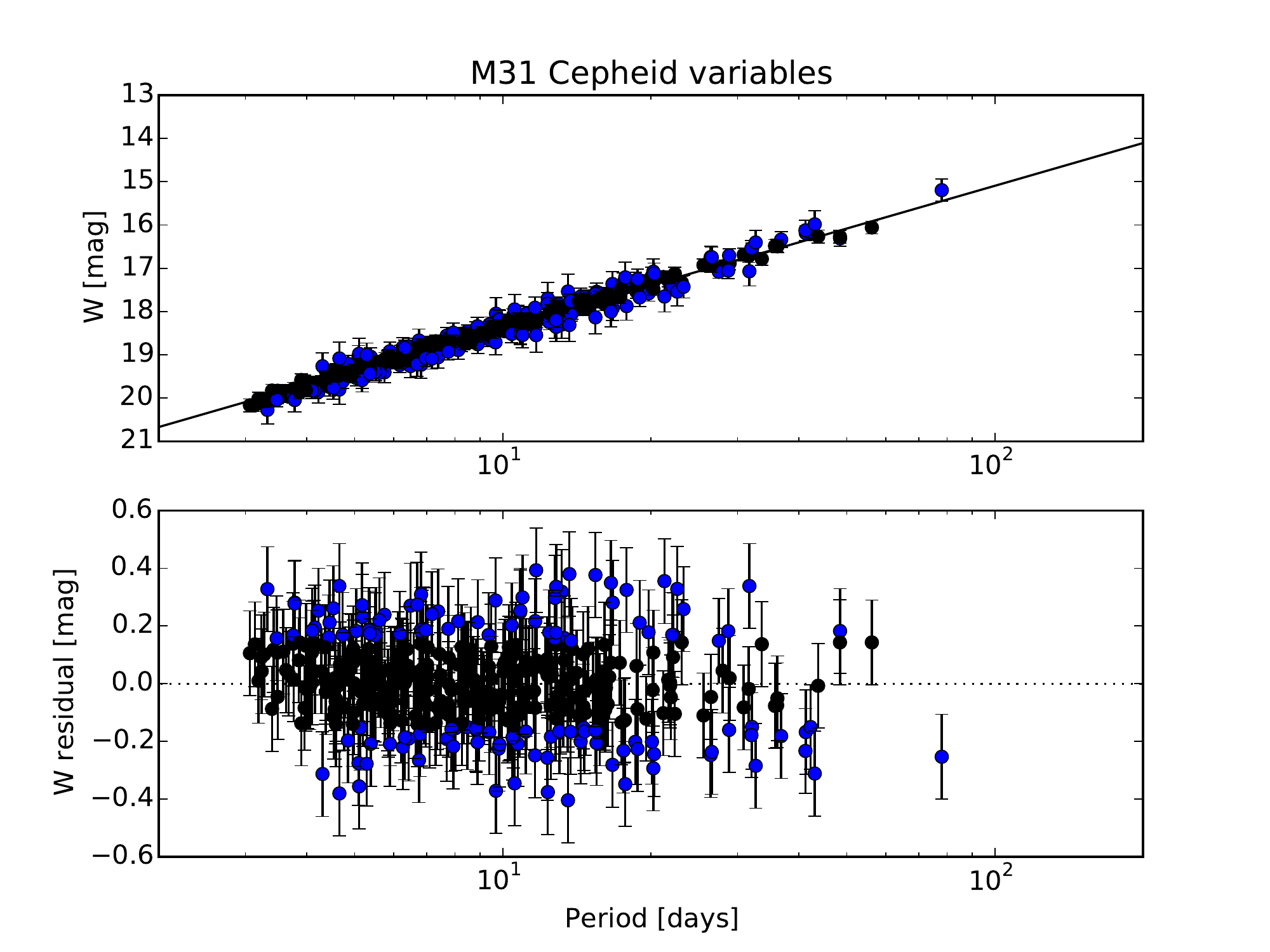}
\caption{Period-luminosity relation (upper panel) and magnitude residuals for the $\MAnd$ Cepheid variables in the R16 data set. The solid line shows the best fit of fit $(44)$. In the upper panel we rescale error bars with HPs, in the lower panel we do not. Data are colour-coded as explained in Figure \ref{Fig:LMC-Cepheid-variables-fit-c}.}
\label{Fig:R16-M31}
\end{figure}

\begin{figure}[hbtp]
\centering
\includegraphics[scale=0.75]{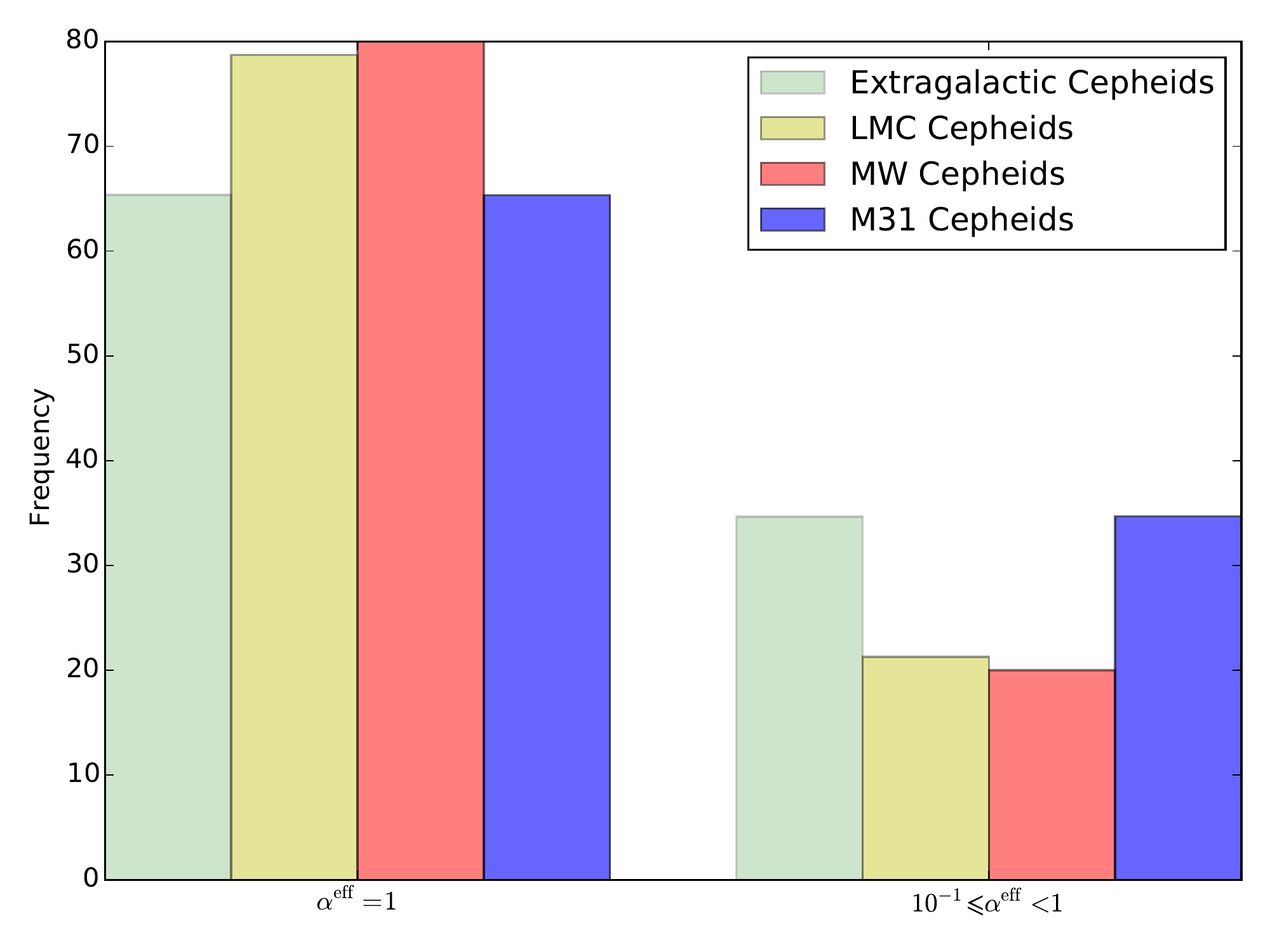}
\caption{Effective hyper-parameters for the R16 Cepheid sample used in fit $(44)$. Out of the $1114$ Cepheid variables in the $19$ $\SNe$ host galaxies and in the $\NGC$ megamaser system, $728$ have $\alpha_{\eff}=1$ and the remaining  $386$ have $10^{-1}\leq \alpha_{\eff} < 1$ (none have $\alpha_{\eff} < 10^{-1}$). Out of the $775$ $\LMC$ Cepheid variables, $610$ have $\alpha_{\eff}=1$ and the remaining $165$ have $10^{-1}\leq \alpha_{\eff} < 1$ (while none have  $\alpha_{\eff} < 10^{-1}$). Out of the $15$ $\MW$ Cepheid stars, $12$ have $\alpha_{\eff}=1$ and the remaining $3$ have  $10^{-1}\leq \alpha_{\eff} < 1$ (compare with Figure \ref{Fig:MW-Cepheid-variables}). Out of the $372$ $\MAnd$ Cepheid stars, $243$ have $\alpha_{\eff}=1$ and the remaining $129$ have $10^{-1}\leq \alpha_{\eff} < 1$ (none have  $\alpha_{\eff} < 10^{-1}$). Overall, $20\%$ of the $\MW$ Cepheids are down-weighted; this fraction reaches $21\%$ and $35\%$  for $\LMC$ Cepheids and $\MAnd$ Cepheids, respectively; as for the Cepheid variables in the $19$ $\SNe$ hosts and the $\NGC$ system the fraction is $35\%$.}
\label{Fig:effective-HP-fit-43}
\end{figure}

Figures \ref{Fig:R16-LMC}--\ref{Fig:R16-M31} show the period-luminosity relation for the best fit of our primary fit $(44)$ for Cepheid stars in galaxies $\LMC,\, \MW,\, \NGC,\,$ and $\MAnd$. In Figure \ref{Fig:effective-HP-fit-43} we show a histogram for HPs in the R16 Cepheid sample used in fit $(44)$. Differently to our analysis in Section \ref{Section:Application-R11}, which used the R11 data set and included outliers (Cepheid stars which did not pass the $2.5\sigma$ outlier criterion in \cite{Riess:2011yx}), there are no Cepheid stars with $\alpha^\eff<0.1$ in the analysis in this section using the R16 data set. Although in the Riess et al. analysis \cite{Riess:2016jrr} outliers  were not released, we find that about $30\%$ of the Cepheid stars in the R16 sample are down-weighted in our analysis (while the fraction of down-weighted Cepheid variables was about $46\%$ in our analysis using the R11 data set). Note that the normalised weight for Cepheids increases from $0.72$ for the primary best fit $(29)$ (with the R11 data set) to $0.85$ for the primary best fit $(44)$ (with the R16 data set), which agrees with the expectations for Gaussian data without outliers.

\begin{figure}[hbtp]
\centering
\includegraphics[width=\textwidth]{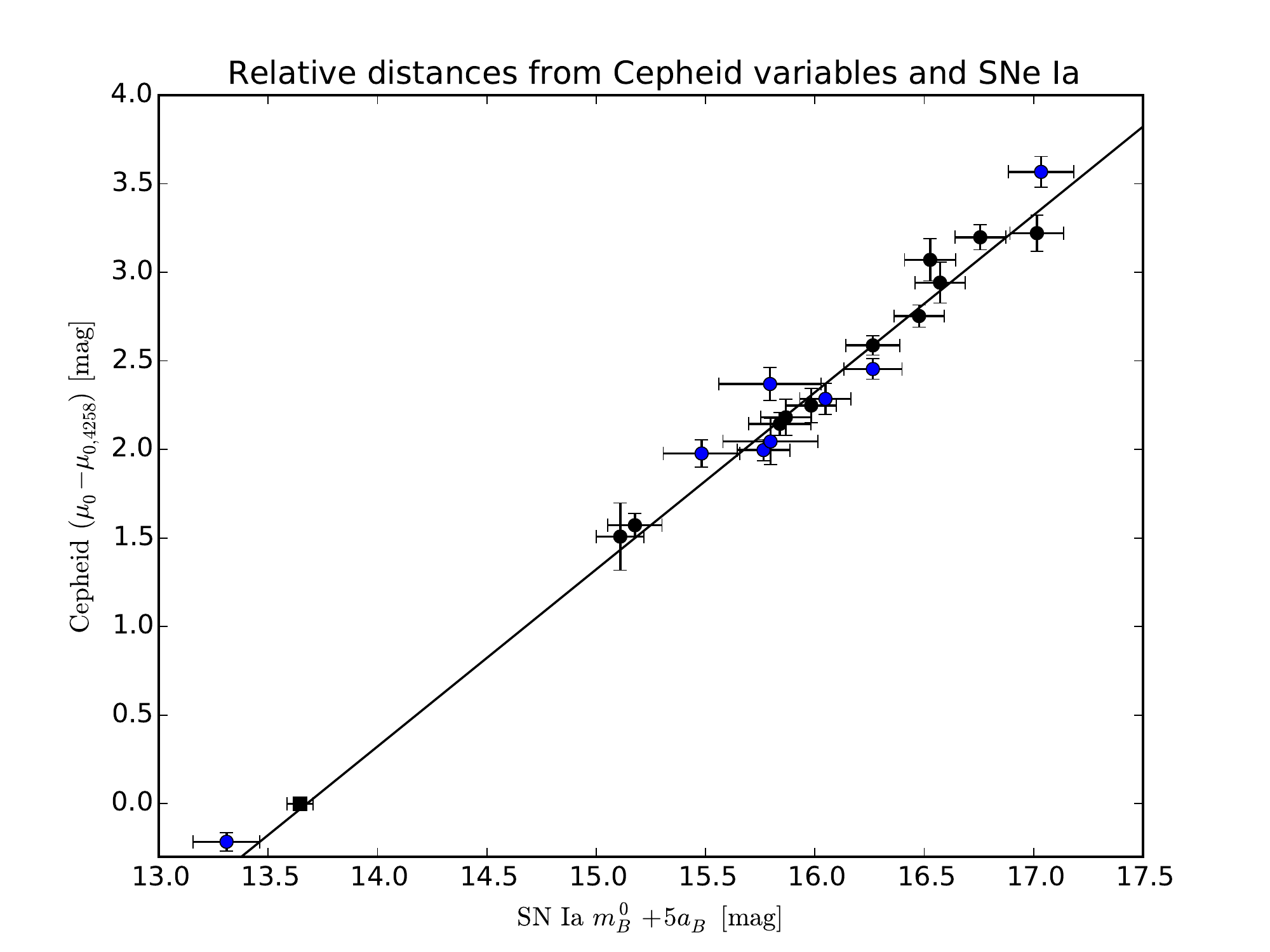}
\caption{Relative distances from Cepheids and $\SNe$. We plot the peak apparent visual magnitudes of each $\SNe$ (from Table 5 in \cite{Riess:2016jrr}) with error bars rescaled by HPs (colour code is the same as in Figure \ref{Fig:LMC-Cepheid-variables-fit-c}) against the relative distances between hosts determined from fit $(44)$ in Tables \ref{Table:details-fits}--\ref{Table:Constraints-main-analysis}. The solid line shows the corresponding best fit. The black square on the left corresponds to the expected reddening-free, peak magnitude of an $\SNe$ appearing in the megamaser system $\NGC$ which is derived from the fit $(44)$.}
\label{Fig:comparison-distances-R16}
\end{figure}

Riess et al. \cite{Riess:2016jrr} used the SALT-II light curve fitter and found no outliers among the $\SNe$ hosts. Although we cannot claim the opposite, we do find that some of the $\SNe$ are down-weighted in our analysis. In Figure \ref{Fig:comparison-distances-R16} we compare the $\SNe$ distances to the approximate, independent Cepheid distances from our primary fit $(44)$. In Table \ref{Table:SNIa-HP-fit-43} we show the distance parameters and the HPs for the $\SNe$ hosts in the R16 data set. The down-weighted $\SNe$ hosts might indicate the presence of unaccounted (or underestimated) systematics in the R16 data set. Whereas our analysis in Section \ref{Section:Application-R11} using the R11 data set showed that three out of the eight host galaxies are down-weighted in the fit, the primary fit $(44)$ using the R16 data set down-weights eight $\SNe$ host galaxies. The two host galaxies n3982 and n5584 are down-weighted in both fit $(29)$ and fit $(43)$. From Tables \ref{Table:details-fits}--\ref{Table:Constraints-main-analysis} we see that the normalised weight for $\SNe$ data is higher for the R16 data set ($0.83$) in comparison to the R11 data set ($0.74$). This indicates an improvement in the compatibility of this kind of data in the R16 sample.

\begin{table}[tbp]
\centering
\begin{tabular}{@{}lccccr}
\hline
\multicolumn{5}{c}{Distance parameters} \\
\hline
Host & $\SNe$ & $\mu_{0,i}-\mu_{0,4258}$ & $\mu_{0,i} $ best & $\alpha_{\eff}$ \\
\hline

 m101 & 2011fe & $-0.215\,(0.051)$&$29.09\,(0.06)$ &$0.59$ \\
 
 n1015 & 2009ig & $3.221\,(0.102)$&$32.53\,(0.10)$ &$1$ \\

 n1309 & 2002fk & $3.198\,(0.070)$& $32.51\,(0.07)$& $ 1$ \\
  
 n1365 & 2012fr & $1.977\,(0.077)$& $31.28\,(0.08)$& $ 0.51$ \\
   
 n1448 & 2001el & $1.996\,(0.059)$& $31.30\,(0.06)$& $ 0.92$ \\   
   
 n2442 & 2015F & $2.144\,(0.065)$& $31.45\,(0.07)$& $ 1$ \\
    
 n3021 & 1995al & $3.070\,(0.119)$& $32.38\,(0.12)$& $ 1$ \\
     
 n3370 & 1994ae & $2.753\,(0.063)$& $32.06\,(0.06)$& $1 $  \\
      
 n3447 & 2012ht & $2.588\,(0.055)$& $31.90\,(0.06)$& $1 $  \\
 
 n3972 & 2011by & $2.286\,(0.087)$& $31.59\,(0.09)$& $0.98 $  \\ 
 
 n3982 & 1998aq & $2.369\,(0.093)$& $31.68\,(0.09)$& $ 0.24$ \\
  
 n4038 & 2007sr & $2.045\,(0.129)$& $31.35\,(0.13)$& $ 0.27$ \\  
  
 n4424 & 2012cg & $1.508\,(0.190)$& $30.82\,(0.19)$& $ 1$ \\  
  
 n4536 & 1981B  & $1.572\,(0.067)$& $30.88\,(0.07)$ &$1$ \\

 n4639 & 1990N & $2.248\,(0.096)$& $31.56\,(0.10)$& $1 $ \\
    
 n5584 & 2007af & $2.454\,(0.058)$& $31.76\,(0.06)$& $ 0.74$ \\
       
 n5917 & 2005cf & $2.942\,(0.116)$& $32.25\,(0.12)$& $ 1$ \\
       
 n7250 & 2013dy & $2.181\,(0.101)$& $31.49\,(0.10)$& $ 1$ \\
        
 u9391 & 2003du & $3.567\,(0.087)$& $32.87\,(0.09)$& $ 0.58$ \\
         
\hline
\end{tabular}
\caption{\label{Table:SNIa-HP-fit-43} Distance parameters for the $\SNe$ hosts corresponding to our primary fit [fit $(44)$] for the R16 data set. Numbers in brackets indicate the standard deviation. The last column contains the effective HP for each $\SNe$ host.}
\end{table}

If we compare the normalised weight of anchors in Tables \ref{Table:details-fits}--\ref{Table:Constraints-main-analysis} for fits $(29)$ and $(44)$ ($0.86$ and  $1$, respectively), we can also see an improvement in the compatibility of this kind of data in the R16 data set. Although this would suggest not to include distance moduli with HPs in the fit (since the HP Gaussian likelihood would add uncertainty because of its slightly wider wings), the normalised weight of anchors depends on the variants of the analysis ($R$, period cut-off, metallicity prior) ranging from $0.62$ to $1$. Hence, there is no strong reason to exclude the use of HPs in this kind of data.

Figure \ref{Fig:H0bestfits} shows the best estimates of $H_0$ by G. Efstathiou \cite{Efstathiou:2013via}, Riess et al. \cite{Riess:2011yx}, Riess et al. \cite{Riess:2016jrr}, and those in this work, fits $(29)$ and $(44)$, along with the indirect determinations of $H_0$ by the WMAP team \cite{Hinshaw:2012aka} and by the Planck collaboration \cite{Ade:2015xua}. Our best estimate using the R11 data set, fit $(29)$, is $75\pm3.9 \km \, \second^{-1} \, \Mpc^{-1}$ and is the most uncertain (a $5.2\%$ measurement) of all presented measurements, but it agrees with all previous direct determinations of $H_0$ and differs by $1.8\sigma$ from the Planck value 2015 and $2 \sigma$ from the updated Planck 2016 value $66.93 \pm 0.62 \,\km\, \second^{-1} \, \Mpc^{-1}$ \cite{Aghanim:2016yuo}. Note that because G. Efstathiou considered only $\NGC$ as an anchor and set a period cut-off of $60$ days, his determination is more uncertain than that of Riess et al. \cite{Riess:2011yx} which used three anchors and no period cut-off. As illustrated in the figure (top panel), our best estimate using the R16 data set, fit $(44)$, also agrees with the previous determinations of $H_0$, while its uncertainty is smaller (a $2.9\%$ measurement) than the one of fit $(29)$. Concerning the indirect determinations of $H_0$, we see that our best estimate, fit $(44)$, agrees within $1\sigma$ with WMAP 2009, but it is 2.6 $\sigma$ larger than the Planck 2015 value of $H_0 = 67.81 \pm 0.92 \,\km\, \second^{-1} \, \Mpc^{-1}$ and about 3.1 $\sigma$ larger than the updated Planck 2016 value $66.93 \pm 0.62 \,\km\, \second^{-1} \, \Mpc^{-1}$ \cite{Aghanim:2016yuo}. This tension could be an indicator of unresolved systematics or new physics (e.g.\ \cite{2016A&A...594A..14P, 2016arXiv160600180A, Bernal:2016gxb, 2013PhRvD..88f3519P}). 


\section{Discussion and Conclusions}
\label{Section:Summary}

In this paper we present a statistical method which enables a comprehensive treatment of available data in order to determine the universe's current expansion $H_0$. The use of Bayesian hyper-parameters avoids the arbitrary removal of data which is implicit in outlier rejection algorithms. Such algorithms have been used in \cite{Riess:2009pu}, \cite{Riess:2011yx}, \cite{Efstathiou:2013via} and in some cases a dependence of the results on the statistical method utilised has been found. We show that the determination of the Hubble constant with Bayesian hyper-parameters is robust against different assumptions in the analysis (e.g., period cut in the Cepheid variables data, prior on the metallicity parameter $Z_W$ of the period-luminosity relation, reddening law) as listed in Tables \ref{Table:details-fits}--\ref{Table:Constraints-main-analysis}. In addition, since the method uses all available data sets, it allows to check how consistent they are with each other and how much weight they are assigned in the fit. Low values of HPs might be due to unrecognised (or underestimated) systematics in the data sets and/or might point to a need for better modelling.

We have shown that, contrary to the usual $\chi^2$ approach, when Cepheid variables are fitted using HPs, the down-weighted data (outlier candidates in an outlier rejection algorithm) do not significantly affect the slope $b_W$ in the period-luminosity relation compared to samples with outliers removed (see Figure \ref{Fig:LMC-Cepheid-variables-fit-c} and Table \ref{Table:LMC-fits}). Note that due to degeneracies in the parameters (e.g., $H_0,\, M_W,\, b_W,\, Z_W,\,$\dots), a bias in the Cepheid parameters can also lead to a bias in the determination of the Hubble constant $H_0$ in the usual $\chi^2$ analysis. Moreover, removing data sets might lead to unnecessary increase of the error bars in the fitting parameters (compare, for instance, Efstathiou \cite{Efstathiou:2013via} and Riess et al. \cite{Riess:2011yx} $H_0$ values in Figure \ref{Fig:H0bestfits}).

Based on the discussion in Subsections \ref{Subsection:LMCR11}--\ref{Subsection:NGC4258} it appears that the three sets of Cepheid variables in the galaxies LMC, MW, and NGC 4258 are consistent with each other ($b_W$ and $M_W$ agree within error bars), thus providing no argument to exclude any of them from the main analysis. In Subsection \ref{Subsection:Zw-dependence} we have studied the period-luminosity relation -- allowing for a metallicity dependence -- of each one of the galaxies in the R11 data set containing Cepheid variables. Table \ref{Table:Zw-dependence-of-PL-relation} shows that at least five galaxies have a slope $b_W$ which differs from that of LMC Cepheid variables by about $2\sigma$. A statistical method combining these data sets without taking those inconsistencies into account could lead to biased results (compare, for instance, $b_W$ for our fits in Tables \ref{Table:details-fits}--\ref{Table:Constraints-main-analysis} with the corresponding fits in Appendix A of \cite{Efstathiou:2013via} which are driven upwards). Our method is able to deal with those data sets without arbitrarily excluding data points.

\begin{figure}[hbtp]
\centering
\includegraphics[width=\textwidth]{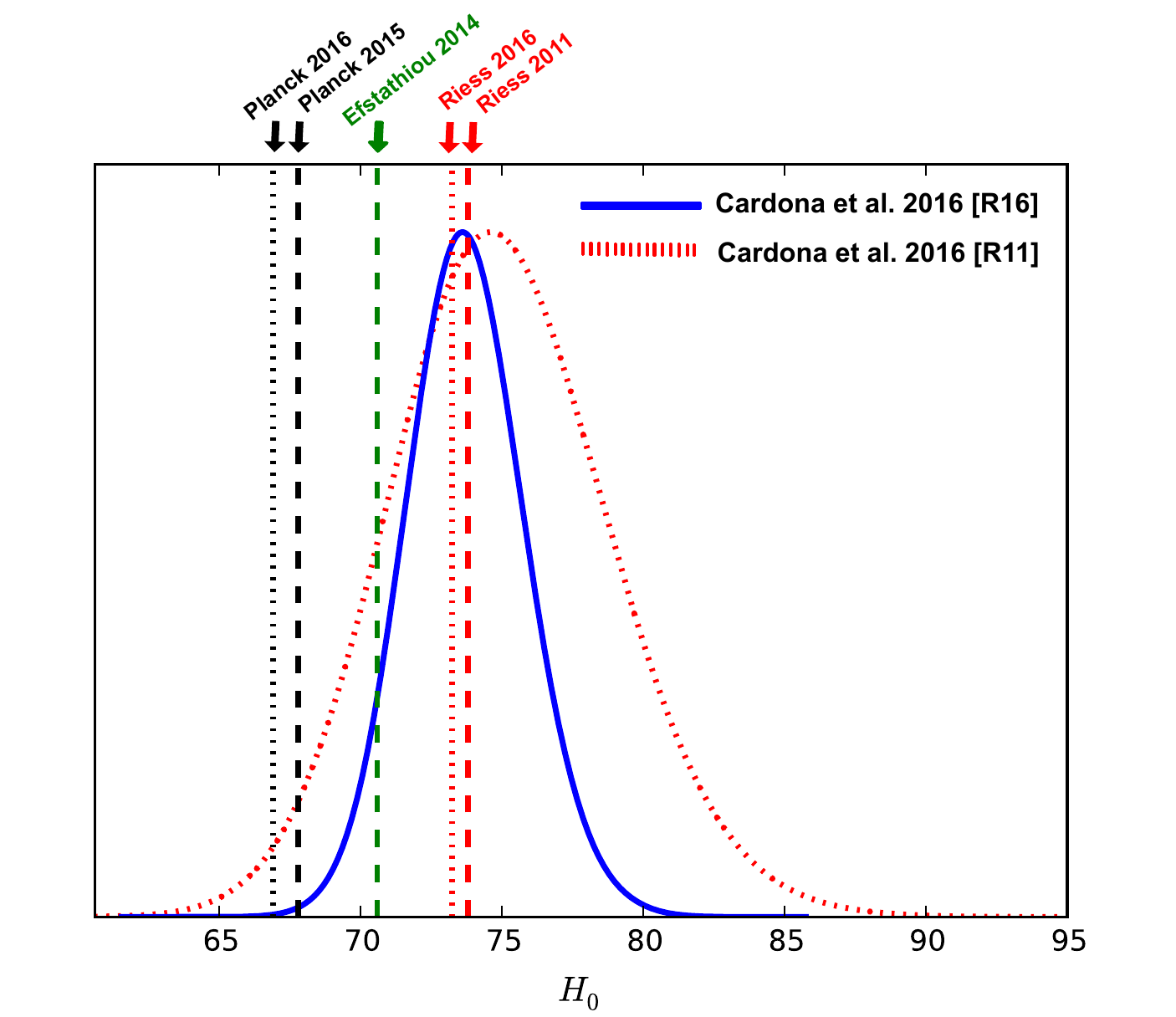}
\caption{Marginalized probability distribution function of the present Hubble parameter $H_0$, as found in our HP analysis, for the two data sets considered in this paper (zoom of lower right corner of Fig. \ref{Fig:Main-analysis-fitM1a}). Here we show our baseline results for the R11 dataset [fit (29) in Tables \ref{Table:details-fits}--\ref{Table:Constraints-main-analysis}, dashed red contours] and for the R16 dataset, which contains more than twice as much data [fit (44), blue contours]. 
The black dotted vertical line indicates the updated Planck 2016 value for the base six-parameter $\Lambda$CDM model \cite{Aghanim:2016yuo}. Black, green, and red dashed vertical lines respectively indicate the values derived by the Planck collaboration for the base six-parameter $\Lambda$CDM model \cite{Ade:2015xua}, Efstathiou's value \cite{Efstathiou:2013via} (based on R11 data) used by Planck collaboration as a prior, and the $3\%$ measurement reported by R11 \cite{Riess:2011yx}; The red dotted vertical line indicates the best estimate from the R16 analysis  \cite{Riess:2016jrr}.
}
\label{Fig:Main-analysis-fitM1a_zoom}
\end{figure}

One of the advantages of using HPs to determine the Hubble constant is that one can assess the compatibility of different data sets. Our best estimates, fits $(29)$ and $(44)$, use HPs to combine all available Cepheid variables (i.e., no period cut), all available independent measurements of distance modulus to $\NGC$, $\LMC$, and $\MAnd$ [only in fit $(44)$], and all available $\SNe$ apparent magnitudes, but we have also performed several variants which are shown in Tables \ref{Table:details-fits}--\ref{Table:Constraints-main-analysis}. 
We have estimated the degree of agreement for different kinds of data in our fits through the normalised weights \eqref{Eq:normalised-weights} and found that, based on this metric, fits $(29)$ and $(44)$ provide the best solution for R11 and R16 data sets, respectively. Our HP-based constraint on $H_0$ using the R16 data is thus
\begin{equation}
\label{final_value}
H_0 = (73.75 \pm 2.11) \, \km \, \second^{-1} \, \Mpc^{-1} \, ,
\end{equation}
which compares to other analysis as illustrated in Figure \ref{Fig:H0bestfits}. In particular, our baseline value for the R16 dataset is about 2.6 $\sigma$ larger than the Planck 2015 value of $H_0 = 67.81 \pm 0.92 \,\km\, \second^{-1} \, \Mpc^{-1}$ and about 3.1 $\sigma$ larger than the updated Planck 2016 value $66.93 \pm 0.62 \,\km\, \second^{-1} \, \Mpc^{-1}$. 
The probability distribution function of the present Hubble parameter $H_0$, as found in our HP analysis, for the two data sets considered in this paper, is shown in Fig.\ \ref{Fig:Main-analysis-fitM1a_zoom}. There we show our baseline results for the R11 dataset [fit (29) in Tables \ref{Table:details-fits}--\ref{Table:Constraints-main-analysis}, dashed red contours] and for the newer R16 dataset, which has more than twice as much data [fit (44), blue contours], after marginalizing over the HPs.
The estimate (\ref{final_value}), Figure \ref{Fig:H0bestfits} and Figure \ref{Fig:Main-analysis-fitM1a_zoom} summarize our main (baseline) result. 

Our analysis shows that there are two main choices influencing the final estimate of $H_0$: the first is the choice of the anchors (in particular the inclusion of MW Cepheids, which pushes $H_0$ to higher values, as discussed in Section \ref{subsection:Anchors}); the second is the  set of data on which HPs are applied (discussed in Section \ref{Subsection:HP-variants} and illustrated in Figure \ref{Fig:H0-values-3-anchors}). 

Our analysis of R11 shows some deviations from the expected HP values for consistent Gaussian data sets. This is not unexpected for the R11 Cepheid data due to known outliers, but also the SNe-Ia data is down-weighted. For R16 the outliers were removed prior to the release of the data set. The HP analysis still down-weights some of the Cepheid stars which passed the outlier rejection algorithm in \cite{Riess:2016jrr}, the normalized weight is however consistent with expectations for Gaussian data, and the same is true for the R16 SNe-Ia. Of course this is an a-posteriori result that could not have been obtained without performing the HP analysis. Overall, our constraints are compatible with the analyses of \cite{Riess:2011yx,Efstathiou:2013via,Riess:2016jrr} when the same distance anchors are used, see Fig.\ \ref{Fig:single-combined-anchor}.

Since our analysis shows down-weighted data points in several data sets, we think an analysis with HPs is generally appropriate. The analysis is safer because it reduces bias in the results in the presence of data points with unreliable error bars. The use of HPs is also conservative because it does not underestimate the error bars on the constraints. Moreover, HPs are useful because they allow to look for the presence of possible underestimated systematics in the data. We conclude that as long as the sum of normalised weights \eqref{Eq:normalised-weights} for the three kinds of data $\sum_j || \alpha^j || \lesssim 2.55$,  HPs offer a more robust approach to measure the universe's expansion rate. 

An important piece of the puzzle concerning the difference between the {\em Planck} inferred value of $H_0$ and the local expansion rate is the impact of the MW Cepheids with parallaxes. In the near future data from the Gaia satellite will vastly improve our knowledge of MW parallaxes 
\cite{2016arXiv160905175C} and either bring the values into better agreement or strengthen the discrepancy.

 
\appendix
\section{$H_0$ and overall table of results}
\label{Section:Appendix}
In Tables \ref{Table:details-fits}--\ref{Table:Constraints-main-analysis} we show the overall table of results presented in the various sections, in order to give an overview of all the cases investigated for this work.

\setlength{\arrayrulewidth}{1mm}
\begin{table}[tbp]
\centering
\resizebox{\textwidth}{9cm}{
{
\rowcolors{4}{green!80!yellow!50}{green!70!yellow!30}
\begin{tabular}{$l^c^c^c^c^c^c^c^c^c^c^c^c^c^c^r}
\hline 
\\
\rowstyle{\bfseries\boldmath} Fit & $\alpha^{\Cepheid}$ & $\alpha^{\SNe}$ & $\alpha^{\Anchors}$ & $P$ & $R$ & $\sigma_{Z_W}$ & $\sigma_{\intt,i}$ & $\sigma_{\intt}^{\LMC}$ & $\sigma_{\intt}^{\MW}$ & CS & $\mu_{0,\NGC}^{\obs}$ & $\mu_{0,\LMC}^{\obs}$ & $\mu_{0,\MAnd}^{\obs}$ & $\MW$ \\ 
\\
\hline
\rowcolor{white} \multicolumn{15}{l}{\bf{R11, NGC4258}}     \\
\rule[-1ex]{0pt}{2.5ex} $1$ & Y & Y & Y & $205$ & $0.410$ & - & V & V & - & R11 & \cite{Humphreys:2013eja} & - & - & - \\ 
\rule[-1ex]{0pt}{2.5ex} $2$ & Y & Y & Y & $60$ & $0.410$ & - & V & V & - & R11 & \cite{Humphreys:2013eja} & - & - & - \\ 
\rule[-1ex]{0pt}{2.5ex} $3$ & Y & Y & Y & $205$ & $0.410$ & $0.02$ & V & V & - & R11 & \cite{Humphreys:2013eja} & - & - & - \\ 
\rule[-1ex]{0pt}{2.5ex} $4$ & Y & Y & N & $205$ & $0.410$ & $0.02$ & V & V & - & R11 & \cite{Humphreys:2013eja} & - & - & - \\ 
\rule[-1ex]{0pt}{2.5ex} $5$ & Y & Y & Y & $60$ & $0.410$ & $0.02$ & V & V & - & R11 & \cite{Humphreys:2013eja} & - & - & - \\ 
\rule[-1ex]{0pt}{2.5ex} $6$ & Y & Y & N & $60$ & $0.410$ & $0.02$ & V & V & - & R11 & \cite{Humphreys:2013eja} & - & - & - \\ 
\midrule
\rowcolor{white} \multicolumn{15}{l}{\bf{R11, LMC}}     \\
\rule[-1ex]{0pt}{2.5ex} $7$ & Y & Y & Y & $205$ & $0.410$ & - & V & V & - & R11 & - & \cite{Pietrzynski:2013gia} & - & - \\ 
\rule[-1ex]{0pt}{2.5ex} $8$ & Y & Y & Y & $60$ & $0.410$ & - & V & V & - & R11 & - & \cite{Pietrzynski:2013gia} & - & - \\ 
\rule[-1ex]{0pt}{2.5ex} $9$ & Y & Y & Y & $205$ & $0.410$ & $0.02$ & V & V & - & R11 & - & \cite{Pietrzynski:2013gia} & - & - \\ 
\rule[-1ex]{0pt}{2.5ex} $10$ & Y & Y & N & $205$ & $0.410$ & $0.02$ & V & V & - & R11 & - & \cite{Pietrzynski:2013gia} & - & - \\ 
\rule[-1ex]{0pt}{2.5ex} $11$ & Y & Y & Y & $60$ & $0.410$ & $0.02$ & V & V & - & R11 & - & \cite{Pietrzynski:2013gia} & - & - \\ 
\rule[-1ex]{0pt}{2.5ex} $12$ & Y & Y & N & $60$ & $0.410$ & $0.02$ & V & V & - & R11 & - & \cite{Pietrzynski:2013gia} & - & - \\
\midrule
\rowcolor{white} \multicolumn{15}{l}{\bf{R11, MW}}     \\
\rule[-1ex]{0pt}{2.5ex} $13$ & Y & Y & - & $205$ & $0.410$ & - & V & V & V & R11 & - & - & - & \cite{vanLeeuwen:2007xw} \\ 
\rule[-1ex]{0pt}{2.5ex} $14$ & Y & Y & - & $60$ & $0.410$ & - & V & V & V & R11 & - & - & - & \cite{vanLeeuwen:2007xw} \\
\rule[-1ex]{0pt}{2.5ex} $15$ & Y & Y & - & $205$ & $0.410$ & $0.02$ & V & V & V & R11 & - & - & - & \cite{vanLeeuwen:2007xw} \\
\rule[-1ex]{0pt}{2.5ex} $16$ & Y & Y & - & $60$ & $0.410$ & $0.02$ & V & V & V & R11 & - & - & - & \cite{vanLeeuwen:2007xw} \\
\midrule
\rowcolor{white} \multicolumn{15}{l}{\bf{R11, NGC4258 + LMC}}     \\
\rule[-1ex]{0pt}{2.5ex} $17$ & Y & Y & Y & $205$ & $0.410$ & - & V & V & - & R11 & \cite{Humphreys:2013eja} & \cite{Pietrzynski:2013gia} & - & - \\
\rule[-1ex]{0pt}{2.5ex} $18$ & Y & Y & Y & $60$ & $0.410$ & - & V & V & - & R11 & \cite{Humphreys:2013eja} & \cite{Pietrzynski:2013gia} & - & - \\
\rule[-1ex]{0pt}{2.5ex} $19$ & Y & Y & Y & $205$ & $0.410$ & $0.02$ & V & V & - & R11 & \cite{Humphreys:2013eja} & \cite{Pietrzynski:2013gia} & - & - \\
\rule[-1ex]{0pt}{2.5ex} $20$ & Y & Y & Y & $60$ & $0.410$ & $0.02$ & V & V & - & R11 & \cite{Humphreys:2013eja} & \cite{Pietrzynski:2013gia} & - & - \\ 
\midrule
\rowcolor{white} \bf{R11, NGC4258 + MW} \\
\rule[-1ex]{0pt}{2.5ex} $21$ & Y & Y & Y & $205$ & $0.410$ & - & V & V & V & R11 & \cite{Humphreys:2013eja} & - & - & \cite{vanLeeuwen:2007xw} \\ 
\rule[-1ex]{0pt}{2.5ex} $22$ & Y & Y & Y & $60$ & $0.410$ & - & V & V & V & R11 & \cite{Humphreys:2013eja} & - & - & \cite{vanLeeuwen:2007xw} \\
\rule[-1ex]{0pt}{2.5ex} $23$ & Y & Y & Y & $205$ & $0.410$ & $0.02$ & V & V & V & R11 & \cite{Humphreys:2013eja} & - & - & \cite{vanLeeuwen:2007xw} \\
\rule[-1ex]{0pt}{2.5ex} $24$ & Y & Y & Y & $60$ & $0.410$ & $0.02$ & V & V & V & R11 & \cite{Humphreys:2013eja} & - & - & \cite{vanLeeuwen:2007xw} \\
\midrule
\rowcolor{white} \bf{R11, LMC + MW} \\
\rule[-1ex]{0pt}{2.5ex} $25$ & Y & Y & Y & $205$ & $0.410$ & - & V & V & V & R11 & - & \cite{Pietrzynski:2013gia} & - & \cite{vanLeeuwen:2007xw} \\ 
\rule[-1ex]{0pt}{2.5ex} $26$ & Y & Y & Y & $60$ & $0.410$ & - & V & V & V & R11 & - & \cite{Pietrzynski:2013gia} & - & \cite{vanLeeuwen:2007xw} \\
\rule[-1ex]{0pt}{2.5ex} $27$ & Y & Y & Y & $205$ & $0.410$ & $0.02$ & V & V & V & R11 & - & \cite{Pietrzynski:2013gia} & - & \cite{vanLeeuwen:2007xw} \\
\rule[-1ex]{0pt}{2.5ex} $28$ & Y & Y & Y & $60$ & $0.410$ & $0.02$ & V & V & V & R11 & - & \cite{Pietrzynski:2013gia} & - & \cite{vanLeeuwen:2007xw} \\
\midrule
\rowcolor{white} \bf{R11, LMC + MW + NGC4258} \\
\rule[-1ex]{0pt}{2.5ex} $\mathbf{29}$ & \textbf{Y} & \textbf{Y} & \textbf{Y} & $\mathbf{205}$ & $\mathbf{0.410}$ & $\mathbf{0.02}$ & \textbf{V} & \textbf{V} & \textbf{V} & \textbf{R11} & \cite{Humphreys:2013eja},\cite{Polshaw:2015ika} & \cite{Pietrzynski:2013gia} & - & \cite{vanLeeuwen:2007xw} \\ 
\rule[-1ex]{0pt}{2.5ex} $30$ & Y & Y & Y & $205$ & $0.410$ & $0.02$ & V & V & V & R11 & \cite{Humphreys:2013eja} & \cite{Pietrzynski:2013gia} & - & \cite{vanLeeuwen:2007xw} \\
\rule[-1ex]{0pt}{2.5ex} $31$ & Y & N & Y & $205$ & $0.410$ & $0.02$ & V & V & V & R11 & \cite{Humphreys:2013eja},\cite{Polshaw:2015ika} & \cite{Pietrzynski:2013gia} & - & \cite{vanLeeuwen:2007xw} \\
\rule[-1ex]{0pt}{2.5ex} $32$ & Y & Y & N & $205$ & $0.410$ & $0.02$ & V & V & V & R11 & \cite{Humphreys:2013eja},\cite{Polshaw:2015ika} & \cite{Pietrzynski:2013gia} & - & \cite{vanLeeuwen:2007xw} \\
\rule[-1ex]{0pt}{2.5ex} $33$ & Y & N & N & $205$ & $0.410$ & $0.02$ & V & V & V & R11 & \cite{Humphreys:2013eja},\cite{Polshaw:2015ika} & \cite{Pietrzynski:2013gia} & - & \cite{vanLeeuwen:2007xw} \\
\rule[-1ex]{0pt}{2.5ex} $34$ & Y & Y & Y & $60$ & $0.410$ & $0.02$ & V & V & V & R11 & \cite{Humphreys:2013eja},\cite{Polshaw:2015ika} & \cite{Pietrzynski:2013gia} & - & \cite{vanLeeuwen:2007xw} \\
\rule[-1ex]{0pt}{2.5ex} $35$ & Y & N & N & $60$ & $0.410$ & $0.02$ & $0.30$ & $0.113$ & $0.10$ & R11 & \cite{Humphreys:2013eja} & \cite{Pietrzynski:2013gia} & - & \cite{vanLeeuwen:2007xw} \\
\rule[-1ex]{0pt}{2.5ex} $36$ & Y & Y & Y & $205$ & $0.410$ & $0.25$ & V & V & V & R11 & \cite{Humphreys:2013eja},\cite{Polshaw:2015ika} & \cite{Pietrzynski:2013gia} & - & \cite{vanLeeuwen:2007xw} \\
\rule[-1ex]{0pt}{2.5ex} $37$ & Y & Y & Y & $60$ & $0.410$ & $0.25$ & V & V & V & R11 & \cite{Humphreys:2013eja},\cite{Polshaw:2015ika} & \cite{Pietrzynski:2013gia} & - & \cite{vanLeeuwen:2007xw} \\
\rule[-1ex]{0pt}{2.5ex} $38$ & Y & Y & Y & $205$ & $0.410$ & - & V & V & V & R11 & \cite{Humphreys:2013eja},\cite{Polshaw:2015ika} & \cite{Pietrzynski:2013gia} & - & \cite{vanLeeuwen:2007xw} \\
\rule[-1ex]{0pt}{2.5ex} $39$ & Y & Y & Y & $60$ & $0.410$ & - & V & V & V & R11 & \cite{Humphreys:2013eja},\cite{Polshaw:2015ika} & \cite{Pietrzynski:2013gia} & - & \cite{vanLeeuwen:2007xw} \\ 
\midrule
\rowcolor{white} \bf{R16, LMC + MW + NGC4258 + M31} \\
\rule[-1ex]{0pt}{2.5ex} $40$ & Y & Y & Y & $205$ & $0.31 $ & $0.25$ & - & - & - & R16 & \cite{Riess:2016jrr} & \cite{Pietrzynski:2013gia} & \cite{Riess:2016jrr} & \cite{Riess:2016jrr} \\ 
\rule[-1ex]{0pt}{2.5ex} $41$ & Y & Y & Y & $205$ & $0.31 $ & $0.02$ & - & - & - & R16 & \cite{Riess:2016jrr} & \cite{Pietrzynski:2013gia} & \cite{Riess:2016jrr} & \cite{Riess:2016jrr} \\
\rule[-1ex]{0pt}{2.5ex} $42$ & Y & Y & Y & $205$ & $0.35 $ & $0.25$ & - & - & - & R16 & \cite{Riess:2016jrr} & \cite{Pietrzynski:2013gia} & \cite{Riess:2016jrr} & \cite{Riess:2016jrr} \\
\rule[-1ex]{0pt}{2.5ex} $43$ & Y & Y & Y & $205$ & $0.39 $ & $0.25$ & - & - & - & R16 & \cite{Riess:2016jrr} & \cite{Pietrzynski:2013gia} & \cite{Riess:2016jrr} & \cite{Riess:2016jrr} \\
\rule[-1ex]{0pt}{2.5ex} $\mathbf{44}$ & \textbf{Y} & \textbf{Y} & \textbf{Y} & $\mathbf{205}$ & $\mathbf{0.47} $ & $\mathbf{0.25}$ & - & - & - & \textbf{R16} & \cite{Riess:2016jrr} & \cite{Pietrzynski:2013gia} & \cite{Riess:2016jrr} & \cite{Riess:2016jrr} \\
\rule[-1ex]{0pt}{2.5ex} $45$ & Y & Y & Y & $205$ & $0.35 $ & $0.02$ & - & - & - & R16 & \cite{Riess:2016jrr} & \cite{Pietrzynski:2013gia} & \cite{Riess:2016jrr} & \cite{Riess:2016jrr} \\
\rule[-1ex]{0pt}{2.5ex} $46$ & Y & Y & Y & $205$ & $0.39 $ & $0.02$ & - & - & - & R16 & \cite{Riess:2016jrr} & \cite{Pietrzynski:2013gia} & \cite{Riess:2016jrr} & \cite{Riess:2016jrr} \\
\rule[-1ex]{0pt}{2.5ex} $47$ & Y & Y & Y & $205$ & $0.47 $ & $0.02$ & - & - & - & R16 & \cite{Riess:2016jrr} & \cite{Pietrzynski:2013gia} & \cite{Riess:2016jrr} & \cite{Riess:2016jrr} \\

\rule[-1ex]{0pt}{2.5ex} $48$ & Y & Y & Y & $60$ & $0.31 $ & $0.25$ & - & - & - & R16 & \cite{Riess:2016jrr} & \cite{Pietrzynski:2013gia} & \cite{Riess:2016jrr} & \cite{Riess:2016jrr} \\ 
\rule[-1ex]{0pt}{2.5ex} $49$ & Y & Y & Y & $60$ & $0.31 $ & $0.02$ & - & - & - & R16 & \cite{Riess:2016jrr} & \cite{Pietrzynski:2013gia} & \cite{Riess:2016jrr} & \cite{Riess:2016jrr} \\
\rule[-1ex]{0pt}{2.5ex} $50$ & Y & Y & Y & $60$ & $0.35 $ & $0.25$ & - & - & - & R16 & \cite{Riess:2016jrr} & \cite{Pietrzynski:2013gia} & \cite{Riess:2016jrr} & \cite{Riess:2016jrr} \\
\rule[-1ex]{0pt}{2.5ex} $51$ & Y & Y & Y & $60$ & $0.39 $ & $0.25$ & - & - & - & R16 & \cite{Riess:2016jrr} & \cite{Pietrzynski:2013gia} & \cite{Riess:2016jrr} & \cite{Riess:2016jrr} \\
\rule[-1ex]{0pt}{2.5ex} $52$ & Y & Y & Y & $60$ & $0.47 $ & $0.25$ & - & - & - & R16 & \cite{Riess:2016jrr} & \cite{Pietrzynski:2013gia} & \cite{Riess:2016jrr} & \cite{Riess:2016jrr} \\
\rule[-1ex]{0pt}{2.5ex} $53$ & Y & Y & Y & $60$ & $0.35 $ & $0.02$ & - & - & - & R16 & \cite{Riess:2016jrr} & \cite{Pietrzynski:2013gia} & \cite{Riess:2016jrr} & \cite{Riess:2016jrr} \\
\rule[-1ex]{0pt}{2.5ex} $54$ & Y & Y & Y & $60$ & $0.39 $ & $0.02$ & - & - & - & R16 & \cite{Riess:2016jrr} & \cite{Pietrzynski:2013gia} & \cite{Riess:2016jrr} & \cite{Riess:2016jrr} \\
\rule[-1ex]{0pt}{2.5ex} $55$ & Y & Y & Y & $60$ & $0.47 $ & $0.02$ & - & - & - & R16 & \cite{Riess:2016jrr} & \cite{Pietrzynski:2013gia} & \cite{Riess:2016jrr} & \cite{Riess:2016jrr} \\
\rule[-1ex]{0pt}{2.5ex} $56$ & N & N & N & $ 205 $ & $ 0.47 $ & $ 0.25 $ & - & - & - & R16 & \cite{Riess:2016jrr} & \cite{Pietrzynski:2013gia} & \cite{Riess:2016jrr} & \cite{Riess:2016jrr} \\
\rule[-1ex]{0pt}{2.5ex} $57$ & N & Y & Y & $ 205 $ & $ 0.47 $ & $ 0.25 $ & - & - & - & R16 & \cite{Riess:2016jrr} & \cite{Pietrzynski:2013gia} & \cite{Riess:2016jrr} & \cite{Riess:2016jrr} \\

\midrule
\rowcolor{white} \bf{R16, NGC4258} \\
\rule[-1ex]{0pt}{2.5ex} $58$ & Y & Y & Y & $ 205 $ & $ 0.47 $ & $ 0.25$ & - & - & - & R16 & \cite{Riess:2016jrr} & - & - & - \\
\rule[-1ex]{0pt}{2.5ex} $59$ & Y & Y & Y & $ 60 $ & $ 0.47 $ & $ 0.25$ & - & - & - & R16 & \cite{Riess:2016jrr} & - & - & - \\

\midrule
\rowcolor{white} \bf{R16, LMC} \\
\rule[-1ex]{0pt}{2.5ex} $60$ & Y & Y & Y & $ 205 $ & $ 0.47 $ & $ 0.25$ & - & - & - & R16 & - & \cite{Pietrzynski:2013gia} & - & - \\
\rule[-1ex]{0pt}{2.5ex} $61$ & Y & Y & Y & $ 60 $ & $ 0.47 $ & $ 0.25$ & - & - & - & R16 & - & \cite{Pietrzynski:2013gia} & - & - \\

\midrule
\rowcolor{white} \bf{R16, MW} \\
\rule[-1ex]{0pt}{2.5ex} $62$ & Y & Y & Y & $ 205 $ & $ 0.47 $ & $ 0.25$ & - & - & - & R16 & - & - & - & \cite{Riess:2016jrr} \\
\rule[-1ex]{0pt}{2.5ex} $63$ & Y & Y & Y & $ 60 $ & $ 0.47 $ & $ 0.25$ & - & - & - & R16 & - & - & - & \cite{Riess:2016jrr} \\


\midrule
\rowcolor{white} \bf{R16, NGC4258 + LMC} \\
\rule[-1ex]{0pt}{2.5ex} $64$ & Y & Y & Y & $ 205 $ & $ 0.47 $ & $ 0.25$ & - & - & - & R16 & \cite{Riess:2016jrr} & \cite{Pietrzynski:2013gia} & - & - \\
\rule[-1ex]{0pt}{2.5ex} $65$ & Y & Y & Y & $ 60 $ & $ 0.47 $ & $ 0.25$ & - & - & - & R16 & \cite{Riess:2016jrr} & \cite{Pietrzynski:2013gia} & - & - \\

\midrule
\rowcolor{white} \bf{R16, NGC4258 + MW} \\
\rule[-1ex]{0pt}{2.5ex} $66$ & Y & Y & Y & $ 205 $ & $ 0.47 $ & $ 0.25$ & - & - & - & R16 & \cite{Riess:2016jrr} & - & - & \cite{Riess:2016jrr} \\
\rule[-1ex]{0pt}{2.5ex} $67$ & Y & Y & Y & $ 60 $ & $ 0.47 $ & $ 0.25$ & - & - & - & R16 & \cite{Riess:2016jrr} & - & - & \cite{Riess:2016jrr} \\

\midrule
\rowcolor{white} \bf{R16, NGC4258 + M31} \\
\rule[-1ex]{0pt}{2.5ex} $68$ & Y & Y & Y & $ 205 $ & $ 0.47 $ & $ 0.25$ & - & - & - & R16 & \cite{Riess:2016jrr} & - & \cite{Riess:2016jrr} & - \\
\rule[-1ex]{0pt}{2.5ex} $69$ & Y & Y & Y & $ 60 $ & $ 0.47 $ & $ 0.25$ & - & - & - & R16 & \cite{Riess:2016jrr} & - & \cite{Riess:2016jrr} & - \\

\midrule
\rowcolor{white} \bf{R16, LMC + MW} \\
\rule[-1ex]{0pt}{2.5ex} $70$ & Y & Y & Y & $ 205 $ & $ 0.47 $ & $ 0.25$ & - & - & - & R16 & - & \cite{Pietrzynski:2013gia} & - & \cite{Riess:2016jrr} \\
\rule[-1ex]{0pt}{2.5ex} $71$ & Y & Y & Y & $ 60 $ & $ 0.47 $ & $ 0.25$ & - & - & - & R16 & - & \cite{Pietrzynski:2013gia} & - & \cite{Riess:2016jrr} \\

\midrule
\rowcolor{white} \bf{R16, LMC + M31} \\
\rule[-1ex]{0pt}{2.5ex} $72$ & Y & Y & Y & $ 205 $ & $ 0.47 $ & $ 0.25$ & - & - & - & R16 & - & \cite{Pietrzynski:2013gia} & \cite{Riess:2016jrr} & - \\
\rule[-1ex]{0pt}{2.5ex} $73$ & Y & Y & Y & $ 60 $ & $ 0.47 $ & $ 0.25$ & - & - & - & R16 & - & \cite{Pietrzynski:2013gia} & \cite{Riess:2016jrr} & - \\

\midrule
\rowcolor{white} \bf{R16, MW + M31} \\
\rule[-1ex]{0pt}{2.5ex} $74$ & Y & Y & Y & $ 205 $ & $ 0.47 $ & $ 0.25$ & - & - & - & R16 & - & - & \cite{Riess:2016jrr} & \cite{Riess:2016jrr} \\
\rule[-1ex]{0pt}{2.5ex} $75$ & Y & Y & Y & $ 60 $ & $ 0.47 $ & $ 0.25$ & - & - & - & R16 & - & - & \cite{Riess:2016jrr} & \cite{Riess:2016jrr} \\

\hline
\end{tabular}}
}
\caption{$\alpha^\Cepheid$: Cepheid stars included with HPs. $\alpha^\SNe$: $\SNe$ magnitudes included with HPs. $\alpha^\Anchors$: distance moduli of anchors included with HPs; `-' stands for no distance moduli included in the fit. In columns $2$--$4$ `Y' stands for `Yes' and `N' stands for `No'. $P$: upper period cutoff. $R$: reddening law. $\sigma_{Z_W}$: standard deviation of the Gaussian prior on the metallicity parameter $Z_W$; `-' stands for a flat, wide prior. $\sigma_{\intt,i}$: internal dispersion for $\SNe$ hosts; `V' stands for varying and marginalised; when the numerical value is given it means fixed internal dispersion was used; `-' stands for no internal dispersion included in the fit. $\sigma_{\intt}^{\LMC}$: $\LMC$ internal dispersion. $\sigma_{\intt}^{\MW}$: $\MW$ internal dispersion. CS: Cepheid sample. Columns $\mu_{0,\NGC}^{\obs}$, $\mu_{0,\LMC}^{\obs}$, and $\mu_{0,\MAnd}^{\obs}$ indicate the references from which these quantities were taken; `-' means that the data was not used in the fit. $\MW$ refers to the reference for $\MW$ Cepheid stars; `-' means that the data was not used in the fit. \label{Table:details-fits}}
\end{table}

\setlength{\arrayrulewidth}{1mm}
\begin{table}[tbp]
\centering
\resizebox{\textwidth}{11cm}{
{
\rowcolors{4}{green!80!yellow!50}{green!70!yellow!30}
\begin{tabular}{$^c^c^c^c^c^c^c^c}
\hline
\\
\rowstyle{\bfseries\boldmath}  Fit & $H_0$ & $M_W$ & $b_W$ & $Z_W$ &$|| \alpha^{\Cepheid}||$ & $|| \alpha^{\SNe}||$ & $|| \alpha^{\Anchors}||$ \\
\\
\hline
\rowcolor{white} \multicolumn{8}{l}{\bf{R11, NGC4258}}     \\
$1$ & $71.2\,(5.4)$& $-3.54\,(1.24)$ & $-3.15\,(0.06)$ & $-0.285\,(0.140) $ & $ 0.72 $ & $ 0.81 $ & $ 1 $ \\
  
$2$ & $72.5\,(5.4)$& $-1.99\,(1.33)$ & $-3.25\,(0.05)$ & $-0.457\,(0.150) $ & $ 0.72 $ & $ 0.82 $ & $ 1 $\\
   
$3$ & $71.1\,(5.5)$& $-6.00\,(0.22)$ & $-3.17\,(0.06)$ & $-0.006\,(0.020) $ & $ 0.72 $ & $ 0.73 $ & $ 1 $ \\

$4$ & $70.8\,(4.2)$& $-6.01\,(0.19)$ & $-3.17\,(0.06)$ & $-0.006\,(0.020) $ & $ 0.72 $ & $ 0.72 $ & - \\

$5$ & $72.7\,(5.7)$& $-5.94\,(0.22)$ & $-3.26\,(0.05)$ & $-0.008\,(0.020) $ & $ 0.72 $ & $ 0.71 $ & $ 1 $ \\

$6$ & $72.1\,(4.2)$& $-5.95\,(0.19)$ & $-3.26\,(0.05)$ & $-0.008\,(0.020) $ & $ 0.72 $ & $ 0.75 $ & - \\
\midrule
\rowcolor{white} \multicolumn{8}{l}{\bf{R11, LMC}}  \\
$7$ & $71.3\,(4.9)$& $-3.48\,(1.16)$ & $-3.15\,(0.06)$& $-0.291\,(0.136)$ & $0.72 $ & $ 0.77 $ & $ 1 $ \\
 
$8$ & $70.1\,(4.5)$& $-2.11\,(1.28)$ & $-3.26\,(0.05)$& $-0.450\,(0.150)$ & $ 0.72 $ & $ 0.87 $ & $ 1 $ \\	
  
$9$ & $74.5\,(4.9)$& $-5.90\,(0.20)$& $-3.17\,(0.06)$& $-0.006\,(0.020)$ & $0.72 $ & $ 0.75 $ & $ 1 $ \\

$10$ & $74.4\,(4.0)$& $-5.90\,(0.18)$& $-3.17\,(0.06)$& $-0.006\,(0.020)$& $ 0.72$ & $ 0.78 $ & - \\

$11$ & $75.0\,(4.8)$& $-5.87\,(0.20)$& $-3.26\,(0.05)$& $-0.008\,(0.020)$ & $ 0.72$ & $ 0.78 $ & $ 1 $ \\
   
$12$ & $74.7\,(3.8)$& $-5.87\,(0.18)$& $-3.26\,(0.05)$& $-0.008\,(0.020)$& $ 0.72$ & $ 0.73 $ & - \\
\midrule
\rowcolor{white} \multicolumn{8}{l}{\bf{R11, MW}}  \\
$13$ & $78.1\,(4.4)$& $-3.44\,(1.25)$ & $-3.16\,(0.06)$& $-0.272\,(0.140)$ & $ 0.73 $ & $ 0.82 $ & - \\
 
$14$ & $78.3\,(4.2)$& $-2.08\,(1.19)$ & $-3.26\,(0.05)$& $-0.426\,(0.133)$ & $ 0.72 $ & $ 0.72 $ & -\\
  
$15$ & $77.4\,(4.4)$& $-5.81\,(0.18)$& $-3.17\,(0.06)$& $-0.006\,(0.020)$ & $ 0.72 $ & $ 0.64 $ & - \\

$16$ & $77.1\,(4.1)$& $-5.81\,(0.18)$& $-3.26\,(0.05)$& $-0.008\,(0.020)$ & $ 0.74 $ & $ 0.77 $ & - \\
   
\midrule
\rowcolor{white} \multicolumn{8}{l}{\bf{R11, NGC4258 + LMC}}  \\
$17$ &$ 71.1\,(4.0)$ & $-3.47\,(1.10)$& $-3.15\,(0.06)$& $-0.293\,(0.128)$ & $ 0.72 $ & $ 0.81 $ & $1 $ \\

$18$ &$ 71.2\,(4.0)$ & $-2.27\,(1.16)$& $-3.25\,(0.05)$& $-0.428\,(0.135)$ & $ 0.71 $ & $ 0.68 $ & $ 1$ \\

$19$ &$73.0\,(4.1)$ & $-5.93\,(0.18)$& $-3.17\,(0.06)$& $-0.007\,(0.020)$ & $ 0.72 $ & $ 0.75 $ & $ 0.84 $ \\

$20$ &$73.9\,(4.0)$ & $-5.89\,(0.18)$& $-3.26\,(0.05)$& $-0.008\,(0.020)$ & $ 0.72 $ & $ 0.78$ & $1 $ \\
\midrule
\rowcolor{white} \multicolumn{8}{l}{\bf{R11, NGC4258 + MW}}  \\
$21$ & $76.4\,(4.2)$&$-3.44\,(1.27)$ &$-3.18\,(0.06) $ &$-0.277\,(0.142) $ & $ 0.73 $ & $ 0.80 $ & $ 0.21 $\\
 
$22$ & $76.9\,(4.0)$&$-2.13\,(1.27)$ &$-3.27\,(0.04) $ &$-0.425\,(0.143) $ & $ 0.73 $ & $ 0.70 $ & $ 0.48 $\\
 
$23$ &$75.6\,(4.2)$ &$-5.85\,(0.18)$ &$-3.20\,(0.05) $ &$-0.006\,(0.020) $ & $ 0.72 $ & $ 0.73 $ & $ 0.41 $\\

$24$ &$75.8\,(3.9)$ &$-5.84\,(0.18)$ &$-3.27\,(0.04) $ &$-0.008\,(0.020) $ & $ 0.72 $ & $ 0.74 $ & $ 0.35 $\\
\midrule
\rowcolor{white} \multicolumn{8}{l}{\bf{R11, LMC + MW}}  \\
$25$ & $76.0\,(4.2)$ &$-4.33\,(1.16)$ &$-3.18\,(0.06) $ &$-0.178\,(0.131) $ & $0.73 $ & $0.75 $ & $ 0.12$\\

$26$ & $76.2\,(4.1)$ &$-3.10\,(1.31)$ &$-3.27\,(0.05) $ &$-0.318\,(0.147) $ & $ 0.72$ & $0.83 $ & $ 0.11$\\
 
$27$ &$76.0\,(4.0)$ &$-5.86\,(0.18)$ &$-3.19\,(0.05) $ &$-0.004\,(0.020) $ & $0.72 $ & $0.74 $ & $ 0.60$\\

$28$ &$76.1\,(3.8)$ &$-5.84\,(0.18)$ &$-3.27\,(0.04) $ &$-0.007\,(0.020) $ & $0.72 $ & $ 0.71$ & $1 $\\

\midrule
\rowcolor{white} \multicolumn{8}{l}{\bf{R11, LMC + MW + NGC4258}} \\
$\mathbf{29}$ & $\mathbf{75.0\,(3.9)}$ & $\mathbf{-5.88\,(0.18)}$ & $\mathbf{-3.20\,(0.05)}$ & $\mathbf{-0.005\,(0.020)} $ & $ \mathbf{0.72} $ & $ \mathbf{0.74} $ & $ \mathbf{0.86}$\\
$30$ & $74.9\,(3.9)$ & $-5.88\,(0.18)$ & $-3.20\,(0.05)$ & $-0.004\,(0.020) $ & $ 0.72 $ & $ 0.74$ & $0.75$\\
$31$ & $73.2\,(2.5)$ & $-5.89\,(0.18)$ & $-3.19\,(0.05)$ & $-0.004\,(0.020) $ & $ 0.72$ & - & $ 0.94$\\
$32$ & $74.1\,(3.7)$ & $-5.89\,(0.18)$ & $-3.21\,(0.05)$ & $-0.005\,(0.020) $ & $ 0.72 $ & $ 0.81$ & -\\
$33$ & $72.4\,(2.2)$ & $-5.90\,(0.17)$ & $-3.20\,(0.05)$ & $-0.004\,(0.020) $ & $ 0.71$ & - & -\\
$34$ & $75.4\,(3.7)$ & $-5.85\,(0.18)$ & $-3.27\,(0.04)$ & $-0.007\,(0.020) $ & $ 0.72 $ & $ 0.73$ & $0.76 $\\
$35$ & $72.6\,(2.4)$ & $-5.90\,(0.18)$ & $-3.26\,(0.07)$ & $-0.005\,(0.020) $ & $ 0.99$ & - & - \\
$36$ & $74.7\,(3.9)$ & $-4.68\,(0.97)$ & $-3.20\,(0.05) $ & $-0.141\,(0.110) $ & $0.72 $ & $ 0.76$ & $0.70 $ \\
$37$ & $ 75.2\,(3.8)$ & $-3.62\,(1.07)$ & $-3.27\,(0.04)$ & $-0.261\,(0.121) $ & $0.72 $ & $ 0.71$ & $0.73 $\\
$38$ & $74.7\,(3.9)$ & $-4.34\,(1.11)$ & $-3.20\,(0.05)$ & $-0.179\,(0.125) $ & $0.72 $ & $0.79 $ & $ 0.76$\\
$39$ & $75.2\,(3.9)$ & $-3.09\,(1.39)$ & $-3.27\,(0.04)$ & $-0.321\,(0.157) $ & $ 0.72 $ & $0.76 $ & $0.72 $\\
\midrule
\rowcolor{white} \multicolumn{8}{l}{\bf{R16, LMC + MW + NGC4258 + M31}} \\
$ 40 $ & $74.11\,(2.13)$ & $-5.12\,(0.72)$ & $-3.23\,(0.02)$ & $-0.08\,(0.08) $ & $0.86 $ & $ 0.83$ & $ 0.75 $\\
$ 41 $ & $74.21\,(2.16)$ & $-5.78\,(0.17)$ & $-3.23\,(0.02)$ & $-0.00\,(0.02) $ & $0.86 $ & $ 0.86$ & $ 0.81 $\\
$ 42 $ & $74.11\,(2.17)$ & $-4.89\,(0.84)$ & $-3.24\,(0.01)$ & $-0.11\,(0.10) $ & $0.86 $ & $ 0.80$ & $ 1$\\
$ 43 $ & $73.91\,(2.15)$ & $-5.49\,(0.50)$ & $-3.25\,(0.01)$ & $-0.05\,(0.06) $ & $0.86 $ & $ 0.78 $ & $0.92 $\\
$ \mathbf{44} $ & $\mathbf{73.75\,(2.10)}$ & $\mathbf{-5.10\,(0.79)}$ & $\mathbf{-3.28\,(0.01)}$ & $\mathbf{-0.10\,(0.09)} $ & $\mathbf{0.85} $ & $ \mathbf{0.83}$ & $ \mathbf{1} $\\
$ 45 $ & $74.06\,(2.12)$ & $-5.83\,(0.18)$ & $-3.24\,(0.01)$ & $-0.00\,(0.02) $ & $0.86 $ & $ 0.80$ & $ 0.76 $\\
$ 46 $ & $73.91\,(2.13)$ & $-5.86\,(0.17)$ & $-3.25\,(0.01)$ & $-0.01\,(0.02) $ & $0.86 $ & $ 0.74$ & $ 0.78$\\
$ 47 $ & $73.76\,(2.09)$ & $-5.94\,(0.18)$ & $-3.28\,(0.01)$ & $-0.00\,(0.02) $ & $ 0.86 $ & $0.81 $ & $ 0.78$\\
$ 48 $ & $73.98\,(2.21)$ & $-4.92\,(0.71)$ & $-3.23\,(0.02)$ & $-0.10\,(0.08) $ & $0.86 $ & $0.83 $ & $ 0.62 $\\
$ 49 $ & $73.83\,(2.17)$ & $-5.79\,(0.18)$ & $-3.23\,(0.02)$ & $-0.00\,(0.02) $ & $ 0.86$ & $0.88 $ & $ 0.86 $\\
$ 50 $ & $73.86\,(2.23)$ & $-4.78\,(0.78)$ & $-3.24\,(0.02)$ & $-0.12\,(0.09) $ & $0.86 $ & $0.85 $ & $0.68 $\\
$ 51 $ & $73.75\,(2.19)$ & $-4.65\,(0.89)$ & $-3.25\,(0.02)$ & $-0.14\,(0.10) $ & $ 0.86$ & $0.83 $ & $ 0.85 $\\
$ 52 $ & $73.50\,(2.20)$ & $-4.90\,(0.87)$ & $-3.28\,(0.02)$ & $-0.12\,(0.10) $ & $ 0.86$ & $0.85 $ & $ 1$\\
$ 53 $ & $73.78\,(2.18)$ & $-5.81\,(0.18)$ & $-3.24\,(0.02)$ & $-0.01\,(0.02) $ & $ 0.86$ & $ 0.84$ & $ 0.75$\\
$ 54 $ & $73.71\,(2.19)$ & $-5.86\,(0.18)$ & $-3.25\,(0.02)$ & $-0.00\,(0.02) $ & $0.86 $ & $ 0.79 $ & $ 0.78 $\\
$ 55 $ & $73.49\,(2.20)$ & $-5.95\,(0.18)$ & $-3.28\,(0.02)$ & $-0.00\,(0.02) $ & $0.86 $ & $0.79 $ & $ 0.79$\\
$ 56 $ & $73.46\,(1.40)$ & $-4.72\,(0.47)$ & $-3.28\,(0.01)$ & $-0.14\,(0.05) $ & - & - & - \\
$ 57 $ & $74.36\,(1.99)$ & $-4.81\,(0.64)$ & $-3.28\,(0.01)$ & $-0.13\,(0.07) $ & - & $0.78$ & $0.75$\\

\midrule
\rowcolor{white} \multicolumn{8}{l}{\bf{R16, NGC4258}} \\
$ 58 $ & $71.20\,(3.77)$ & $-4.84\,(0.84)$ & $-3.28\,(0.01)$ & $-0.14\,(0.10) $ & $0.85$ & $0.80$ & $1$\\
$ 59 $ & $70.33\,(4.01)$ & $-4.50\,(1.03)$ & $-3.28\,(0.02)$ & $-0.18\,(0.11) $ & $0.86$ & $0.83$ & $1$\\

\midrule
\rowcolor{white} \multicolumn{8}{l}{\bf{R16, LMC}} \\
$ 60 $ & $72.41\,(3.50)$ & $-4.73\,(0.88)$ & $-3.28\,(0.01)$ & $-0.15\,(0.10) $ & $0.85$ & $0.82$ & $1$\\
$ 61 $ & $72.68\,(3.97)$ & $-4.49\,(0.89)$ & $-3.28\,(0.02)$ & $-0.17\,(0.10) $ & $0.86$ & $0.84$ & $1$\\

\midrule
\rowcolor{white} \multicolumn{8}{l}{\bf{R16, MW}} \\

$ 62 $ & $76.16\,(2.76)$ & $-4.88\,(0.89)$ & $-3.28\,(0.01)$ & $-0.12\,(0.10) $ & $0.85$ & $0.84$ & -\\
$ 63 $ & $76.10\,(2.76)$ & $-4.48\,(1.05)$ & $-3.28\,(0.02)$ & $-0.16\,(0.12) $ & $0.86$ & $0.79$ & -\\

\midrule
\rowcolor{white} \multicolumn{8}{l}{\bf{R16, NGC4258 + LMC}} \\

$ 64 $ & $71.78\,(2.46)$ & $-4.88\,(0.80)$ & $-3.28\,(0.01)$ & $-0.13\,(0.09) $ & $0.85$ & $0.80$ & $1$\\
$ 65 $ & $71.60\,(3.27)$ & $-4.42\,(0.82)$ & $-3.28\,(0.02)$ & $-0.18\,(0.09) $ & $0.86$ & $0.80$ & $1$\\

\midrule
\rowcolor{white} \multicolumn{8}{l}{\bf{R16, NGC4258 + MW}} \\

$ 66 $ & $74.40\,(2.60)$ & $-4.85\,(0.89)$ & $-3.28\,(0.01)$ & $-0.13\,(0.10) $ & $0.86$ & $0.83$ & $0.55$\\
$ 67 $ & $74.37\,(2.53)$ & $-4.65\,(0.88)$ & $-3.28\,(0.02)$ & $-0.15\,(0.10) $ & $0.86$ & $0.77$ & $1$\\

\midrule
\rowcolor{white} \multicolumn{8}{l}{\bf{R16, NGC4258 + M31}} \\

$ 68 $ & $72.13\,(3.07)$ & $-4.94\,(0.80)$ & $-3.28\,(0.01)$ & $-0.12\,(0.09) $ & $0.85$ & $0.82$ & $0.90$\\
$ 69 $ & $71.85\,(3.21)$ & $-4.76\,(0.97)$ & $-3.28\,(0.02)$ & $-0.14\,(0.11) $ & $0.86$ & $0.79$ & $0.77$\\

\midrule
\rowcolor{white} \multicolumn{8}{l}{\bf{R16, LMC + MW}} \\

$ 70 $ & $74.56\,(2.44)$ & $-4.83\,(0.81)$ & $-3.28\,(0.01)$ & $-0.13\,(0.09) $ & $0.86$ & $0.81$ & $1$\\
$ 71 $ & $74.55\,(2.47)$ & $-4.51\,(0.85)$ & $-3.28\,(0.02)$ & $-0.16\,(0.10) $ & $0.86$ & $0.77$ & $1$\\

\midrule
\rowcolor{white} \multicolumn{8}{l}{\bf{R16, LMC + M31}} \\

$ 72 $ & $73.11\,(2.88)$ & $-4.69\,(0.76)$ & $-3.28\,(0.01)$ & $-0.15\,(0.09) $ & $0.85$ & $0.79$ & $1$\\
$ 73 $ & $72.92\,(2.78)$ & $-4.74\,(0.94)$ & $-3.28\,(0.02)$ & $-0.14\,(0.11) $ & $0.86$ & $0.85$ & $1$\\

\midrule
\rowcolor{white} \multicolumn{8}{l}{\bf{R16, MW + M31}} \\

$ 74 $ & $75.78\,(2.54)$ & $-4.70\,(0.78)$ & $-3.28\,(0.01)$ & $-0.14\,(0.09) $ & $0.86$ & $0.83$ & $1$\\
$ 75 $ & $75.66\,(2.58)$ & $-4.33\,(0.89)$ & $-3.28\,(0.02)$ & $-0.18\,(0.10) $ & $0.86$ & $0.80$ & $1$\\

\hline
\end{tabular}}}
\caption{\label{Table:Constraints-main-analysis} Constraints for fits in Table \ref{Table:details-fits}. Numbers in brackets indicate the standard deviation.}
\end{table}
\acknowledgments

We are very grateful to Adam Riess for assistance with the R16 data. We thank Filipe Abdalla for advice during the first stages of the project, at the SuperJedi Workshop 2013 organized by Bruce Bassett.
We further thank the programme `mobility in international research collaborations' funded by the German excellence initiative for travel support.
Part of the numerical work presented in this publication used the Baobab cluster of the University of Geneva.
VP acknowledges the DFG TransRegio TRR33 grant on The Dark Universe. WC is supported by the Departamento Administrativo de Ciencia, Tecnolog\'{i}a e Innovaci\'{o}n (Colombia). MK acknowledges support by the Swiss NSF.


\bibliography{H0_refs}
\bibliographystyle{JHEP}

\end{document}